\documentclass[11pt,twoside]{article}
\usepackage{fancyhdr}
\pdfoutput=1
\usepackage[colorlinks,citecolor=blue,urlcolor=blue,linkcolor=blue,bookmarks=false]{hyperref}
\usepackage{amsfonts,epsfig,graphicx}
\usepackage{afterpage}
\usepackage{pgfplots}
\usepackage{tikz}
\usetikzlibrary{arrows.meta,calc,angles}
\usepackage{nicefrac}
\usepackage{comment}
\usepackage{bbm}
\usepackage{amsmath,amssymb,amsthm} 
\usepackage{fullpage}
\usepackage[T1]{fontenc} 
\usepackage{epsf} 
\usepackage{cancel}
\newcommand{\ind}{\perp\!\!\!\!\perp}
\usepackage{graphics} 
\usepackage{amsfonts,amsmath}
\usepackage[round]{natbib}
\usepackage{psfrag,xspace}
\usepackage{color,etoolbox}
\usepackage{subcaption} 
\usepackage{listings}
\usepgfplotslibrary{groupplots}
\usepackage[noabbrev,capitalize, nameinlink]{cleveref}

\usepgfplotslibrary{fillbetween}

\def\ind{\perp\!\!\!\perp}

\newcommand{\E}{\mathbb{E}}

\usepackage{pgfplots}
\pgfmathdeclarefunction{poiss}{1}{%
  \pgfmathparse{(#1^x)*exp(-#1)/(x!)}%
}

\DeclareSymbolFont{bbold}{U}{bbold}{m}{n}
\DeclareSymbolFontAlphabet{\mathbbold}{bbold}

\newtheorem{theorem}{Theorem}
\newtheorem{lemma}{Lemma}
\newtheorem{corollary}{Corollary}
\newtheorem{proposition}{Proposition}

\theoremstyle{definition}

\theoremstyle{remark}

\newtheorem{remark}{Remark}

\definecolor{dkgreen}{rgb}{0,0.6,0}
\definecolor{gray}{rgb}{0.5,0.5,0.5}
\definecolor{mauve}{rgb}{0.58,0,0.82}
\definecolor{brightblue}{HTML}{00BFC4}

\newcommand\blfootnote[1]{%
  \begingroup
  \renewcommand\thefootnote{}%
  \footnote{#1}%
  \addtocounter{footnote}{-1}%
  \endgroup
}

\pgfplotsset{compat=1.18}
\begin{document}

\def\spacingset#1{\renewcommand{\baselinestretch}%
{#1}\small\normalsize} \spacingset{1}

\raggedbottom
\allowdisplaybreaks[1]


  \title{\vspace*{-.4in} {Causal Geodesy: Counterfactual Estimation Along the Path Between Correlation and Causation}}
  \author{\\ $\text{Kyle Schindl}^{\S}$, $\text{Larry Wasserman}^{\dagger \ddag}$ \\ \\
    $^{\S}$Department of Statistics \\
    Iowa State University \\
    \texttt{kschindl@iastate.edu} \\ \\ 
    $^\dagger$Department of Statistics \& Data Science \\
    $^\ddag$Machine Learning Department \\
    Carnegie Mellon University \\
    \texttt{larry@stat.cmu.edu}
\date{}
    }

  \maketitle
  \blfootnote{Accompanying \texttt{R} code is available via \href{https://github.com/kyleschindl/causal-geodesy}{https://github.com/kyleschindl/causal-geodesy}}
  \thispagestyle{empty}

\begin{abstract}
{ We introduce \textit{causal geodesy}, a framework for studying the landscape of stochastic interventions that lie between the two extremes of performing no intervention and performing a sharp intervention that sets an exposure equal to a specific value. We define this framework by constructing paths of distributions that smoothly interpolate between the treatment density and a point mass at the target intervention. Thus, each path starts at a purely observational (or correlational) quantity and moves into a counterfactual world. Of particular interest are paths that correspond to geodesics in some metric, i.e.\ the shortest path. We show that stochastic interventions defined by the Wasserstein geodesic possess many favorable statistical properties. In particular, unlike other paths (such as those defined by exponentially tilted intervention distributions), the Wasserstein geodesic does not introduce path-parameterization artifacts; its curvature is inherited from the outcome regression itself. We use this to develop higher-order bias-corrected inference for sharp interventions. Finally, we explore geodesics defined by the spherical Hellinger geometry to show how the interpretation, estimation, and statistical difficulty of the problem is governed by the underlying geometry.}
\end{abstract}


\bigskip 
\spacingset{1.1}

\section{Introduction}

The goal of causal inference is to understand the effect of an intervention on a stochastic system. Typically, we observe an outcome $Y$, a (possibly continuous) exposure $A$, and covariates $X$ with joint density $p(x,a,y) = p(x) \pi(a \! \mid \! x) p(y \! \mid \! x,a)$. We say $A \mid X = x \sim \pi( \, \cdot \mid x)$ is the natural treatment value \citep{YoungHernanRobins+2014+1+19}, i.e.\ the observed treatment level assigned conditional on $X = x$. Practitioners are then often interested in understanding the counterfactual outcome $Y(a)$, which corresponds to the \textit{sharp intervention} under which we deterministically set $A = a$; conceptually, this replaces the natural treatment density $\pi(\, \cdot \mid x)$ with a point mass at $a$ \citep{rubin1974estimating, ROBINS19861393}. Under the standard identifiability conditions of consistency, exchangeability, and positivity, the dose-response curve is identified as
\begin{align*}
    \mathbb{E}[Y(a)] = \int_{\mathcal{X}}\mu(x,a)dP_X(x),
\end{align*}
where $\mu(x, a) = \mathbb{E}[Y \mid X = x, A = a]$. However, in many practical applications, enforcing a single fixed exposure for every unit can be too rigid or unrealistic. Consequently, it is becoming increasingly common to instead consider \textit{stochastic interventions} that replace the natural treatment density $\pi(a \! \mid \! x)$ with an intervention density, $\rho(a \! \mid \! x)$ \citep{hubbard2008, diaz2012}. Then, instead of deterministically setting treatment such that $A = a$, we suppose $A \mid X = x \sim \rho( \, \cdot \mid x)$ in order to explore the counterfactual world under which treatment was drawn from an alternate assignment mechanism.  Then, the counterfactual quantity of interest is
\begin{align} \label{stochastic_expectation}
   \psi_\rho  = \int_{\mathcal{X}} \int_{\mathcal{A}}  \mathbb{E}[Y(a) \mid X = x] \rho(a \! \mid \! x)da \, d P_X(x).
\end{align}
Typically, such stochastic interventions require weaker identifying assumptions (particularly with respect to positivity) since they represent a less extreme counterfactual intervention than the sharp intervention \citep{kennedy2019nonparametric}. In fact, if we perform no intervention, then $\rho(a \! \mid \! x) = \pi(a \! \mid \! x)$ and $\psi_\rho$ reduces to the mean of $Y$, which can be estimated without any causal assumptions. Alternatively, we can recover the sharp intervention $\mathbb{E}[Y(a)]$ by taking $\rho( \, \cdot \mid x)$ to be a point mass at $a$. Thus, we can think of stochastic interventions as interventions that describe the space \textit{between} the null intervention and the sharp intervention ``set $A=a$,'' that is, the space between correlation and causation. 

In this paper, we introduce a framework called {\em causal geodesy} for studying the landscape between these two extremes. In the sciences, ``geodesy'' refers to the study of fundamental properties of the Earth, such as its shape, orientation, and gravitational field. Analogously, we view each intervention distribution as a point on a metric space of probability measures \citep{ambrosio2008gradient, Villani2009OptimalTransport, amari2016information}, which allows us to use its underlying geometry to study the corresponding causal effects. We construct paths of stochastic interventions,
\begin{align*}
    \rho_t(a \! \mid \! x):\ 0 \leq t < 1,
\end{align*}
such that $\rho_0(a \! \mid \! x) = \pi(a \! \mid \! x)$ and $\rho_t(\, \cdot \! \mid \! x)$ converges to the point mass $\delta_{a_\star}$ as $t \to 1$, for some chosen intervention $A = a_\star$. Each path defines a smooth flow in the space of distributions, carrying us from pure observation to a sharp causal intervention. Of particular interest are paths that correspond to geodesics with respect to some metric, i.e. the shortest path between the two extremes. We show that geodesic interventions possess many favorable properties in the estimation and interpretation of causal effects. 

The remainder of the paper is organized in the following manner: In \cref{notation_section} we discuss the causal assumptions used and provide a review of stochastic interventions, geodesics, and optimal transport theory. In \cref{causal_geodesy_section}, under the Wasserstein geometry, we formally establish the causal geodesy framework and its properties. Specifically, in \cref{curvature_section} we explore the curvature of incremental effects across $t$; we show that some paths introduce artificial curvature due to the way they are parameterized, but the Wasserstein geodesic does not introduce any parameterization artifacts. In \cref{efficiency_theory_section} we derive the efficient influence function for our estimand and related efficiency theory; we show that as $t \to 1$ stochastic interventions act as a kernel-smoothed dose-response estimator. In \cref{asymptotics_section} we establish asymptotic theory and rates of convergence for our estimator that allow for $t \to 1$, even under an exploding variance. Furthermore, we explore the transition between parametric and nonparametric behavior of incremental effects under continuous exposures and establish a minimax rate that depends on the smoothness of the regression function. In \cref{endpoint_section} we derive higher-order bias-corrected estimators of sharp interventions by leveraging the curvature properties established in \cref{curvature_section}. Then, in \cref{hellinger_section} we explore geodesics in other geometries by considering effects defined by the Hellinger geodesic between probability distributions. Finally, in \cref{simulation_section} we validate our theoretical findings via simulation and a real-world application and in \cref{conclusion_section} we discuss our results and possible directions for future work.

\section{Background} \label{notation_section}

In this section we provide a brief overview of the causal assumptions used in this manuscript and a review of stochastic interventions, paths and geodesics, and optimal transport theory in order to provide a proper background for establishing the causal geodesy framework.

\subsection{Causal Inference and Stochastic Interventions} \label{causal_background}
Let $(Z_1, \ldots, Z_n) \overset{iid}{\sim} P$ denote a sample of independent and identically distributed observations, where $Z_i = (X_i, A_i, Y_i)$. Here, $X_i \in \mathcal{X}$ are the covariates, $A_i$ is some continuous exposure with compact support $\mathcal{A} \subset \mathbb{R}$, and $Y_i \in \mathbb{R}$ is the outcome. For simplicity, throughout the paper we assume $\mathcal{A} = [0, 1]$; however, this is not essential for our results. It simply allows us to avoid making tedious tail conditions; analogous results hold for $\mathcal{A} = \mathbb{R}$ under suitable regularity conditions. We use the potential outcomes framework, such that $Y(a)$ denotes the outcome that would be observed under $A = a$ \citep{rubin1974estimating}. Let $\pi(a \! \mid \! x)$ be the conditional density of $A$ given $X = x$; this can be thought of as a generalized propensity score, in the spirit of \cite{imai_vandyk2004} and \cite{hirano2004propensity}. As opposed to traditional causal inference estimands, such as the dose-response curve $\mathbb{E}[Y(a)]$, which are defined by static, deterministic interventions in which every subject receives a fixed treatment level $A = a$ \citep{kennedy2017cte}, we consider causal effects defined by stochastic interventions. Broadly speaking, stochastic interventions fall into a larger literature of causal interventions that depend on or modify the natural treatment value, such as modified treatment policies \citep{haneuse2013, Diaz03042023} or natural-value interventions \citep{YoungHernanRobins+2014+1+19}. The following proposition, analogous to the $g$-formula \citep{ROBINS19861393}, establishes the conditions under which we may identify the stochastic intervention effect.
\begin{proposition}[Identification] \label{identification_proposition}
    Suppose the following assumptions hold: $(i)$ \textit{Consistency:} $Y = Y(a)$ if $A = a$, $(ii)$ \textit{Exchangeability:} $A \ind Y(a) \mid X$ for $a \in \mathcal{A}$, and $(iii)$ \textit{Conditional intervention support:} For $P_X$-almost every $x$, the intervention distribution $\rho(\,\cdot\mid x)$ is absolutely continuous with respect to the observational treatment distribution $P_{A\mid X=x}$. Then, \cref{stochastic_expectation} is identified as
\begin{align*}
    \psi_\rho =  \int_{\mathcal{X}} \int_{\mathcal{A}}  \mu(x, a) \rho(a \! \mid \! x) da \,  dP_X(x),
\end{align*}
where $\mu(x, a) = \mathbb{E}[Y \mid X = x, A = a]$ is the regression function.
\end{proposition}

Stochastic interventions have a number of practical benefits over traditional static interventions. First, as shown by \cref{identification_proposition}, they require weaker assumptions for identification when $\rho(a \! \mid \! x)$ is absolutely continuous with respect to $\pi(a \! \mid \! x)$. In particular, deterministically setting a treatment level for all subjects requires the positivity assumption: that is, the natural treatment density must be positive at all treatment levels of interest, i.e. $\pi(a \! \mid \! x)  \geq \varepsilon > 0$ for all $(x, a) \in \mathcal{X} \times \mathcal{A}$. In practice positivity is often violated, especially when treatment is continuous, and even near violations can cause the variance of causal estimators to blow up \citep{busso2014new, branson2023causal}. Second, many interventions are naturally stochastic. For example, when considering continuous exposures such as air pollution, medical dosages, or economic policies, an intervention that modifies the distribution of exposure is more realistic than an intervention that assigns a single exposure level to all units \citep{Diaz03042023, huang2026multivariateincrementaleffectscontinuous}. Third, stochastic interventions are often highly interpretable and can generalize sharp interventions \citep{kennedy2019nonparametric}. When the exposure $A$ is continuous, a commonly chosen stochastic intervention is the exponentially tilted intervention distribution,
\begin{align} \label{exponential_tilt_def}
    q_\delta(a \! \mid \! x) = \frac{\exp(\delta a) \pi(a \! \mid \! x)}{\int_\mathcal{A} \exp(\delta a) \pi(a \! \mid \! x) da},
\end{align}
where now $\delta \in (-\infty, \infty)$. Notably, $q_\delta(a \! \mid \! x) = 0$ whenever $\pi(a \! \mid \! x) = 0$, so it does not assign mass outside of the natural treatment support and does not require a strong form of positivity for identification. Because of its many convenient properties, the exponentially tilted intervention distribution has recently been explored in mediation analysis \citep{diaz2020causal}, for continuous exposures \citep{schindl2024incremental}, multivariate exposures \citep{huang2026multivariateincrementaleffectscontinuous}, local effects \citep{rakshit2024localeffectscontinuousinstruments}, sensitivity analysis \citep{levis2024stochastic}, and difference-in-differences \citep{jetsupphasuk2026differenceindifferencesstochasticpolicyshifts}.

More generally, the exponentially tilted intervention distribution $q_\delta(a \! \mid \! x)$ falls under the causal geodesy framework because it can be viewed as a path between distributions. For example, when $\delta = 0$ then no intervention is performed and (when the treatment support is bounded) as $\delta \to \infty$  then $q_\delta(a \! \mid \! x)$ becomes a point-mass at the edge of the treatment support. Furthermore, as shown in \cite{schindl2024incremental}, one can construct a path to interior points by defining a ``reflected'' exponential tilt that combines negative and positive tilts. However, this path is just one of many we could take. In the next section, we review paths and geodesics in the space of probability measures --- this will provide us with the language and framework necessary to distinguish the costs and benefits of each type of intervention path.

\subsection{Review of Paths and Geodesics} \label{geodesic_introduction}

Suppose that $P$ and $Q$ are two probability distributions with support $\Omega \subseteq \mathbb{R}^d$. We may describe the family of absolutely continuous paths between $P$ and $Q$ by a time-dependent density $\rho_t(x)$, where $t \in [0, 1]$, such that $\rho_0 = P$ and $\rho_1 = Q$. A path is said to be absolutely continuous if there exists a function
$m\in L_1$ such that, for all $0 \leq s \leq t \leq 1$, $W_2(\rho_s,\rho_t)\leq \int_s^t m(u) du$, where $W_2$ is the 2-Wasserstein distance; that is, $W^2_2(P, Q)$ is the minimum of $\E[\|Y - X\|^2_2]$ over all joint distributions for $(X,Y)$ such that $X\sim P$ and $Y\sim Q$ \citep{Villani2009OptimalTransport}. Furthermore, an absolutely continuous path must satisfy the continuity equation
\begin{align} \label{continuity_equation}
    \frac{\partial}{\partial t} \rho_t + \nabla \cdot (\rho_t \vec{v}_t) = 0
\end{align}
for some velocity field $\vec{v}_t(x)$, which describes how the probability mass moves at time $t$ \citep{ambrosio2008gradient}.  The continuity equation ensures that probability mass is neither created nor destroyed as it is transported from $P$ to $Q$. In particular, we focus on paths that minimize the distance between $P$ and $Q$ with respect to some metric, i.e. a geodesic.
\begin{figure}[h]
    \centering
\begin{tikzpicture}[scale=0.8, every node/.style={font=\small}]

\fill[gray!20, opacity=0.4] 
    plot[smooth cycle, tension=0.5] 
    coordinates {(0.2, -0.2) (3.7, 2.05) (6.45, 3.55) (2.2, 5.3) (-3.8, 2.3) (-2.3, 1.3)} 
    [shift={(0.2, -0.2)}];

\fill[white, thick, draw=black!70] 
    plot[smooth cycle, tension=0.5] 
    coordinates {(0, 0) (3.5, 2.25) (6.25, 3.75) (2, 5.5) (-4, 2.5) (-2.5, 1.5)};

\node[font=\Large] at (-0.1, 0.9) {$\mathcal{S}$};

  \filldraw[black] (-1, 2.8) circle (2pt) node[above, xshift=-7pt, yshift=1pt] {$P$};
  \filldraw[black] (4,4) circle (2pt) node[right] {$Q$};

\draw[thick, black, ->] (-1, 2.8) .. controls (0, 3.5) and (2, 4.5) .. (3.9, 4.01) 
    node[midway, above, yshift=2pt] {$\nu_t$};

\fill[white, thick, draw=black!70] 
    (-3.2, 7.5)
    .. controls (-2.9, 7.6) and (-1.0, 7.6) .. (-0.7, 7.5)  
    .. controls (-0.5, 7.3) and (-0.5, 5.2) .. (-0.7, 5.0)
    -- (-1.3, 5.0)
    -- (-1.3, 3.6)
    -- (-2, 5.0)
    -- (-3.2, 5.0)
    .. controls (-3.4, 5.3) and (-3.4, 7.3) .. (-3.2, 7.5)
    -- cycle;

\draw[thick, black, ->] (-3.2, 5.3) -- (-0.7, 5.3) node[above, xshift=-4pt] {$x$};
\draw[thick, black, -> ] (-2.8, 5.3) -- (-2.8, 7.4) node[right, yshift=-3pt] {$p(x)$};

\draw[thick, blue] plot[domain=-3.2:-0.7, samples=100] 
    (\x, {5.3 + 1.6 * (exp(-(\x + 2.5)^2 / 0.1) + 0.6 * exp(-(\x + 1.5)^2 / 0.125))});

\fill[white, thick, draw=black!70] 
    (4.3, 8.0)  
    .. controls (4.6, 8.1) and (6.5, 8.1) .. (6.8, 8.0)  
    .. controls (7.0, 7.8) and (7.0, 5.7) .. (6.8, 5.5)  
    -- (5.95, 5.5)  
    -- (4.65, 4.4)  
    -- (5.15, 5.5)  
    -- (4.3, 5.5)   
    .. controls (4.1, 5.8) and (4.1, 7.8) .. (4.3, 8.0)  
    -- cycle;

\draw[thick, black, ->] (4.3, 5.8) -- (6.8, 5.8) node[above, xshift=-4pt] {$x$};  
\draw[thick, black, ->] (4.7, 5.8) -- (4.7, 7.9) node[right, yshift=-3pt] {$q(x)$};  

\draw[thick, blue] plot[domain=4.3:6.8, samples=50] 
    (\x, {5.8 + 1.6 * (1 / (2 * 3.14 * 0.15)^(1/2)) * exp(-(\x - 5.5)^2 / (2 * 0.15))});  

\end{tikzpicture}
    \caption{Geodesic between two distributions on a statistical manifold $\mathcal{S}$.}
    \label{fig:enter-label}
\end{figure}

Geodesics are curves that define the shortest path between two points on a manifold with respect to a chosen metric. In the context of probability distributions, distances are typically measured using the Wasserstein, Fisher--Rao, or Hellinger metrics --- each of which defines a different notion of shortest path. These are special paths of shortest length which we focus on in this paper. To provide intuition for geodesics in probability spaces, we consider those induced by the Wasserstein distance, which arises in optimal transport theory and describes how probability mass can be reallocated in the most efficient way. A fundamental result from \cite{benamou2000computational} shows that the squared Wasserstein distance may also be written as
\begin{align*}
W^2_2(P, Q) = \inf_{\rho_t,\vec v_t} \left\{ \int^1_0 \|\vec{v}_t\|^2_{\rho_t} dt : 
\frac{\partial}{\partial t} \rho_t + \nabla \cdot (\rho_t \vec{v}_t) = 0,  \rho_0 = P, \rho_1 = Q\right\}.
\end{align*}
This variational formulation shows that the squared Wasserstein distance can be written as the minimal total kinetic energy required to transport $P$ to $Q$ along some continuous flow $\rho_t(x)$, which, importantly, must satisfy \cref{continuity_equation}. When the following limit exists in $L_2(\rho_t)$, the velocity field can be represented as
\begin{align*}
\vec{v}_t(x) = \underset{\varepsilon \to 0}{\text{lim}} \left\{ \frac{T_\varepsilon(x) - x}{\varepsilon} \right\},
\end{align*}
where $T_\varepsilon$ is the optimal transport map from $\rho_t(x)$ to $\rho_{t + \varepsilon}(x)$. It can be shown that $W^2_2(P, Q)$ is minimized, yielding the shortest path between $P$ and $Q$ in a Wasserstein space. We say a transport map $T$ is an \emph{optimal transport map} if $\int |T(x)-x|^2_2 dP_X(x) = W_2^2(P,Q)$. As shown by \cite{mccann1997convexity}, this path is defined by the interpolation
\begin{align} \label{wasserstein_geodesic_definition}
\nu_t := \big((1 - t) \text{id} + t T \big)_{\#} P,
\end{align}
where $\text{id}(\cdot)$ is the identity map. Here, for a measurable function $f : \Omega \to \Omega$,  $f_{\#} P$ is the \textit{push-forward}. That is, for all measurable $S \subseteq \Omega$,  $(f_{\#}P)(S) = P(f^{-1}(S))$, which denotes the distribution of $f(X)$ where $X\sim P$. We refer the reader to \cite{ambrosio2008gradient} and \cite{chewi2024statistical} for more detail regarding geodesics in the space of probability measures.

In the geodesy framework, for $t \in [0, 1)$ we let $\rho_t(a \! \mid \! x)$ represent an arbitrary path of stochastic interventions such that $\rho_0(a \! \mid \! x) = \pi(a \! \mid \! x)$ and $\rho_t(\, \cdot  \! \mid \! x)$ converges to the point mass $\delta_{a_{\star}}$ as $t \to 1$. However, when a path is specifically a geodesic with respect to some metric, we use $\nu_t(a \! \mid \! x)$ to denote it. Now that we have covered geodesics in Wasserstein spaces and optimal transport maps, in the next section we describe how these concepts can be used to describe causal interventions.

\section{Causal Geodesy} \label{causal_geodesy_section}

In general, consider a path of densities that smoothly evolve over time, beginning at the treatment density $\pi(a \! \mid \! x)$ and gradually transitioning toward a point mass at $a_{\star}$. Every such path can be thought of as a set of stochastic interventions that interpolates between ``no intervention'' and the target intervention. We refer to the study of such paths as {\em causal geodesy}. Of particular interest are the shortest paths between our baseline and target distribution, i.e. geodesics. Specifically, we focus on the Wasserstein geodesic; following the discussion and definitions outlined in \cref{geodesic_introduction}, when the target distribution is a point mass, the Wasserstein geodesic has a simple form, which we define in the following lemma. Henceforth, we refer to this as the \textit{Wasserstein intervention}.

\begin{lemma}[Wasserstein Geodesic] \label{wasserstein_geodesic_density_point_mass}
Let $\mathcal{A}_t = \left\{(1-t)u+ta_\star:u\in\mathcal{A} \right\}$ denote the transported intervention support and let $\delta_{a_{\star}}$ be our target distribution. Then, for $t \in [0, 1)$, 
\begin{align*}
    \nu_t(a \! \mid \! x) = \frac{1}{1 - t} \pi \! \left( \frac{a - t a_{\star}}{1 - t} \mid x\right) \mathbb{I}(a\in\mathcal{A}_{t}).
\end{align*}
This intervention
corresponds to moving a subject
from their natural value $A$
to $(1-t) A + t a_{\star}$.
\end{lemma}

Throughout the rest of the manuscript we extend $\pi(\, \cdot\mid x)$ by zero outside $\mathcal{A}$ so that $\nu_t(a \mid x) = 0$ whenever $a\notin\mathcal{A}_{t}$; consequently, we suppress the support indicator in the definition of $\nu_t(a \mid x)$ when no ambiguity can arise. The Wasserstein intervention has a number of properties that make it convenient to use. Similar to exponentially tilted intervention distributions, $\nu_t(a \! \mid \! x)$ is a smooth function of $\pi(a \! \mid \! x)$ and provides a smooth interpolation between the observational and interventional regimes, as illustrated in \cref{geodesic_comparison}. However, the Wasserstein intervention provides a more interpretable parametrization of the transition between distributions.  Under exponentially tilted intervention distributions, the tilting parameter $\delta$ represents the rate of change in log-likelihood ratios, i.e.
\begin{align*}
    \delta = \frac{\partial}{\partial a} \left\{ \text{log}\left(\frac{q_\delta(a \! \mid \! x)}{\pi(a \! \mid \! x)} \right) \right\},
\end{align*}
which can be difficult to interpret as an intervention on the treatment itself \citep{schindl2024incremental}. Meanwhile, the Wasserstein intervention provides a consistent interpretation where $t \in [0, 1)$ represents moving each unit a fraction $t$ of the way from the observed treatment value $A$ to the target value $a_{\star}$. However, this cleaner interpretation comes with a drawback; under the Wasserstein intervention, $\nu_t(a \! \mid \! x)$ will place mass on zero-density portions of $\pi(a \! \mid \! x)$ if they exist. Consequently, global positivity is required for identification.
\begin{proposition}[Identification of the Incremental Wasserstein Effect] \label{identification_wass_effect} Suppose that conditions $(i)$ and $(ii)$ of \cref{identification_proposition} hold. Furthermore, suppose that for $P_X$-almost every $x$, the transported treatment distribution is absolutely continuous with respect to $P_{A\mid X=x}$. Equivalently, in terms of the treatment density, $\pi( \frac{a - ta_\star}{1 - t} \mid x )>0 \Longrightarrow \pi(a\mid x)>0$ for Lebesgue-almost every $a\in\mathcal{A}_t$. Then, the Wasserstein incremental effect is identified as
\begin{align} \label{continuous_wass_effect}
    \psi_\nu(t) = \int_{\mathcal{X}} \int_{\mathcal{A}} \mu(x, a) \nu_t(a \! \mid \! x)  da \, dP_X(x).
\end{align}
\end{proposition}

The overlap condition in \cref{identification_wass_effect} is the minimal positivity condition needed for identification of $\psi_\nu(t)$. However, throughout the rest of the manuscript we often assume the global lower bound $\pi(a\mid x)\geq\pi_{\min}>0$ for all $(x,a)\in\mathcal{X}\times\mathcal{A}$, which is stronger than necessary for identification, but greatly simplifies many proofs and results.

\begin{figure}[ht]
    \centering
    \begin{tikzpicture}
        \definecolor{brightblue}{HTML}{00BFC4}

        \pgfplotstableread[col sep=comma]{files/wasserstein_altA_faint_light.csv}\faintdata
        \pgfplotstableread[col sep=comma]{files/wasserstein_altA_selected_light.csv}\seldata

        \begin{axis}[
            width=0.78\textwidth,
            height=0.50\textwidth,
            xmin=0, xmax=1,
            ymin=0, ymax=26,
            xtick={0,0.25,0.50,0.75,1.00},
            ytick={0,5,10,15,20,25},
            grid=both,
            grid style={line width=0.1pt, draw=gray!20, opacity=0.5},
            xlabel={$a$},
            ylabel={$\nu_t(a \! \mid \! x)$},
            ylabel style={yshift=-0cm},
            xtick pos=bottom,
            ytick pos=left,
            legend cell align=left,
            legend image post style={sharp plot,-, very thick},
            legend style={
                at={(0.98,0.98)},
                anchor=north east,
                fill=white,
                fill opacity=0.95,
                text opacity=1,
                draw=black!40,
                font=\small
            }
        ]

        \addplot[color=gray!65, opacity=0.75, line width=0.8pt, forget plot]
            table[x="a", y="t_10", col sep=comma] {\faintdata};

        \addplot[color=gray!65, opacity=0.75, line width=0.8pt, forget plot]
            table[x="a", y="t_30", col sep=comma] {\faintdata};

        \addplot[color=gray!65, opacity=0.75, line width=0.8pt, forget plot]
            table[x="a", y="t_50", col sep=comma] {\faintdata};

        \addplot[color=gray!65, opacity=0.75, line width=0.8pt, forget plot]
            table[x="a", y="t_70", col sep=comma] {\faintdata};

        \addplot[color=gray!65, opacity=0.75, line width=0.8pt, forget plot]
            table[x="a", y="t_85", col sep=comma] {\faintdata};

        \addplot[color=gray!65, opacity=0.75, line width=0.9pt, forget plot]
            table[x="a", y="t_20", col sep=comma] {\seldata};

        \addplot[color=gray!65, opacity=0.75, line width=0.9pt, forget plot]
            table[x="a", y="t_40", col sep=comma] {\seldata};

        \addplot[color=gray!65, opacity=0.75, line width=0.9pt, forget plot]
            table[x="a", y="t_60", col sep=comma] {\seldata};

        \addplot[color=gray!65, opacity=0.75, line width=0.9pt, forget plot]
            table[x="a", y="t_80", col sep=comma] {\seldata};

        \addplot[color=black, line width=1.15pt]
            table[x="a", y="t_00", col sep=comma] {\seldata};
        \addlegendentry{$t = 0$}

        \addplot[color=gray!65, opacity=0.75, line width=0.9pt]
            table[x="a", y="t_50", col sep=comma] {\faintdata};
        \addlegendentry{$t\in(0,0.90)$}

        \addplot[color=brightblue, line width=1.4pt]
            table[x="a", y="t_90", col sep=comma] {\seldata};
        \addlegendentry{$t = 0.90$}

        \addplot[dashed, black, line width=0.8pt, forget plot]
            coordinates {(0.27,0) (0.27,26)};
        \node[anchor=west] at (axis cs:0.278,25.2) {};

        \end{axis}
    \end{tikzpicture}
    \caption{
    Wasserstein geodesic from $\pi(a \! \mid \! x)$ to the interior point mass $\delta_{a_{\star}}$.
    }
    \label{geodesic_comparison}
\end{figure}

Intuitively, $\psi_\nu(t)$ represents the average outcome in the counterfactual world where, conditional on $X=x$, treatment is drawn from the Wasserstein intervention density $\nu_t(\, \cdot\mid x)$ rather than from the natural treatment density $\pi(\, \cdot\mid x)$. Since the post-intervention treatment value is $(1-t)A+t a_{\star}$, when $t = 0$, no intervention is performed, and thus $\psi_\nu(0)= \mathbb{E}[\mu(X,A)]= \mathbb{E}[Y]$. Then, as $t\to 1$, the intervention distribution concentrates at $a_{\star}$, so that
\begin{align*}
    \psi_\nu(t) = \mathbb{E}[\mu(X,(1-t)A+ta_{\star})] \longrightarrow \mathbb{E}[\mu(X,a_{\star})] = \mathbb{E}[Y(a_{\star})],
\end{align*}
assuming continuity of $a \mapsto \mu(x,a)$. Therefore, the curve $t \mapsto \psi_\nu(t)$ interpolates between the purely observational mean and the sharp causal intervention, with intermediate values of $t$ corresponding to partial movement toward the target treatment level. One important benefit of this framework is that the intervention defines an entire effect curve across $t \in [0, 1)$, rather than a single contrast. Consequently, understanding the curvature of $\psi_\nu(t)$ can lead to a deeper understanding of the intervention effect, which we explore in the following section.

\subsection{Path-Induced Curvature} \label{curvature_section}

Under a continuous exposure, average derivative effects are often of particular interest because they describe the local behavior of a causal estimand, rather than comparing two fixed counterfactual worlds such as a contrast between treatment levels. In many applications, a small change in treatment may be more realistic than the large deterministic interventions defined by dose-response curves. Classical approaches typically summarize the local sensitivity of the outcome regression with respect to treatment; this framework has a long history in semiparametric econometric theory \citep{hardle_stoker_89, powell89, newey93}. More recently, average derivative effects have been characterized by an infinitesimal shift intervention, where we evaluate average outcomes under the intervention $A_\varepsilon = A + \varepsilon$ for some $\varepsilon > 0$. Then, one may consider the causal effect defined by
\begin{align} \label{avg_derivative_effect}
    \left. \frac{\partial}{\partial \varepsilon} \Big\{ \mathbb{E}[\mu(X, A + \varepsilon)] \Big\} \right|_{\varepsilon=0} = \mathbb{E}\! \left[ \frac{\partial}{\partial a}\mu(X,A) \right].
\end{align}
These types of average derivative effects have been explored extensively; interested readers should refer to \cite{rothenhausler2020incrementalcausaleffects, athey2021, mcclean2024nonparametric, zhou2025marginalinterventionaleffects, zhang2025doublyrobustinferencecausal, hines26} and \cite{klyne2026}. In this section, we take a complementary perspective: we consider derivatives along the entire incremental effect curve, as indexed by $t$.

Let $\rho_t$ denote an arbitrary path and let $\psi_\rho(t) = \int_{\mathcal{X}} \int_{\mathcal{A}} \mu(x, a) \rho_t(a \! \mid \! x) da \, dP_X(x)$ denote the associated incremental effect. For paths whose densities can be differentiated under the integral, we consider derivative effects of the form
\begin{align} \label{incremental_curve_derivative}
    \frac{\partial^k}{\partial t^k} \big\{ \psi_\rho(t) \big\} =  \mathbb{E}\! \left[ \int_{\mathcal{A}} \mu(X,a) \left\{ \frac{\partial^k}{\partial t^k}\rho_t(a \! \mid \! X) \right\} da \right],
\end{align}
for $k \in \mathbb{N}$. There is an important distinction between the derivative effects defined by \cref{avg_derivative_effect} and \cref{incremental_curve_derivative}: namely, $\psi^\prime_\rho(t)$ describes a derivative with respect to the \textit{intervention path}, not necessarily treatment itself. This can be troublesome because in practice it is common to interpret the curvature along $t$ as some kind of intermediate causal effect. However, it may be the case that some curvature is induced by the path itself, and not the treatment effect (as measured by changes in the outcome regression). For example, if $\mu(x, a)$ is affine in $a$, then we would hope that $\psi_\rho(t)$ is also affine, although this is not necessarily true. In the following theorem, we establish a curvature decomposition for the incremental effect curve under arbitrary paths $\rho_t$.

\begin{theorem}[Curvature Decomposition] \label{curvature_decomposition}

For $P_X$-almost every $x$ and $t \in (0, 1)$, let $\rho_t(\, \cdot\mid x)$ be a path of probability densities that satisfy the continuity equation
\begin{align*}
    \frac{\partial}{\partial t} \rho_t(a \! \mid \! x) + \frac{\partial}{\partial a} \big\{ \rho_t(a \! \mid \! x) \vec{v}_t(a, x) \big\} = 0,
\end{align*}
where $\vec{v}_t(a, x)$ is a scalar velocity field. Furthermore, suppose that the no-flux condition holds on $\mathcal{A} = [0, 1]$, i.e.\ $ \rho_t(0 \! \mid \! x) \vec{v}_t(0, x) = \rho_t(1 \! \mid \! x) \vec{v}_t(1, x) = 0$, which ensures no probability mass flows outside the support boundaries. Next, suppose that $a \mapsto \mu(x, a)$ is twice continuously differentiable, that $(t, a) \mapsto \vec{v}_t(a, x)$ and $(t, a) \mapsto \rho_t(a \! \mid \! x)$ are continuously differentiable in $t$ and $a$, and that integrable envelopes exist for all functions. Then,
\begin{align*}
    \psi''_\rho(t) = \underbrace{\mathbb{E}_{\rho_t} \! \left[ \vec{v}_t(A, X)^2 \frac{\partial^2 }{\partial u^2} \mu(X, u) \Big|_{u = A} \right]}_{\text{response surface curvature}}+ \underbrace{\mathbb{E}_{\rho_t} \!\left[ \vec{a}_t(A, X) \frac{\partial }{\partial u} \mu(X, u) \Big|_{u = A} \right]}_{\text{path-induced curvature}},
\end{align*}
where $\vec{a}_t(a, x) = \frac{\partial}{\partial t} \vec{v}_t(a, x) + \vec{v}_t(a, x) \frac{\partial}{\partial a} \vec{v}_t(a, x)$ is the acceleration of the path and for an integrable function $f$, we define $\mathbb{E}_{\rho_t}[f(X, A)] = \int_{\mathcal{X}}\int_{\mathcal{A}} f(x, a)\rho_t(a\mid x) da\,dP_X(x)$.
\end{theorem}

In \cref{curvature_decomposition} we let $t \in (0, 1)$ only to avoid dealing with one-sided derivatives at the boundaries. Furthermore, we use the vector notation $\vec{v}_t$ for the velocity field even though it is a scalar in order to visually distinguish it from the Wasserstein geodesic $\nu_t$. By \cref{curvature_decomposition} we can see that there are two important sources of curvature along $\psi_\rho(t)$: that which is induced by the outcome regression and that which is induced by the choice of path itself. The latter is driven by the acceleration of the path through the treatment space, $\vec{a}_t(a, x)$. Importantly, some choices of the treatment path may induce \textit{artificial acceleration}, that is, nonzero acceleration that is an artifact of the path parameterization. For example, consider the exponentially tilted intervention distribution of \cite{diaz2020causal, schindl2024incremental}; suppose we have a model with no covariates, $A \sim \text{Uniform}(0, 1)$ and a linear outcome model $\mu(a) = a$. Then, 
\begin{align*}
    q_\delta(a) = \frac{\delta \exp(\delta a)}{\exp(\delta) - 1}
\end{align*}
and it can be shown that
\begin{align*}
    \psi''_{q}(\delta) = \left\{\frac{\exp( \delta)(\exp(\delta) +1)}{(\exp(\delta)-1)^3} -\frac{2}{\delta^3} \right\}.
\end{align*}
Consequently, even though $\mu(a) = a$ is linear, the exponential-tilt-based incremental effect curve is nonlinear in the tilt parameter $\delta$. This nonlinearity in $\delta$ is an artifact of the parameterization, rather than the curvature of the response surface itself. Thus, if a practitioner were to try to causally interpret intermediate values of $\psi(t)$ along $t$, they would not be able to distinguish if the curvature they are observing is due to a change in outcomes through $\mu(x, a)$ or if it is simply due to $\delta$. Notably, the Wasserstein geodesic is special because it moves treatment values linearly toward the target dose through the treatment space. As a result, it introduces no artificial acceleration in treatment space, such that $\vec{a}_t(a, x) = 0$ almost surely. Consequently, the curvature of the Wasserstein incremental effect can be traced directly to curvature of the outcome regression. We formalize this idea in the following lemma.

\begin{lemma}[Wasserstein Effect Curvature] \label{effect_curvature}
Let $k \geq 1$ be an integer. For $P_X$-almost every $x$, suppose that $a \mapsto \mu(x,a)$ extends to a $k$-times continuously differentiable function on an open neighborhood of $\mathcal{A}$. Furthermore, suppose there exists a measurable function $M:\mathcal{X}\to[0,\infty)$ satisfying $\mathbb{E}[M(X)]<\infty$ and
\begin{align*}
    \max_{0\leq j\leq k} \sup_{a\in\mathcal{A}} \left| \frac{\partial^j}{\partial a^j} \mu(x,a) \right| \leq M(x)
\end{align*}
for $P_X$-almost every $x$, where the derivative of order zero is understood to be $\mu$ itself. Then, $t\mapsto\psi_\nu(t)$ is $k$-times continuously
differentiable on $(0,1)$ and, for every
$j\in\{1,\ldots,k\}$,
\begin{align*}
    \psi_\nu^{(j)}(t) = \mathbb{E} \! \left[ (a_\star-A)^j \left. \frac{\partial^j}{\partial u^j} \mu(X,u) \right|_{ u=(1-t)A+ta_\star} \right].
\end{align*}
Moreover, these derivatives extend continuously to
$t=0$ and $t=1$, where they are interpreted as right- and
left-hand derivatives, respectively.
\end{lemma}

By \cref{effect_curvature} we find a stronger result than no artificial acceleration: the Wasserstein geodesic does not introduce any path parameterization artifacts at any derivative order. Now, it is easy to see that the sign of $\psi'_\nu(t)$ depends on the alignment between the velocity and the treatment-effect gradient; if mass is moving in a direction where $\mu(x, a)$ increases then we can expect an increasing effect curve. Similarly, if mass is moving against the gradient then the curve will be decreasing. Furthermore, it now immediately follows that the curvature of $\psi_\nu(t)$ is directly inherited from the outcome regression $\mu(x, a)$. For example, if $\mu(x, a)$ is convex then $\psi_\nu(t)$ must also be convex; if $\mu(x, a)$ is affine in $a$, then $\psi_\nu(t)$ will be affine in $t$. More generally, if $\mu(x, a)$ is a polynomial of degree at most $r$ in $a$, then $\psi_\nu(t)$ is a polynomial of degree at most $r$ in $t$. This is the central interpretability advantage of the Wasserstein geodesic: its higher-order behavior is inherited directly from $\mu(x, a)$, rather than being generated by acceleration or nonlinear parameterization of the intervention path.

\subsection{Efficiency Theory} \label{efficiency_theory_section}

In this section we develop the usual semiparametric efficiency theory for the geodesy framework by deriving the efficient influence function and one-step estimator for $\psi_\nu(t)$ \citep{bickel1998efficient, Chernozhukov2018, kennedy2024semiparametric}. Here, we take a nonparametric perspective in the sense that for each fixed $t\in[0,1)$, we estimate $\psi_\nu(t)$ without imposing a structural model on the shape of the incremental effect curve $t\mapsto\psi_\nu(t)$. This will allow for flexible estimation of the nuisance functions $\mu$ and $\pi$, and, moreover, we show it yields a doubly robust estimator of the incremental Wasserstein effect. Throughout this section, we work in a dominated nonparametric model and consider regular parametric submodels satisfying the standard conditions required to interchange differentiation and integration. In the following theorem we establish the efficient influence function for $\psi_\nu(t)$. 

\begin{theorem}[Efficient Influence Function] \label{eif_theorem}
Fix $t\in[0,1)$ and suppose that $\nu_t(\,\cdot\mid x)\ll\pi(\,\cdot\mid x)$ for $P_X$-almost every $x$. Furthermore, let $\sigma^2(x,a)=\mathbb{V}(Y\mid X=x,A=a)$ and assume $\mathbb{E}\!\left[ \mu(X, (1-t)A+ta_\star)^{2} \right]<\infty$ and
\begin{align*}
    \mathbb{E}\!\left[\left(\frac{\nu_t(A\mid X)}{\pi(A\mid X)}\right)^{2} \sigma^2(X,A) \right]<\infty.
\end{align*}
We define density ratios to be zero where both densities vanish. Then, under the submodel regularity conditions stated above, the efficient influence function of $\psi_\nu(t)$ under a nonparametric model is given by
\begin{equation}\label{eq::if}
\varphi_\nu(Z; t) =  \frac{\nu_t(A \mid X)}{\pi(A \mid X)}\big(Y - \mu(X, A) \big) +  \mu\big(X, (1 - t)A + t a_{\star}\big) - \psi_\nu(t).
\end{equation}
\end{theorem}

\cref{eif_theorem} shows that the efficient influence function decomposes into three pieces. The first term behaves much like an inverse-propensity-weighted residual that upweights or downweights each outcome by how much the Wasserstein intervention deviates from the natural assignment. The second term, $\mu(X, (1 - t)A + t a_{\star})$, smoothly interpolates the regression function between the observed exposure $A$ (when $t = 0$) and the target exposure $a_{\star}$ (when $t = 1$). Finally, subtracting $\psi_\nu(t)$ centers the expression such that $\varphi_\nu(Z; t)$ is mean zero. Notably, when $t = 0$ then $\varphi_\nu(Z; t)$ reduces to the efficient influence function for the population mean, $Y - \mathbb{E}[Y]$. As $t \to 1$, then $\varphi_\nu(Z; t)$ acts analogously to an influence function for a kernel-smoothed dose-response parameter. For example, let $h = 1 - t$ act as a bandwidth; then $\varphi_\nu(Z; 1 - h)$ can be written as
\begin{align} \label{bandwidth_framing}
    \frac{1}{h} \frac{\pi\left(a_\star + \frac{A - a_\star}{h} \mid X \right)\mathbb{I}(A \in \mathcal{A}_{1 - h})}{\pi(A \mid X)} \big(Y - \mu(X, A) \big) +  \mu\big(X, a_\star + h(A - a_\star)\big) - \psi_\nu(1-h).
\end{align}
As $h \to 0$, we have a localized inverse-density residual, a smoothed regression plug-in, and the target, respectively. This is similar in structure to the influence functions explored in \cite{bibaut2017dataadaptivesmoothingoptimalrateestimation} and more generally related to \cite{kennedy2017cte} and \cite{Colangelo02012026}, who also explore kernel-localized doubly robust estimators of the dose-response curve. After debiasing \cref{eq::if}, we obtain the one-step estimator,
\begin{align*}
    \widehat{\psi}_\nu(t) = \frac{1}{n} \sum^n_{i=1}\left(\frac{\widehat{\nu}_t(A_i \mid X_i)}{\widehat{\pi}(A_i \mid X_i)}\Big(Y_i - \widehat{\mu}(X_i, A_i) \Big) +  \widehat{\mu}\big(X_i, (1-t)A_i + ta_{\star}\big) \right).
\end{align*}
The estimator $\widehat{\psi}_\nu(t)$ has a very straightforward interpretation. Namely, when $t = 0$ then $\widehat{\psi}_\nu(t)$ is simply the sample average of the outcomes, $\frac{1}{n} \sum^n_{i=1} Y_i$, a purely correlational quantity. As $t \to 1$, then $\widehat{\psi}_\nu(t) \approx \frac{1}{n} \sum^n_{i=1} \widehat{\mu}(X_i, a_{\star})$, i.e. the sample average of the regression function evaluated at $A_i = a_{\star}$, our sharp intervention. We discuss flexible estimation of $\mu$ and $\pi$ along with regularity conditions and convergence rates in \cref{asymptotics_section}.

One important point highlighted by \cref{bandwidth_framing} is that as $h \to 0$ (i.e.\ as $t \to 1$), the intervention concentrates its mass within an increasingly narrow neighborhood of $a_\star$; i.e.\ on $\mathcal{A}_{1-h} = \left\{a_\star+h(u-a_\star):u\in\mathcal{A} \right\}$. When $\mathcal{A}=[0,1]$, the width of $\mathcal{A}_{1-h}$ is exactly $h$. This shrinking window is a source of the increasing statistical uncertainty. The ratio of densities is nonzero only when the naturally observed treatment $A$ falls in $\mathcal{A}_{1-h}$, which occurs with probability proportional to $h$. Thus, only an $O(h)$ fraction of observations provide direct information about an intervention at a given $h$. At the same time, because $\nu_{1-h}(\, \cdot\mid x)$ must integrate to one while placing all of its mass in a window of width $h$, its density, and hence the corresponding likelihood-ratio weight, must be of order $h^{-1}$ within that window, which is exploding as $h \to 0$. Consequently, the one-step estimator has variance of order $(nh)^{-1}$ and standard error of order $(nh)^{-1/2}$. The following corollary formalizes this localization argument and identifies the leading constant in the efficiency bound.

\begin{corollary}[Nonparametric Efficiency Bound] \label{var_corollary}
    Let $\sigma^2_\nu(t) = \mathbb{V}(\varphi_\nu(Z; t))$. For $t \in [0, 1)$, suppose that $0 < \pi_{\min} \leq \pi(a \! \mid \! x) \leq \pi_{\max} < \infty$ and $\sigma^2(x, a) \geq \sigma^2_{\min} > 0$ and $\mathbb{E}[Y^2 \mid X = x, A = a] \leq \sigma^2_{\max}$ for all $x \in \mathcal{X}$ and $a \in \mathcal{A}$.  Then, 
    \begin{align*}
    \sigma^2_\nu(t) = \mathbb{E}\! \left[ \left(\frac{\nu_t(A\mid X)}{\pi(A\mid X)}\right)^2 \sigma^2(X, A)  \right] + \mathbb{V}\big(\mu\big(X, (1 - t)A + ta_\star\big)\big) \asymp \frac{1}{1-t} .
    \end{align*}
    Furthermore, suppose for $P_X$-almost every $x$ that $a \mapsto \pi(a \! \mid \! x)$ and $a \mapsto \mathbb{V}(Y \mid X = x, A=a)$ are continuous at $a_\star$. Then, as $t \to 1$,
    \begin{align*}
        (1-t)\sigma^2_\nu(t) \longrightarrow  \mathbb{E}\! \left[\frac{\sigma^2_\star(X)}{\pi_\star(X)} \| \pi(\, \cdot \mid X) \|^2_{L_2(\mathcal{A})} \right],
    \end{align*}
where we say $\sigma^2_\star(X) = \mathbb{V}(Y \mid X, A = a_\star)$ and $\pi_{\star}(X) = \pi(a_\star \mid X)$.
\end{corollary}
Note that by applying the law of total variance we can further expand the variance of the transported outcome regression as
\begin{align*}
    \mathbb{V}\big(\mu\big(X, (1 - t)A + ta_\star\big)\big) = \mathbb{E}\left[ \mathbb{V}_{\nu_t}(\mu(X,A)\mid X)\right] +\mathbb{V}\left[ \mathbb{E}_{\nu_t}(\mu(X,A)\mid X) \right].
\end{align*}
The rate in $(1-t)^{-1}$ obtained in \cref{var_corollary} is important for a few reasons.  First, for each fixed $t \in [0, 1)$, $\sigma_\nu^2(t)$ is the nonparametric efficiency bound and therefore lower bounds the asymptotic variance of any regular estimator of $\psi_\nu(t)$ \citep{van2000asymptotic}. Second, the natural stochastic error is not $n^{-1/2}$ uniformly in $t$, but rather
\begin{align*}
    \frac{\sigma_\nu(t)}{\sqrt{n}}
    \asymp
    \frac{1}{\sqrt{n(1-t)}}.
\end{align*}
Consequently, treating $t$ as an inconsequential fixed parameter could undersell the difficulty in estimation. Thus, analogous to \cite{schindl2024incremental}, we develop the conditions under which asymptotic normality holds uniformly across $t$, even as $t \to 1$; we formalize this in the following section.

\subsection{Asymptotic Distribution \& Convergence Rate} \label{asymptotics_section}

In this section, we establish asymptotic normality for the incremental Wasserstein effect. For every fixed $t \in [0, 1)$ this is trivial --- $\psi_\nu(t)$ is pathwise differentiable and the one-step estimator is asymptotically linear at the usual $\sqrt{n}$ rate. However, this regularity is not uniform if we allow $t \to 1$. Consequently, we would like to make the dependence on the path more explicit in our analysis, such that any choice of $t$ is supported. Thus, moving forward, we index $t$ by $n$ to allow it to grow with the sample size. Importantly, under a continuous exposure, asymptotic normality for incremental effects typically requires stronger than usual control of nuisance function estimation error. For example, under the exponentially tilted intervention distribution $q_\delta(a \! \mid \! x)$ (as defined in \cref{exponential_tilt_def}), for some measurable function $f(x, a)$, \cite{schindl2024incremental} require mixed $L_2(P)$-sup norms of the form 
\begin{align*}
     \|f \|^2_{L^2_x, L^2_\infty} =  \int_{\mathcal{X}} \left\{\underset{a \in \mathcal{A}}{\text{sup}} \,  f(x, a)^2 \right\} dP_X(x)
\end{align*}
for nuisance estimation. Requiring a uniform bound across the treatment density takes the sting out of the dependence on $\delta$ by requiring $f$ to be estimated well across its entire treatment support. With this type of constraint in place, even as $\delta \to \infty$, the remainder from the von Mises expansion can still be controlled. In the case of the Wasserstein geodesic, we do \textit{not} require treatment-uniform control. Instead, as $\psi_\star$ is approached, the errors are localized to the shrinking support of the intervention. We define the local $L_2(P)$ seminorm
\begin{align} \label{wasserstein_local_norm}
    \|f\|^2_{t_n} =\frac{1}{1-t_n}\mathbb{E}[ f(X,A)^2\mathbb{I}(A\in\mathcal{A}_{t_n})].
\end{align}
When $t_n=0$ and no intervention is performed, then $\mathcal{A}_0=\mathcal{A}$ and \cref{wasserstein_local_norm} is exactly the usual squared $L_2(P)$ norm. The factor $(1-t_n)^{-1}$ normalizes for the width of the shrinking interval $\mathcal{A}_{t_n}$;
thus, $\|f\|_{t_n}$ measures the normalized mean-square error in the treatment region on which the Wasserstein intervention is supported. It does not impose a supremum over treatment, but it does prevent global $L_2(P)$ consistency from concealing poor estimation in a neighborhood of $a_\star$ whose width shrinks with $t_n$. We show later that global $L_2(P)$ norms can be used, but these result in slightly conservative nuisance estimation rates.

In what follows, we formally establish the asymptotic distribution and convergence rate for the incremental Wasserstein effect estimator. Throughout, we assume $\widehat{\psi}_\nu(t_n)$ denotes the $K$-fold cross-fitted version of the one-step estimator in order to avoid imposing Donsker-type restrictions on the complexity of the function class \citep{Chernozhukov2018}; however, for notational simplicity we suppress the dependence on fold $k \in \{1, \ldots, K \}$. 

\begin{theorem} \label{wasserstein_asymptotics}
    Suppose that $|Y| \leq C$ with probability one, the conditions of \cref{var_corollary} hold, and that $0 < \pi_{\min} \leq \widehat{\pi}(a \! \mid \! x) \leq \pi_{\max}$ for all $a \in \mathcal{A}$ and $x \in \mathcal{X}$. Then, as $t_n \to 1$, if
    \begin{enumerate}
        \item[$(i)$] $\sqrt{n(1-t_n)} \to \infty$.
        \item[$(ii)$]  $ \|\widehat{\pi} - \pi\|_2 + \|\widehat{\pi} - \pi\|_{t_n} = o_{P}(1)$ and  $\| \widehat{\mu} - \mu \|_{t_n} =  o_{P}(1)$.
        \item[$(iii)$] $ \| \widehat{\mu} - \mu \|_{t_n} \Big( \|\widehat{\pi} - \pi\|_{t_n} + \|\widehat{\pi} - \pi\|_2 \Big)   = o_{P}\left((n(1-t_n))^{-1/2}\right)$
    \end{enumerate}
    then as $n \to \infty$, it follows that
    \begin{align*}
        \frac{\sqrt{n}}{\sigma_\nu(t_n)}\left(\widehat{\psi}_\nu(t_n) - \psi_\nu(t_n) \right)\overset{d}{\longrightarrow}N(0,1),
    \end{align*}
where $\sigma^2_\nu(t_n) = \mathbb{V}(\varphi_\nu(Z;t_n))$ is the nonparametric efficiency bound defined in \cref{var_corollary}. 
\end{theorem}
For any fixed $t \in [0, 1)$, \cref{wasserstein_asymptotics} reduces to the 
usual $n^{-1/2}$ asymptotic normality result for a regular, pathwise differentiable parameter. 
In this case, the local seminorm is related to the ordinary $L_2(P)$ norm through
\begin{align*}
    \|f\|_t^2 = \frac{1}{1-t} \mathbb{E}[f(X,A)^2\mathbb{I}(A\in\mathcal{A}_t)] \leq \frac{1}{1-t}\|f\|_2^2.
\end{align*}
Thus, for fixed $t$, global $L_2(P)$ consistency implies local consistency, while the converse need not hold because $\|f\|_t$ ignores errors outside $\mathcal{A}_t$. When $t=t_n\to1$, however, the factor $(1-t_n)^{-1}$ diverges, so global consistency alone need not ensure accurate estimation within the shrinking region that identifies the intervention. The local seminorm is therefore best viewed as targeted rather than uniformly stronger or weaker: it requires accuracy where the intervention places mass without requiring uniform accuracy over the entire treatment support. Finally, note that the one-step estimator is doubly robust in the usual remainder sense, since 
\begin{align*}
    |R_2(\widehat P,P)| \lesssim \|\widehat\mu-\mu\|_{t_n} \Big\{ \|\widehat\pi-\pi\|_{t_n} + \|\widehat\pi-\pi\|_2 \Big\},
\end{align*}
where $ R_2(\widehat{P}, P) = \psi_{\nu}(t_n; \widehat{P}) - \psi_{\nu}(t_n; P) + \int \varphi_{\nu}(z; t_n, \widehat{P}) dP_Z(z)$ is the remainder from the von Mises expansion. Consequently, the remainder vanishes if either the outcome regression or the treatment density is correctly estimated, and asymptotic normality only requires their errors to satisfy a product-rate condition. For fixed $t$, this is the usual $o_P(n^{-1/2})$ requirement; near the endpoint, it becomes an effective-sample-size requirement of order $o_P\{(n(1-t_n))^{-1/2}\}$.

\subsubsection{Minimax Elbow}

An interesting statistical phase transition occurs along the path to the sharp intervention. At first sight, there appears to be an abrupt statistical discontinuity: for every fixed $t \in [0, 1)$, the incremental effect $\psi_\nu(t)$ is a regular, pathwise differentiable functional admitting $n^{-1/2}$ rate estimation. Meanwhile, at $t=1$, the intervention collapses to a point mass and the target becomes the point-evaluation functional $\psi_\star=\mathbb{E}[\mu(X,a_\star)]$, which is nonregular under a continuous treatment and cannot generally be estimated at the parametric rate \citep{diaz2013targeted, kennedy2017cte}. However, this regularity for $t < 1$ holds pointwise in $t$, not uniformly; as $t_n \to 1$ we observe increasingly nonparametric behavior, e.g.\ the exploding variance described in \cref{var_corollary}. Notably, this deterioration persists along the path until $1 - t_n$ reaches some critical bandwidth $b_n$, which represents the smallest scale at which H\"older-admissible local variation can be distinguished from sampling noise; then the problem becomes fully nonparametric in nature. In what follows, we formalize this statistical elbow phenomenon.

Consider a Gaussian known-design model with no covariates. Let $A\in\mathcal{A}\subset\mathbb{R}^q$ have a known density $\pi$. In this example, we establish an explicit dependence on the treatment dimension $q$, although throughout the remainder of the paper we let $q=1$. Then, suppose $Y=\mu(A)+\varepsilon$ where $\varepsilon\sim N(0,\sigma^2)$ and $\varepsilon\ind A$ and let $P_\mu$ denote the joint law of $(Y, A)$ induced by the regression function $\mu$. From here, fix $t_0\in[0,1)$, $a_\star\in\mathcal{A}$, and let $K$ be a bounded probability density with compact support $\mathcal{U}$. Suppose that $a_\star+(1-t)\mathcal{U}\subseteq\mathcal{A}$ for every $t\in[t_0,1)$. This ensures that the intervention is supported on the treatment space and when $a_\star$ lies on its boundary, it permits $K$ to have appropriately one-sided support. Then, for $t\in[t_0,1)$, we define $\psi_t(\mu) = \int_{\mathcal{A}}\mu(a)\rho_t(a)da$ under the local density
\begin{align*}
    \rho_t(a) &= \frac{1}{(1-t)^q} K\left(\frac{a-a_\star}{1-t}\right).
\end{align*}
Next, fix $0 < \beta \leq 1$ and $L > 0$, and let $\mathcal{H}_q^\beta(L)$ denote the bounded isotropic H\"older class,
\begin{align*}
     \mathcal{H}_q^\beta(L)
    =
    \left\{
        \mu:\mathcal{A}\to\mathbb{R}:
        \begin{array}{l}
        \mu \text{ is measurable},\
        \|\mu\|_{\infty,\mathcal{A}}\leq L,\\[0.05in]
        |\mu(a)-\mu(a')|
        \leq
        L\|a-a'\|_2^\beta
        \text{ for every }a,a'\in\mathcal{A}
        \end{array}
    \right\}.
\end{align*}
When $\beta=1$, this is the bounded Lipschitz class \citep{tsbakov2009}. For absolute-error loss, define the risk
\begin{align*}
\mathcal{R}_n(t) = \inf_{\widehat\psi_t} \sup_{\mu\in\mathcal{H}_q^\beta(L)} \mathbb{E}_\mu \! \left[\left| \widehat\psi_t -  \psi_t(\mu) \right| \right],
\end{align*}
where the infimum is over estimators measurable with respect to the observed sample and we write $\mathbb{E}_\mu$ for the expectation under the $n$-sample law $P_\mu^{\otimes n}$. With these definitions in place, the following theorem characterizes the transition from regular estimation to the sharp intervention.

\begin{theorem}[Minimax Elbow]
\label{minimax_elbow}
Under the preceding setup, suppose there exists $r_0>0$ such that $0<\pi_{\min} \leq \pi(a) \leq \pi_{\max} <\infty$ for Lebesgue-almost every $a\in\mathcal{A}\cap B(a_\star,r_0)$, where $B(a, r)$ denotes the open Euclidean ball of radius $r$ centered at $a$. Further suppose that $t_0$ is chosen so that $a_\star+(1-t)\mathcal{U} \subseteq \mathcal{A}\cap B(a_\star,r_0)$ for every $t\in[t_0,1)$. Let $b_n=n^{-1/(2\beta+q)}$. Then, for all sufficiently large $n$ and every $t\in[t_0,1]$,
\begin{align*} 
\mathcal{R}_n(t)
\asymp
\frac{1}{\sqrt{n(1-t)^q}}
\wedge
n^{-\beta/(2\beta+q)}.
\end{align*}
\end{theorem}

\cref{minimax_elbow} reveals two distinct statistical regimes along the path to $\psi_\star$. As $t_n \to 1$, risk exhibits the elbow
\begin{align*}
\mathcal{R}_n(t_n) \asymp
\begin{cases}
\left\{n(1-t_n)^q\right\}^{-1/2}
& 1-t_n\gtrsim b_n \\[0.05in]
n^{-\beta/(2\beta+q)}
& 1-t_n\lesssim b_n.
\end{cases}
\end{align*}
Thus, when $1 - t_n \gtrsim b_n$, the intervention distribution is sufficiently diffuse, so the target behaves as a regular linear functional. In the known-design model, the direct estimator
\begin{align*}
    \widehat{\psi}(t) = \frac{1}{n}\sum_{i=1}^n \frac{\rho_t(A_i)}{\pi(A_i)}Y_i
\end{align*}
is unbiased and has standard deviation of order $(n(1-t)^q)^{-1/2}$. The factor $n(1-t)^q$ can be interpreted as a local effective sample size in the sense that only an $O((1-t)^q)$ fraction of the observations lie in the region on which the intervention is concentrated. Meanwhile, once $(1-t_n) \lesssim b_n$, the same estimator remains unbiased but its variance becomes unnecessarily large; how far along the path we travel before reaching $b_n$ depends on the smoothness of $\mu$. Importantly, the known-design experiment is a submodel of the full causal problem: one may take $X$ to be degenerate, hold the treatment density fixed and known, and vary only the outcome regression. Consequently, the lower bound in \cref{minimax_elbow} automatically lower bounds estimation in the larger causal model.

\subsection{Inference for Sharp Interventions} \label{endpoint_section}

In \cref{asymptotics_section} we established asymptotic normality for the incremental Wasserstein effect estimator $\widehat{\psi}_\nu(t_n)$. Now, in this section we establish inference for $\psi_\star$. Inference for the sharp intervention can be done using the usual nonparametric smoothing methods \citep{kennedy2017cte, takatsu2025}. However, geodesy leads to a related but different approach with the added benefit that it is interpretable as a stochastic intervention for every choice of tuning parameter. To begin, observe that our estimator admits the decomposition 
\begin{align*}
    \sqrt{n(1-t_n)} \left(\widehat\psi_\nu(t_n)- \psi_\star \right) &= \underbrace{
        \sqrt{n(1-t_n)} \left( \widehat\psi_\nu(t_n)- \psi_\nu(t_n) \right)}_{\text{estimation error}} + \underbrace{ \sqrt{n(1-t_n)} \Big(\psi_\nu(t_n)-\psi_\star
        \Big)}_{\text{approximation bias}}.
\end{align*}
Thus, for valid inference on $\psi_\star$, the approximation bias must be negligible for the centered Gaussian limit to hold; typically, this requires either some kind of undersmoothing requirement on $1 - t_n$ or bias-correction \citep{calonico2014, Calonico2018}. For example, under the smoothness assumptions outlined in \cref{effect_curvature} for $k=2$, a Taylor expansion gives us $\psi_\nu(t_n)-\psi_\star = (1-t_n) B_\star +O ((1-t_n)^2)$ where $B_\star = \mathbb{E}\left[(A-a_\star)\frac{\partial}{\partial u}\mu(X,u)|_{u=a_\star}\right]$.
Consequently,
\begin{align*}
    \sqrt{n(1-t_n)} \left( \psi_\nu(t_n)-\psi_\star\right) = \sqrt{n(1 - t_n)^3}B_\star + O\left(\sqrt{n(1-t_n)^5}\right).
\end{align*}
Therefore, if $n(1-t_n)^3\to0$, the approximation bias vanishes and the limit is $N(0,V_\star)$. When $B_\star \neq 0$, balancing the squared approximation bias against the variance gives the bandwidth choice $(1-t_n)\asymp n^{-1/3}$; however, the first-order bias is of the same order as the standard error, so an uncorrected confidence interval for $\psi_\nu(t_n)$ need not have nominal coverage for $\psi_\star$.

First-order bias correction relaxes the undersmoothing requirement. Since $\psi_\star = \psi_\nu(t_n) + (1-t_n)\psi^{\prime}_\nu(t_n) + O((1-t_n)^2)$, the bias-corrected estimator
\begin{align*}
    \widehat{\psi}_\nu^{\,\mathrm{BC}}(t_n) = \widehat{\psi}_\nu(t_n) + (1-t_n)\widehat{\psi}^{\prime}_\nu(t_n)
\end{align*}
removes the leading $(1 - t_n) B_\star$ term. Its remaining bias is $O((1-t_n)^2)$ and therefore has scaled order of $\sqrt{n(1 - t_n)^5}$. Consequently, first-order bias correction relaxes the rate condition  from $n(1 - t_n)^3\to0$ to $n(1 - t_n)^5 \to 0$. Although $n(1- t_n)^5 \to 0$ remains an undersmoothing condition relative to the second-order MSE-balancing choice $(1-t_n)\asymp n^{-1/5}$, it is substantially weaker than the uncorrected requirement. This idea extends naturally to higher-order bias correction. By Taylor's theorem, it follows that
\begin{align*}
    \psi_\star - \sum_{k=0}^{p} \frac{(1 - t_n)^k}{k!}\psi_\nu^{(k)}(t_n) = \frac{1}{p!} \int_{t_n}^{1} (1-s)^p\psi_\nu^{(p+1)}(s)ds.
\end{align*}
Thus, the remaining approximation bias is $O((1-t_n)^{p+1})$ and is negligible at the $\sqrt{n(1-t_n)}$ scale whenever $n(1 - t_n)^{2p+3}\to0$; the undersmoothing rate required depending on the approximation order $p$ is summarized in \cref{table_of_rates}. The convenient curvature properties of the incremental Wasserstein effect developed in \cref{effect_curvature} make it a natural candidate for higher-order approximations, when sufficient smoothness of the regression function exists. 
\begin{table}[h]
\centering
\begin{tabular}{l|cc}
$p$ & Bias & Rate \\ \hline \hline
$0$ & $O(1-t)$ &  $n(1-t)^3 \to 0$  \\
$1$ & $O((1-t)^2)$ & $n(1-t)^5 \to 0$  \\
$2$ & $O((1-t)^3)$ &  $n(1-t)^7 \to 0$
\end{tabular}
\caption{Approximation bias and bias-negligibility condition after a $p$th-order correction.}
\label{table_of_rates}
\end{table}

Now, our goal is to derive the efficient influence function for $\psi^{(k)}_\nu$ in order to construct a higher-order bias-corrected estimator for $\psi_\star$. However, before we do, we must impose an additional regularity condition: namely, we require a no-flux condition on the boundaries of the treatment support. To illustrate this condition, consider the first derivative of the incremental Wasserstein effect, evaluated at $t = 0$, i.e.\ the average derivative effect. Integration by parts yields
\begin{align*}
    \psi^\prime_\nu(0) &= \mathbb{E}\! \left[\int^1_0 (a_\star - a) \frac{\partial }{\partial a} \mu(X, a) \pi(a \mid X) da \right] \\
    &= \mathbb{E} \!\left[\left[(a_\star - a) \mu(X, a) \pi(a \mid X) \right]^1_0 \right] - \mathbb{E} \! \left[\int^1_0 \mu(X, a) \frac{\partial}{\partial a} \{ (a_\star - a) \pi(a \mid X) \} da \right].
\end{align*}
When $a_\star \in (0, 1)$, the weight $(a_\star - a)$ is non-zero at both endpoints. Thus, unless the corresponding boundary terms vanish, the pathwise derivative depends on point evaluations of $\mu(x, 0)$ and $\mu(x, 1)$ and consequently, an $L_2(P)$ influence-function representation in the unrestricted nonparametric model does not exist. As a result, to derive the efficient influence function for $\psi^\prime_\nu(t)$ we must assume $ \lim_{a \downarrow 0} \left\{ \pi(a\mid x) \right\} = \lim_{a \uparrow 1} \left\{ \pi(a\mid x) \right\} = 0$, and for higher-order derivatives, its $k$th derivative analogue. That is, for $P_X$-almost every $x$, we assume
\begin{align} \label{no_flux_assumption}
    \lim_{a\downarrow 0} \frac{\partial^j}{\partial a^j} \left\{ (a_\star-a)^k\pi(a\mid x) \right\} =  \lim_{a\uparrow 1} \frac{\partial^j}{\partial a^j} \left\{ (a_\star-a)^k\pi(a\mid x) \right\} = 0
\end{align}
for $j=0,\ldots,k-1$.  In \cref{jackknife_supplement}, we explore alternate estimation methods that avoid the no-flux condition through a generalized jackknife approach. Notably, this condition conflicts with the global positivity assumption we have been working with, i.e.\ $\pi(a \mid x) \geq \pi_{\min} > 0$ for all $(x, a) \in \mathcal{X} \times \mathcal{A}$, since that disallows a tapering density at the boundaries. Instead, we can assume a localized positivity condition,
\begin{align} \label{localized_positivity}
    \inf_{x\in\mathcal{X}} \inf_{a\in\mathcal{A}_{t_0}} \pi(a\mid x)>0
\end{align}
for some $t_0>0$, because $\mathcal{A}_{t_0}$ lies strictly inside the original support and contains every $\mathcal{A}_t$ with $t\geq t_0$. This weakened positivity condition suffices when we are interested in taking $t_n \to 1$; however, our analysis is then restricted to $t \in [t_0, 1)$. Note that this localized positivity is used to establish results like asymptotic normality and inequalities for the nonparametric efficiency bound; for identification purposes we still only require the conditions of \cref{identification_proposition} to hold. In what follows, we derive the efficient influence function for the $k$th Wasserstein derivative.

\begin{theorem}[Efficient Influence Function for Wasserstein Derivatives] \label{derivative_EIF}

Let $k\geq1$ be a fixed integer and let $t \in [0, 1)$. Suppose that, for $P_X$-almost every $x$, $a\mapsto\mu(x,a)$ is $k$ times weakly differentiable and for $u \in \mathcal{A}_t$, $u \mapsto \nu_t(u\mid x)(\frac{a_\star-u}{1-t})^k$ has $k$ derivatives on $\mathcal{A}_t$. Furthermore, assume the no-flux boundary condition of \cref{no_flux_assumption} holds. Then the efficient influence function of $\psi^{(k)}_\nu(t)$ under a nonparametric model is given by
\begin{align*}
       \varphi_{\nu,k}(Z;t) =  \frac{\eta_{k, t}(X, A)}{\pi(A \mid X)}\{Y-\mu(X,A)\} + (a_\star-A)^k \frac{\partial^k}{\partial u^k}\mu(X,u) \Big|_{u=(1-t)A + ta_\star} - \psi^{(k)}_\nu(t),
\end{align*}
where 
\begin{align*}
    \eta_{k,t}(x,a) = \frac{\mathbb{I}(a\in\mathcal{A}_t)}{(1-t)^{k+1}} \sum_{j=0}^{k} \binom{k}{j} \frac{k!}{j!} \left( \frac{a-a_\star}{1-t}\right)^j \pi^{(j)}\left(\frac{a-ta_\star}{1-t}\mid x\right).
\end{align*}
Furthermore, the nonparametric efficiency bound is given by
\begin{align*}
    \mathbb{V}(\varphi_{\nu,k}(Z;t)) = \mathbb{E} \! \left[ \int_{\mathcal{A}_t} \frac{\eta_{k,t}(X, a)^2}{\pi(a\mid X)} \sigma^2(X,a)da \right] + \mathbb{V} \!\left[(a_\star-A)^k \left. \frac{\partial^k}{\partial u^k}\mu(X,u) \right|_{u=(1-t)A + t a_\star} \right].
\end{align*}
    
\end{theorem}

There are a few key implications of \cref{derivative_EIF}. First, let $\theta_{\nu,p}(t) = \sum_{k=0}^{p}\frac{(1-t)^k}{k!}\psi_\nu^{(k)}(t)$ be the order-$p$ Taylor target. Then, for a correction of order $p$, by the linearity of the Gateaux derivative, the combined influence function is
\begin{align*}
    \Phi_{\nu, p}(Z; t) = \sum_{k=0}^p \frac{(1-t)^k}{k!} \varphi_{\nu,k}(Z;t).
\end{align*}
Consequently, to construct an order-$p$ one-step estimator we may write $ \Phi_{\nu,p}(Z;t) =\phi_{\nu,p}(Z;t)- \theta_{\nu,p}(t)$ where $\phi_{\nu,p}(z;t) = \frac{\omega_{p,t}(x,a)}{\pi(a\mid x)}(y-\mu(x,a))+\Gamma_{p,t}(x,a)$ and
\begin{align*}
    \omega_{p,t}(x,a) &= \sum_{k=0}^{p}\frac{(1-t)^k}{k!}\eta_{k,t}(x,a), \\
    \Gamma_{p,t}(x,a) &= \sum_{k=0}^{p}\frac{(1-t)^k}{k!}(a_\star-a)^k \mu^{(k)}\! \big(x,(1-t)a + t a_\star\big).
\end{align*}
Thus, our bias-corrected estimator for $\psi_\star$ is simply $  \widehat\psi_{\nu,p}^{\,\mathrm{BC}}(t) =\frac{1}{n}\sum_{i=1}^{n} \widehat{\phi}_{\nu,p}^{\, \mathrm{BC}}(Z_i;t)$. Later, we discuss estimation of the nuisance functions $\mu^{(k)}$ and $\pi^{(k)}$. Second, analogously to the nonparametric efficiency bound developed in \cref{var_corollary}, it follows that as $t_n \to 1$, under the localized positivity assumption of \cref{localized_positivity},
\begin{align*}
    \mathbb{V}(\varphi_{\nu,k}(Z;t_n)) \asymp (1-t_n)^{-(2k+1)},
\end{align*}
and consequently, $\mathbb{V}(\widehat{\psi}_{\nu, p}^{\,\mathrm{BC}}(t_n)) \asymp (n(1 - t_n))^{-1}$. Thus, if one were to balance the squared bias $(1 - t_n)^{2p+2}$ against the variance of the bias-corrected estimator, they would obtain the mean-squared error balancing bandwidth choice of $(1-t_n)\asymp n^{-1/(2p+3)}$. Notably, this is the minimax optimal rate for nonparametric estimation of one-dimensional $s$-smooth regression functions \citep{stone80, stone82}. To see this, note that when the map $a\mapsto\mu(x,a)$ belongs to a H\"older ball of smoothness $s=p+1$, then 
\begin{align*}
    (1 - t_n) \asymp n^{-1/(2s+1)} =n^{-1/(2p+3)}.
\end{align*}
When $p=0$, this recovers the familiar $n^{-1/3}$ rate for a Lipschitz function. With these derivations in place, we now establish asymptotic normality for the bias-corrected estimator and the undersmoothing rates required for inference. Again, we assume cross-fitting has been employed to learn all nuisance functions, but for notational simplicity we suppress the dependence on the fold.

\begin{theorem}[Asymptotic Normality of the Bias-Corrected Estimator] \label{higher_order_normality}
    Fix an integer $p\geq0$ and let $t_n \to 1$. Then, suppose the following conditions hold:
\begin{enumerate}
    \item[$(i)$] \textit{Rate Conditions:} Both $n(1 - t_n) \to \infty$ and $n (1 - t_n)^{2p+3} \to 0$.
    \item[$(ii)$] \textit{Smoothness:} The smoothness and no-flux conditions of \cref{effect_curvature} and \cref{derivative_EIF} hold through order $p+1$ and through order $p$, respectively. Moreover, there exist $\delta>0$, $C<\infty$, and a measurable $M_\mu\in L_{2+\delta}(P_X)$ such that, $P_X$-almost surely,
    \begin{align*}
        \max_{0\leq k\leq p+1}\sup_{a\in\mathcal A}
        |\mu^{(k)}(X,a)| \leq M_\mu(X), \qquad 
        \max_{0\leq j\leq p}\sup_{a\in\mathcal A}
        |\pi^{(j)}(a\mid X)| &\leq C,
    \end{align*}
    and $\sup_{a\in\mathcal A_{t_0}}
        \mathbb{E}\!\left[
            |Y-\mu(X,A)|^{2+\delta}\mid X,A=a
        \right] \leq C$ for some $t_0 < 1$.

    \item[$(iii)$] \textit{Regularity:} The localized positivity condition in \cref{localized_positivity} holds for some $t_0<1$, and, with probability tending to one, the cross-fitted treatment-density estimators are bounded away from zero on $\mathcal{A}_{t_0}$. For $P_X$-almost every $x$, $a\mapsto\pi(a\mid x)$ and $a\mapsto \mathbb{V}(Y\mid X=x,A=a)$ are continuous at $a_\star$. Furthermore, $\sigma_\star^2(X)=\mathbb{V}(Y\mid X,A=a_\star)\geq\sigma_{\min}^2>0$ almost surely.

    \item[$(iv)$] \textit{Nuisance Estimation:} To control the empirical process term we require
    \begin{align*}
        \|\widehat\mu-\mu\|_{t_n} + \sqrt{1 - t_n} \left( \, \left\| \frac{\widehat\omega_{p,t_n}}{\widehat\pi} -\frac{\omega_{p,t_n}}{\pi} \right\|_2 + \|\widehat\Gamma_{p,t_n} - \Gamma_{p,t_n}\|_2 \right) = o_P(1),
    \end{align*}
    and to control the von Mises remainder we require
    \begin{align*}
        \|\widehat\mu-\mu\|_{t_n} \left( \sqrt{1 - t_n}\left\| \frac{\widehat\omega_{p,t_n}}{\widehat\pi} -\frac{\omega_{p,t_n}}{\pi} \right\|_2 \, \right) = o_P\left((n(1-t_n))^{-1/2}\right).
    \end{align*}
\end{enumerate}
Let $\sigma_{\nu,p}^2(t_n) =\mathbb{V}(\Phi_{\nu, p}(Z; t_n))$.
Then
\begin{align*}
    \frac{\sqrt{n}}{\sigma_{\nu,p}(t_n)} \left(\widehat\psi_{\nu,p}^{\,\mathrm{BC}}(t_n)-\psi_\star\right) \overset{d}{\longrightarrow}N(0,1).
\end{align*}
Furthermore, define the order-$p$ kernel $ K_p(u\mid x) =\sum_{j=0}^{p} \binom{p+1}{j+1}\frac{(u-a_\star)^j}{j!} \pi^{(j)}(u\mid x)$. Then,
\begin{align*}
    \sqrt{n(1-t_n)} \left(\widehat{\psi}_{\nu,p}^{\,\mathrm{BC}}(t_n)-\psi_\star\right) \overset{d}{\longrightarrow}N(0,V_{p,\star}),
\end{align*}
where $V_{p,\star} =\mathbb{E} \! \left[ \frac{\sigma_\star^2(X)}{\pi_\star(X)} \| K_p(\, \cdot \mid X) \|^2_{L_2(\mathcal{A})} \right]$.
\end{theorem}

There are several features of \cref{higher_order_normality} that are worth emphasizing. First, we can immediately obtain the expansion
\begin{align*}
     \widehat{\psi}_{\nu,p}^{\,\mathrm{BC}}(t_n)-\psi_\star = O_P\!\left((n(1-t_n))^{-1/2}\right) +O((1-t_n)^{p+1}),
\end{align*}
so $n(1 - t_n)$ plays the role of the effective sample size in the shrinking treatment region. The requirement $n(1 - t_n) \to\infty$ prevents $t_n$ from approaching one so quickly that too little information remains, whereas $n(1 - t_n)^{2p+3} \to 0$ ensures that the residual $(p+1)$st-order approximation bias is negligible relative to the standard error. Equivalently, if we say $(1 - t_n) = n^{-\alpha}$, valid centered inference then requires
\begin{align*}
    \frac{1}{2p+3}<\alpha<1,
\end{align*}
or equivalently, $n^{-1} \ll (1 - t_n) \ll n^{-1/(2p+3)}$. It is important to note: higher-order correction is not costless; it requires additional smoothness and more demanding nuisance estimation, and it changes the equivalent kernel and hence the asymptotic variance constant $V_{p,\star}$.

\section{Paths and Overlap} \label{hellinger_section}

Now that we have established the statistical properties of the Wasserstein intervention, we want to consider paths that avoid overlap problems. In other words, we seek a geodesic $\nu_t$ such that $\nu_t(a \! \mid \! x) = 0$ whenever $\pi(a \! \mid \! x) = 0$ for all $(x, a) \in \mathcal{X} \times \mathcal{A}$ in order to relax the global positivity required by the Wasserstein intervention. To achieve this, we need to change our geometric perspective. Under the Wasserstein metric, the distance between two distributions is measured by how far mass must physically travel. However, this is by no means the only way to measure how close two distributions are. Another interesting perspective to consider is square-root geometry induced by the Hellinger and Fisher--Rao distances \citep{Hellinger1909, amari2016information}.

Let $P$ and $Q$ be probability measures on a common measurable space, and let $\lambda$ be a measure that dominates both $P$ and $Q$ (e.g. $\lambda = P + Q$). Then, the squared Hellinger distance is given by
\begin{align*}
    H^2(P, Q) = \frac{1}{2} \int \left( \sqrt{\frac{dP}{d\lambda}} - \sqrt{\frac{dQ}{d\lambda}} \right)^2d\lambda = 1 -  \int \left(\sqrt{\frac{dP}{d\lambda}} \sqrt{\frac{dQ}{d\lambda}} \right) d\lambda = 1 - B(P, Q)
\end{align*}
where we say $B(P, Q)$ is the Hellinger affinity or Bhattacharyya coefficient \citep{bhattacharyya1943measure}. Consider the geometry induced by the square-root mappings $P \mapsto \sqrt{ dP / d\lambda}$ and $Q \mapsto \sqrt{ dQ / d\lambda}$. Let
\begin{align*}
    s_P = \sqrt{\frac{dP}{d\lambda}} \qquad \text{and} \qquad s_Q = \sqrt{\frac{dQ}{d\lambda}}
\end{align*}
define the square-root representation of $P$ and $Q$ and note that because both $P$ and $Q$ are valid probability measures, $\| s_P \|^2_{L^2(\lambda)} = \| s_Q \|^2_{L^2(\lambda)} = 1$. Thus, the square-root mapping identifies the probability measures dominated by $\lambda$ within the positive orthant of the unit sphere \citep{Srivastava2007riemannian, srivastava2016functional}, i.e.\ 
\begin{align*}
    \mathbb{S}_+(\lambda) = \left\{ s\in L^2(\lambda): s\geq 0,\  \|s\|_{L^2(\lambda)}=1 \right\}.
\end{align*}
It is important to note that the Hellinger distance is proportional to the Euclidean distance between our two square roots, $ H(P,Q) = \frac{1}{\sqrt{2}} \|s_P-s_Q\|_{L^2(\lambda)}$. Thus, the ordinary Hellinger distance is an \emph{extrinsic}, or chordal, distance on the probability sphere \citep{mielke2025noteshellingerdistancevarious}. This instrinsic-extrinsic distinction matters when it comes to defining a geodesic between $P$ and $Q$: if we simply take the linear interpolation in square-root coordinates $s_t^{\mathrm{H}} = (1-t)s_P+t s_Q$, then it can be shown that
\begin{align*}
    \left\|s_t^{\mathrm{H}}\right\|_{L^2(\lambda)}^2 = 1-2t(1-t)H^2(P,Q) < 1
\end{align*}
when $P \neq Q$, for $t \in (0, 1)$. Hence, the straight Hellinger chord passes through the \textit{interior} of the $L^2$ sphere and corresponds to a path of subprobability measures --- this intuition is visualized in \cref{fig:hellinger_vs_fisherrao}. Although this is a valid geodesic on the cone of finite nonnegative measures, it does not generally define a path of stochastic intervention distributions.

To obtain a path that remains within the space of probability measures, we must constrain the square root to remain on $\mathbb{S}_+(\lambda)$ at every intermediate value of $t$. Intuitively, the shortest path on the unit sphere is the shorter great-circle arc, as shown in \cref{fig:hellinger_vs_fisherrao}. To describe this path, we define the \textit{Spherical Hellinger} distance to be the angle
\begin{align} \label{theta_definition}
    \theta(P,Q) = \arccos \left\{ \left\langle s_P,s_Q\right\rangle_{L^2(\lambda)} \right\} = \arccos\{B(P,Q)\}.
\end{align}
Notably, this definition of the Spherical Hellinger distance coincides up to constants with the Fisher--Rao Riemannian metric on a manifold of smooth positive densities, so both  produce the same geodesic \citep{Srivastava2007riemannian, kurtek2015}. Note that because $s_P$ and $s_Q$ are nonnegative, their affinity belongs to $[0,1]$, and hence $\theta(P,Q)\in[0,\pi/2]$. Thus, the great-circle arc between them remains in the positive orthant of the sphere. We formalize the Spherical Hellinger geodesic in the following proposition.

\begin{figure}[h]
\centering
\begin{tikzpicture}[scale=1.8, >=Latex]

    \def\r{1.55}
    \def\angP{122}
    \def\angQ{32}
    \def\tpar{0.56}

    \coordinate (O) at (0,0);
    \coordinate (P) at ({\r*cos(\angP)},{\r*sin(\angP)});
    \coordinate (Q) at ({\r*cos(\angQ)},{\r*sin(\angQ)});
    \coordinate (H) at ($(P)!\tpar!(Q)$);
    \pgfmathsetmacro{\angT}{(1-\tpar)*\angP + \tpar*\angQ}
    \coordinate (G) at ({\r*cos(\angT)},{\r*sin(\angT)});

    \fill[white] (O) circle (\r);
    \draw[thick] (O) circle (\r);

    \draw[thin,gray!55] (O) -- (P);
    \draw[thin,gray!55] (O) -- (Q);

    \draw[very thick, blue!70, dashed] (P) -- (Q);

    \draw[very thick, red!80!black]
        ({\angP}:\r) arc[start angle=\angP,end angle=\angQ,radius=\r];

    \pic[draw,->,angle radius=0.38cm,angle eccentricity=1.18]
        {angle=Q--O--P};
    \node at (0.14,0.42) {$\theta(P,Q)$};

    \fill (P) circle (0.018);
    \fill (Q) circle (0.018);
    \fill[blue!70] (H) circle (0.0);
    \fill[red!80!black] (G) circle (0.0);

    \node[above left=2pt] at (P) {$s_P$};
    \node[right=3pt] at (Q) {$s_Q$};
    \node[below left=0.15pt, blue!70] at (H) {$s_t^{\mathrm{H}}$};
    \node[above right=0.5pt, red!80!black] at (G) {$s_t^{\mathrm{SH}}$};

    \node[align=center] at (0,-0.36)
        {};


\end{tikzpicture}
\caption{
Square-root geometry for probability measures. The dashed chord represents the ordinary Hellinger geodesic and the solid arc represents the spherical Hellinger geodesic.}
\label{fig:hellinger_vs_fisherrao}
\end{figure}

\begin{proposition}[Spherical Hellinger geodesic]
\label{fr_geodesic}
Let $P$ and $Q$ be probability measures such that $P \neq Q$, let $\lambda$ dominate both measures, and let $\theta = \theta(P, Q)$ be defined as in \cref{theta_definition}. For $t\in[0,1]$, define
\begin{align*}
    \alpha_t(\theta) = \frac{\sin\{(1-t)\theta\}}{\sin(\theta)}, \qquad \beta_t(\theta) = \frac{\sin(t\theta)}{\sin(\theta)}.
\end{align*}
Then, the constant-speed Spherical Hellinger geodesic between $P$ and $Q$ is the probability measure $\nu^{\mathrm{SH}}_t$ satisfying
\begin{align*} 
    \sqrt{\frac{d\nu^{\mathrm{SH}}_t}{d\lambda}}
    =
    \alpha_t(\theta)
    \sqrt{\frac{dP}{d\lambda}}
    +
    \beta_t(\theta)
    \sqrt{\frac{dQ}{d\lambda}}.
\end{align*}
\end{proposition}

One benefit of formulating \cref{fr_geodesic} for probability measures, rather than only for Lebesgue densities, is that we may take the target distribution to be a point mass directly. Although $\delta_{a_\star}$ has no density with respect to Lebesgue measure, it has an ordinary Radon--Nikodym derivative with respect to a common dominating measure such as $\lambda_x=\Pi_x+\delta_{a_\star}$, 
where $\Pi_x$ is the distribution of $A \mid X = x$. Thus, the measure-level square-root representation remains well defined. Under this framework, we can construct an analogous version of \cref{wasserstein_geodesic_density_point_mass} for the $L^2$ sphere. 
\begin{corollary}[Spherical Hellinger geodesic to a point mass] \label{fr_pointmass}
    Suppose that $\Pi_x(\{a_\star\})=0$ for $P_X$-almost every $x$ (for example, suppose $\Pi_x$ is absolutely continuous with respect to Lebesgue measure), and let $\delta_{a_\star}$ be the target measure. Then, the angle between $\Pi_x$ and $\delta_{a_\star}$ is $\pi/2$ and the geodesic is
\begin{align*}
    \nu_t^{\mathrm{SH}}(\,\cdot\mid x) = \cos^2\left(\frac{\pi t}{2}\right)\Pi_x + \sin^2\left(\frac{\pi t}{2}\right)\delta_{a_\star}.
\end{align*}
\end{corollary}

The path defined in \cref{fr_pointmass} has a very different interpretation from the Wasserstein geodesic. Under the Wasserstein intervention, every natural treatment value is moved continuously toward $a_\star$. Under the Spherical Hellinger intervention, no treatment value is moved. Instead, the intervention gradually removes mass from the entire natural treatment law and creates an atom at $a_\star$. In this sense,  the square-root geometry defines a reweighting (or reaction) geometry rather than a transport geometry \citep{liero2016reaction}. A second, related consequence is that Spherical Hellinger geometry does not encode the physical location of the target treatment level. If $\Pi_x$ is atomless, then $\Pi_x$ is mutually singular with $\delta_a$ for every $a\in\mathcal{A}$. Hence, every point mass is the same Spherical Hellinger distance from the natural treatment distribution, regardless of whether a particular $a \in \mathcal{A}$ is near the center of the treatment distribution or at the edge of its support. Third, we can see that under the definition of \cref{fr_pointmass}, $\nu^{\mathrm{SH}}_t( \, \cdot \mid x) \not \ll \Pi_x$ for all $t \in (0, 1]$. Thus, identification of the incremental Spherical Hellinger effect requires some form of positivity. Specifically, whenever $\psi_\star = \mathbb{E}[\mu(X,a_\star)]$ is identified (e.g.\ under consistency, exchangeability, and appropriate local support around $a_\star$), as well as under regularity of the response
surface $\mu(x, a)$, then it follows that
\begin{align} \label{fr_effect}
    \psi_{\mathrm{SH}}(t) =  \cos^2\left(\frac{\pi t}{2}\right)\mathbb{E}[Y] + \sin^2\left(\frac{\pi t}{2}\right)\psi_\star.
\end{align}
The simple form of \cref{fr_effect} is also noteworthy when it comes to estimation. In an unrestricted nonparametric model with a continuous treatment, the pointwise dose-response functional $\psi_\star$ is generally not pathwise differentiable
\citep{diaz2013targeted, kennedy2017cte}. Consequently, for every fixed $t\in(0,1]$, $\psi_{\mathrm{SH}}(t)$ is also not pathwise differentiable and does not possess a nonparametric efficient influence function \citep{bickel1998efficient,van2000asymptotic}. Finally, \cref{fr_effect} highlights an important point regarding the curvature of the Spherical Hellinger intervention. If we reparameterize the path by
\begin{align*}
    u = \sin^2\left(\frac{\pi t}{2}\right),
\end{align*}
then the effect curve becomes $\psi_{\mathrm{SH}}(u) = (1-u)\mathbb{E}[Y]+u\psi_\star,$ which is affine in $u$. Thus, even though $\nu^{\mathrm{SH}}_t$ represents a geodesic in the $L^2$ sphere, it still introduces path-induced curvature. 

\begin{remark}[Support-Adaptive Interventions] The Spherical Hellinger geodesic appears to have several apparent drawbacks due to requiring positivity, nonregularity of its estimator, and its path-induced curvature. However, an interesting direction for future work could be to develop support-adaptive interventions; by changing the target from a sharp intervention to some kind of localized intervention, we can address many of these limitations. For example, let $g_h(x, a)$ be a known, nonnegative localization function centered at $a_{\star}$ where $h > 0$ is some chosen bandwidth, e.g., a Gaussian bump
\begin{align*}
    g_h(x, a) = \exp\left\{-\frac{1}{2}\left(\frac{a-a_{\star}}{h}\right)^2\right\},
\end{align*}
a compactly supported kernel $K((a-a_{\star})/h)$, or the target-window indicator $\mathbb{I}(|a-a_{\star}|\leq h)$. Assuming $0<\mathbb{E}[g_h(X,A)\mid X=x]<\infty$ for $P_X$-almost every $x$, we define the support-adaptive local endpoint to be
\begin{align} \label{adaptive_endpoint}
    q_{h}(a\!\mid\!x) = \frac{g_h(x, a)\pi(a\!\mid\!x)}{\mathbb{E}[g_h(X, A) \mid X = x]}.
\end{align}
This assignment-adaptive interpretation mimics the incremental propensity score and exponential tilt interventions, which also depend on $\pi(a \! \mid \! x)$ \citep{kennedy2019nonparametric, schindl2024incremental}. This structure guarantees $q_{h}(a\!\mid\!x) = 0$ whenever $\pi(a\!\mid\!x)=0$ (under the Spherical Hellinger geodesic) and thereby allows us to avoid global positivity. Notably, there are now two levers to pull: how far along the path you are (as indexed by $t$), and how localized the intervention is (as indexed by $h$).
    
\end{remark}

\section{Simulations \& Application} \label{simulation_section}

In this section we validate our theoretical findings via simulation and an applied data example. In particular, following \cref{curvature_section}, we compare incremental effects induced by the Wasserstein and exponential-tilt-based intervention distributions. We choose an outcome regression that is affine in treatment assignment to illustrate how the intervention path and its parameterization affect the shape of the effect curve. Finally, we then analyze the U.S. Job Corps training data, using length of training as a continuous exposure.

\subsection{Simulations} \label{simulation_subsection}

Let $X_1,\ldots,X_4 \sim \operatorname{Uniform}(-1,1)$. Then, define $L_1(a)=2a-1$ and $L_2(a)=6a^2-6a+1$ for treatment $a \in [0, 1]$. We define the conditional treatment density as 
\begin{align*}
    \pi(a\mid x)&=1+\theta_1(x)L_1(a)+\theta_2(x)L_2(a)
\end{align*}
where $\theta_1(x)=0.20x_1-0.12x_2+0.08x_1x_2$ and $\theta_2(x)=0.08x_3+0.06(x_4^2-1/3)$. We then generate the outcomes $Y=\mu(X,A)+(0.8+0.2A+0.1X_1^2)\varepsilon$ where
\begin{align*}
    \mu(x,a)&=1+0.5x_1-0.4x_2+0.3x_3^2+0.2x_4
        +(1+0.3x_1-0.2x_2)a,
\end{align*}
and where $\varepsilon\sim\operatorname{Unif}(-\sqrt{3},\sqrt{3})$ is independent of $(X,A)$. Thus, we can see that $\mu(x, a)$ is affine in the treatment assignment $a$. Here, we target the sharp intervention $a_\star=1$. To plot the exponentially tilted intervention on the same axis as the Wasserstein intervention, we index $q_\delta(a \mid x)$, as defined in \cref{exponential_tilt_def}, by $\delta(t)=t/(1-t)$. In our simulation we use 2,500 repetitions at each $n\in\{500, 5000\}$, with five-fold cross-fitting. We estimate $\mu$ by a regression forest using treatment and all covariates, and $\pi$ by conditional kernel density estimation with bandwidths selected within each training fold. These learners allow nonlinear treatment effects and covariate interactions. Finally, we compute the one-step estimators and nominal $95\%$ pointwise Wald intervals using the sample variances of their cross-fitted score contributions.

\begin{figure}[h]
    \centering

    \definecolor{fanorange}{HTML}{E76F51} 
    \definecolor{fanblue}{HTML}{00BFC4}

    \pgfplotstableread[col sep=comma]
        {files/wasserstein_exponential_fan_plot.csv}\fanresults
    \pgfplotstableread[col sep=comma]
        {files/wasserstein_exponential_truth.csv}\fantruth

    \newcommand{\plotFanPanel}[1]{%
        \addplot[name path=fanWlower#1, draw=none, forget plot]
            table[x=t,y=wasserstein_n#1_average_ci_lower]{\fanresults};
        \addplot[name path=fanWupper#1, draw=none, forget plot]
            table[x=t,y=wasserstein_n#1_average_ci_upper]{\fanresults};
        \addplot[fill=fanblue, fill opacity=0.16, draw=none, forget plot]
            fill between[of=fanWlower#1 and fanWupper#1];

        \addplot[name path=fanElower#1, draw=none, forget plot]
            table[x=t,y=exponential_tilt_n#1_average_ci_lower]{\fanresults};
        \addplot[name path=fanEupper#1, draw=none, forget plot]
            table[x=t,y=exponential_tilt_n#1_average_ci_upper]{\fanresults};
        \addplot[fill=fanorange, fill opacity=0.16, draw=none, forget plot]
            fill between[of=fanElower#1 and fanEupper#1];

        \addplot[color=fanblue, thick, no marks, forget plot]
            table[x=t,y=wasserstein_truth]{\fantruth};
        \addplot[color=fanorange, thick, dashed, no marks, forget plot]
            table[x=t,y=exponential_tilt_truth]{\fantruth};

        \addplot[color=fanblue, only marks, mark=*, mark size=1.3pt,
            mark options={fill=white, line width=0.4pt}, forget plot]
            table[x=t,y=wasserstein_n#1_estimate]{\fanresults};
        \addplot[color=fanorange, only marks, mark=triangle*, mark size=1.8pt,
            mark options={fill=white, line width=0.4pt}, forget plot]
            table[x=t,y=exponential_tilt_n#1_estimate]{\fanresults};
    }

    \begin{tikzpicture}
    \begin{groupplot}[
group style={
    group size=2 by 1,
    horizontal sep=0.055\textwidth,
    ylabels at=edge left,
    yticklabels at=all
},
        width=0.525\textwidth,
        height=0.395\textwidth,
        xmin=0, xmax=1,
        xtick={0,0.2,0.4,0.6,0.8,1},
        scaled ticks=false,
        ylabel={Incremental effect},
        tick label style={font=\small},
        title style={font=\small\bfseries},
        label style={font=\small},
        grid=major,
        grid style={line width=0.1pt, draw=gray!20},
        xtick pos=bottom,
        ytick pos=left,
        clip=true,
        legend cell align=left,
        legend columns=2,
        legend style={draw=none, font=\small,
            /tikz/every even column/.append style={column sep=1em}}
    ]
        \nextgroupplot[title={$n=500$},
          ymin=1, ymax=3,
    ytick={1,1.5,2,2.5,3},
    legend to name=simulationfanlegend]
        \plotFanPanel{500}
        \addlegendimage{color=fanblue, thick, mark=*, mark size=1.3pt,
            mark options={fill=white}}
        \addlegendentry{Wasserstein}
        \addlegendimage{color=fanorange, thick, dashed,
            mark=triangle*, mark size=1.8pt, mark options={fill=white}}
        \addlegendentry{Exponential tilt}

        \nextgroupplot[title={$n=5{,}000$},
          ymin=1.5, ymax=2.25]
        \plotFanPanel{5000}
    \end{groupplot}

    \node[anchor=north, font=\small] at
        ($(group c1r1.south)!0.5!(group c2r1.south)+(0,-0.65cm)$)
        {Path index, $t$};
    \node[anchor=north, inner sep=0pt] at
        ($(group c1r1.south)!0.5!(group c2r1.south)+(0,-1.25cm)$)
        {\pgfplotslegendfromname{simulationfanlegend}};
    \end{tikzpicture}

    \caption{Incremental effects for $n=500$ and $n=5{,}000$.}
    \label{wasserstein_vs_exp_tilt}
\end{figure}

\cref{wasserstein_vs_exp_tilt} highlights a number of important points. First, we can see that the endpoints under both interventions are the same: both interpolate between $\mathbb{E}[Y]$ and $\mathbb{E}[Y(1)]$. Thus, the sharp intervention itself is invariant to the choice of path. However, there is a clear difference between the two interventions in the intermediate regime that lies between these two endpoints. Clearly, the exponential tilt produces a curved effect due to the parameterization choice of $\delta$, whereas the Wasserstein intervention produces a linear effect curve, as inherited from the affine-in-$a$ outcome regression. Second, we can see that there is increasing uncertainty as $t \to 1$, as outlined in \cref{var_corollary}. The confidence intervals of both estimators greatly widen toward the end of the path; in particular, there is a dramatic increase around $t \approx 0.9$. This may point to the elbow phenomenon outlined in \cref{minimax_elbow}, as the difficulty in estimation transitions toward nonparametric behavior.

\subsection{Empirical Application} \label{application_subsection}

In this section we analyze data from the U.S. Job Corps training program, an experiment conducted in the 1990s in which access to vocational training was randomly assigned to low-income youth. Participants in the study received 1,200 hours of training, on average, across an eight-month time period. \cite{schochet2008} report a positive impact, finding increased educational attainment, lower criminality, and increased employment after the program was conducted. There have been many subsequent analyses exploring the effects of the program beyond participation \citep{flores2012, frolich2017, huber2020, Colangelo02012026}; in particular, \cite{Colangelo02012026} estimate the dose-response relationship between training hours and employment. They find a complex, non-monotone dose-response relationship between training hours and employment; however, there are generally diminishing returns across hours.

We use the public replication data of \citet{Colangelo02012026}. Here, the exposure is the total number of hours spent in academic and vocational instruction during the first year after assignment, and $Y$ is the percentage of weeks employed during the second year. We adjust for the supplied baseline demographic, educational, household, and labor-market variables, including missingness indicators. Following \citet{Colangelo02012026}, we analyze the 4{,}024 individuals who have completed at least 40 hours of training. Note that randomization only admits access to the Job Corps, and therefore we assume conditional exchangeability of training intensity given the baseline covariates (along with consistency and positivity). We choose $a_\star \in \{500,1800\}$ as our target outcomes. Since roughly 1{,}200 hours of training were received, on average, 500 hours represents a ``minimal'' intervention, and 1{,}800 would be a significant ramp-up in the intensity of treatment. We estimate the incremental Wasserstein effect curves toward using five-fold cross-fitting. The outcome regression $\mu$ is estimated using Super Learner \citep{vanderLaan2007SuperLearner, polley2019package}, combining a sample-mean benchmark, LASSO, and regression forests, with weights obtained by normalized nonnegative least squares. The treatment density $\pi$ is estimated using an ensemble of conditional histograms fitted with probability forests, with weights chosen by cross-validated likelihood. Forests are implemented using the \texttt{ranger} package \citep{wright2017}. We compute the cross-fitted one-step estimator for each $t$ and construct nominal $95\%$ pointwise Wald intervals. Estimates and intervals are reported on the percentage scale.

\begin{figure}[h]
    \centering

    \definecolor{joborange}{HTML}{E76F51} 
    \definecolor{jobblue}{HTML}{00BFC4}

    \pgfplotstableread[col sep=comma,ignore chars={"}]
        {files/job_corps_incremental_curve.csv}\jobresults
    \pgfplotstablegetelem{0}{observed_mean}\of\jobresults
    \edef\jobBaseline{\pgfplotsretval}

    \pgfplotsset{
        keep target/.style={
            restrict expr to domain={\thisrow{target_hours}}{#1:#1},
            unbounded coords=discard
        }
    }

    \begin{tikzpicture}
    \begin{axis}[
        width=0.90\textwidth,
        height=0.42\textwidth,
        xlabel={Incremental path ($t$ increases away from the center)},
        ylabel={Second-year weeks employed (\%)},
        xmin=-1.04, xmax=1.04,
        ymin=35, ymax=60,
        xtick={-1,-0.5,0,0.5,1},
        xticklabels={
            $\mathbb{E}[Y(500)]$,$t=0.5$,$\mathbb{E}[Y]$,
            $t=0.5$,$\mathbb{E}[Y(1800)]$
        },
        ytick={30,40,50,60},
        scaled ticks=false,
        tick label style={font=\small},
        label style={font=\small},
        grid=major,
        grid style={line width=0.1pt, draw=gray!20},
        axis lines=box,
        axis line style={draw=black, line width=0.4pt},
        tick align=inside,
        tick style={draw=gray!70, line width=0.4pt},
        xtick pos=bottom, ytick pos=left,
        filter discard warning=false,
        clip=true,
        legend columns=2,
        legend cell align=left,
        legend style={
            at={(0.5,-0.23)}, anchor=north,
            draw=none, font=\small,
            /tikz/every even column/.append style={column sep=1em}
        }
    ]
        \addplot[name path=jobLow500, draw=none, forget plot,
            keep target=500]
            table[x=signed_path,y=conf_low]{\jobresults};
        \addplot[name path=jobHigh500, draw=none, forget plot,
            keep target=500]
            table[x=signed_path,y=conf_high]{\jobresults};
        \addplot[fill=jobblue, fill opacity=0.16, draw=none, forget plot]
            fill between[of=jobLow500 and jobHigh500];

        \addplot[name path=jobLow1800, draw=none, forget plot,
            keep target=1800]
            table[x=signed_path,y=conf_low]{\jobresults};
        \addplot[name path=jobHigh1800, draw=none, forget plot,
            keep target=1800]
            table[x=signed_path,y=conf_high]{\jobresults};
        \addplot[fill=joborange, fill opacity=0.16, draw=none, forget plot]
            fill between[of=jobLow1800 and jobHigh1800];

        \addplot[gray, dashed, thick, no marks, forget plot]
            coordinates {(-1.04,\jobBaseline) (1.04,\jobBaseline)};
        \draw[gray!65, densely dotted]
            (rel axis cs:0.5,0) -- (rel axis cs:0.5,1);

        \addplot[color=jobblue, thick, mark=*, mark size=1.3pt,
            mark options={fill=white, line width=0.4pt}, keep target=500]
            table[x=signed_path,y=estimate]{\jobresults};
        \addlegendentry{Toward 500 hours}

        \addplot[color=joborange, thick, mark=*, mark size=1.3pt,
            mark options={fill=white, line width=0.4pt}, keep target=1800]
            table[x=signed_path,y=estimate]{\jobresults};
        \addlegendentry{Toward 1,800 hours}

        \addplot[only marks, mark=*, mark size=1.8pt,
            mark options={draw=gray!80, fill=white, line width=0.7pt},
            forget plot] coordinates {(0,\jobBaseline)};
    \end{axis}
    \end{tikzpicture}

    \caption{Estimated employment under Wasserstein interventions toward
    500 and 1,800 training hours. Shading represents nominal 95\% pointwise
    confidence intervals; the dashed horizontal line marks the observed
    mean employment.}
    \label{job-corps-incremental}
\end{figure}

In \cref{job-corps-incremental} we find that concentrating training hours toward 500 hours generally increases estimated second-year employment, whereas concentration toward 1,800 hours does not produce any meaningful change in employment. This suggests limited additional employment returns to greater training intensity: concentrating hours around a moderate duration may yield employment gains without the additional time required by a more intensive intervention. Importantly, note that each intervention increases hours for participants below its target and decreases hours for those above it, so our comparison concerns the redistribution of training hours across the analysis population, rather than uniformly increasing or decreasing training. This interpretation complements the non-monotone dose-response estimates of \cite{Colangelo02012026}, whose dose-response estimates exhibit local peaks near 500 and 1,000–1,200 hours and declining employment around 1,500 hours.

\section{Discussion \& Conclusion} \label{conclusion_section}

In this paper we introduced {\em causal geodesy}, a framework for defining smooth paths of stochastic interventions that interpolate between the observed treatment density and a point-mass. Each of these paths defines a causal effect curve that moves from a purely correlational quantity to a sharp causal intervention. In particular, we considered the shortest paths between these distributions, i.e. \textit{geodesics}, with respect to the Wasserstein and Spherical Hellinger distances. Our analysis show that the choice of path governs both the interpretation of the resulting effect curve and the statistical difficulty of estimating it.

In particular, through our curvature decomposition we separated the variation inherited from the outcome regression from curvature introduced by acceleration along the intervention path; we showed that the Wasserstein intervention introduces no parameterization artifacts. Next, we developed inference for the incremental Wasserstein effect with an explicit dependence on the path parameter. As $t_n\to1$, the intervention concentrates onto a shrinking region, and the effective sample size is of order $n(1-t_n)$ rather than $n$. Our asymptotic theory accounts for this localization through combined global and local $L_2$ conditions on nuisance estimation, establishing asymptotic normality even as the efficiency bound diverges. In a Gaussian known-design submodel, we characterized a minimax elbow at which estimation of the increasingly concentrated intervention reaches the nonparametric rate for point evaluation; these results describe how statistical difficulty changes along the path. We also showed how this geometric structure provides a way to conduct inference for sharp interventions. By estimating derivatives of the incremental effect curve, we develop an order-$p$ correction that reduces the approximation bias to $O((1-t_n)^{p+1})$. We then established asymptotic normality for the corrected estimator when $n(1-t_n)\to\infty$ and $n(1-t_n)^{2p+3}\to 0$. Finally, our comparison with spherical Hellinger geometry illustrates that these properties do not follow from geodesicity alone. We show the incremental Spherical Hellinger effect can exhibit path-induced curvature despite being a geodesic. Thus, choosing a shortest path does not by itself ensure favorable overlap, regularity, or interpretability.

Within the causal geodesy framework, there are many open questions and avenues for future work. First, in this paper we only consider a single intervention; it would be mathematically interesting and useful to extend this framework to longitudinal or time-varying interventions. A second direction is to extend the framework to multivariate treatments. In more than one dimension, treatment values may be able to move around regions of zero density rather than cross them; optimal transport with obstacle-induced distances \citep{cavalletti2012} suggests a way to construct interventions that respect positivity. Finally, it would be interesting to construct marginal structural models \citep{robins2000} on the incremental effect curve that pool information across values of $t$. Our curvature results suggest an interesting basis for such a model: polynomial structure in the outcome regression induces polynomial structure along the Wasserstein path. Thus, one could fit a model in the low-variance regions of the path and extrapolate to estimate sharp interventions.

\section*{Acknowledgments}

GPT 5.6 Sol was used to write the \texttt{R} code supporting the simulation study and application. \medskip

\noindent This article/paper is a product of the Iowa Agriculture and Home Economics Experiment Station, Ames, Iowa. Project No. IOW05806 is supported by USDA/NIFA and State of Iowa funds. Any opinions, findings, conclusions, or recommendations expressed in this publication are those of the authors and do not necessarily reflect the views of the U.S. Department of Agriculture.

\section*{References}
\vspace{-1cm}
\bibliographystyle{abbrvnat}
\bibliography{references}

\setlength{\parindent}{0cm}
\appendix

\begin{center}
{\large\bf SUPPLEMENTARY MATERIAL}
\end{center}

\begin{description}

\item \cref{jackknife_supplement}: Contains a discussion of a generalized jackknife estimator for derivative effects.

\item \cref{sec:proofs}: Contains all proofs from the main text, including:
\begin{description}
    \item \cref{wasserstein_geodesic_density_point_mass_proof}: Proof of \cref{wasserstein_geodesic_density_point_mass}.
    \item \cref{curvature_decomposition_proof}: Proof of \cref{curvature_decomposition}.
    \item \cref{effect_curvature_proof}: Proof of \cref{effect_curvature}.
    \item \cref{eif_theorem_proof}: Proof of \cref{eif_theorem}.
    \item \cref{var_corollary_proof}: Proof of \cref{var_corollary}.
    \item \cref{wasserstein_asymptotics_proof}: Proof of \cref{wasserstein_asymptotics}.
    \item \cref{derivative_EIF_proof}: Proof of \cref{derivative_EIF}.
    \item \cref{minimax_elbow_proof}: Proof of \cref{minimax_elbow}.
    \item \cref{higher_order_normality_proof}: Proof of \cref{higher_order_normality}.
    \item \cref{fr_geodesic_proof}: Proof of \cref{fr_geodesic}.
    \item \cref{fr_pointmass_proof}: Proof of \cref{fr_pointmass}.

\end{description}
\end{description}

\section{Generalized Jackknife Estimator for Derivative Effects} \label{jackknife_supplement}

An alternative way to remove endpoint bias is to replace the path derivatives by finite differences of the original incremental Wasserstein effect estimator. This approach uses only estimators of $\psi_\nu(s)$ at regular points $s<1$ and therefore avoids both direct estimation of $\mu^{(k)}$ and $\pi^{(k)}$ and the no-flux condition needed to derive influence functions for the path derivatives. For example, for any fixed $ \omega >1$ such that $\omega(1-t_n)\leq1$, we can approximate the sharp endpoint with
\begin{align*}
\psi_{\nu,1}^{\, \mathrm{R}}(t_n;\omega) = \frac{ \omega\psi_\nu(t_n) - \psi_\nu(1-\omega(1-t_n))
}{\omega - 1}.
\end{align*}
Thus, the estimated $\widehat{\psi}_{\nu,1}^{\, \mathrm{R}}(t_n;\omega)$ can be thought of as a generalized jackknife estimator \citep{Cattaneo2013_jackknife} or a two-scale Richardson extrapolation \citep{richardson27, bach2021}. For higher-order bias corrections one can choose $1=\omega_0<\omega_1<\cdots<\omega_p$ satisfying $\omega_p (1-t_n)\leq1$ to be distinct positive constants, which yields a weighted estimator
\begin{align*}
\widehat\psi_{\nu,p}^{\,\mathrm{R}}(t_n) = \sum_{j=0}^p \alpha_j\widehat\psi_\nu(1-\omega_j(1-t_n)).
\end{align*}
Here, the weights $\alpha_0,\ldots,\alpha_p$ are chosen to satisfy $\sum_{j=0}^p \alpha_j=1$ and $\sum_{j=0}^p \alpha_j \omega_j^\ell=0$, for $\ell=1,\ldots,p$. This removes the first $p$ bias terms since, after applying a Taylor expansion, it follows that
\begin{align*}
    \psi_{\nu,p}^{\,\mathrm R}(t_n)-\psi_\star &= \sum_{\ell=1}^{p} b_\ell (1-t_n)^\ell \underbrace{\left(\sum_{j=0}^{p}\alpha_j\omega_j^\ell \right)}_{= \, 0} + b_{p+1}(1-t_n)^{p+1} \sum_{j=0}^{p}\alpha_j\omega_j^{p+1} +o((1-t_n)^{p+1}),
\end{align*}
where $b_j =  \frac{1}{j!} \mathbb{E}[(A-a_\star)^j  \frac{\partial^j}{\partial a^j}\mu(X,a)|_{a=a_\star}]$. Thus, the approximation bias under the Richardson extrapolation is $O((1-t_n)^{p+1})$. Specifically, the weights $\alpha_0, \ldots, \alpha_p$ are uniquely determined by the Vandermonde system
\begin{align} \label{vandermonde}
    \begin{pmatrix}
        1 & 1 & \cdots & 1 \\
        \omega_0 & \omega_1 & \cdots & \omega_p \\
        \vdots & \vdots & & \vdots \\
        \omega_0^p & \omega_1^p & \cdots & \omega_p^p
    \end{pmatrix}
    \begin{pmatrix}
        \alpha_0\\ \alpha_1\\ \vdots\\ \alpha_p
    \end{pmatrix}
    =
    \begin{pmatrix}
        1\\0\\ \vdots\\0
    \end{pmatrix}.
\end{align}
Distinctness of the $\omega_j$ makes this matrix nonsingular. Because the weights are deterministic, the corresponding estimator has the simple influence function 
\begin{align*}
    \varphi_{\nu,p}^{\,\mathrm R}(Z;t) = \sum_{j=0}^{p}\alpha_j \varphi_\nu(Z;1-\omega_j(1-t)).
\end{align*}
Consequently, under the regularity conditions established in \cref{wasserstein_asymptotics}, it follows that
\begin{align*}
    \widehat\psi_{\nu,p}^{\,\mathrm R}(t_n)-\psi_\star = O_P\!\left(\{n(1-t_n)\}^{-1/2}\right) +O\!\left((1-t_n)^{p+1}\right).
\end{align*}
Consequently, $\widehat\psi_{\nu,p}^{\,\mathrm R}(t_n)$ has the same bias order, convergence rate, and endpoint undersmoothing condition $n(1-t_n)^{2p+3}\to0$ as the derivative-based $p$th-order correction.

Despite the convenience of the higher-order Richardson extrapolation, it is often both statistically and numerically unstable. As the order of the approximation increases, the solution of the Vandermonde system in \cref{vandermonde} becomes poorly conditioned \citep{bach2021}. Furthermore, $\widehat\psi_{\nu,p}^{\,\mathrm R}(t_n)$ is sensitive to the choice of $\omega_0, \ldots, \omega_p$. This problem is noticible even for a first-order approximation. Writing $s_\omega=1-\omega(1-t)$, $\varphi_t=\varphi_\nu(Z;t)$, $\sigma_t^2=\mathbb{V}(\varphi_t)$, and $\sigma_{t,s_\omega}=\mathrm{Cov}(\varphi_t,\varphi_{s_\omega})$, we have
\begin{align*}
    \mathbb{V}(\varphi_{\nu,1}^{\,\mathrm R}(Z;t)) = \frac{\omega^2\sigma_t^2+\sigma_{s_\omega}^2-2\omega\sigma_{t,s_\omega}}{(\omega-1)^2}.
\end{align*}
It is easy to see that the resulting variance is sensitive to the choice of $\omega$, especially as $\omega \to 1$. The choice of $\omega$ creates a clear tradeoff: if $\omega$ is close to one, the variance can blow up. However, if $\omega$ is large, $\omega(1 - t)$ is farther from the endpoint and consequently the remaining approximation bias is larger. Thus, we recommend this approach only when there is reason to believe that the no-flux condition does not hold at the boundaries of the treatment support.

\section{Proofs}
\label{sec:proofs}

\subsection{Proof of \cref{wasserstein_geodesic_density_point_mass}} \label{wasserstein_geodesic_density_point_mass_proof}

\begin{proof}[\textbf{Proof:}] The proof of \cref{wasserstein_geodesic_density_point_mass} follows directly from the standard change-of-variables formula for pushforward measures. In particular, by Lemma 6.2 of \cite{chewi2024statistical}, if $\mu$ is a density on $\mathbb{R}^d$, $T : \mathbb{R}^d \to \mathbb{R}^d$ is a diffeomorphism, and $\nu = T_{\#}\mu$, then the density of $\nu$ is given by
\begin{align*}
    \nu(T(x)) = \frac{\mu(x)}{|\text{det} \nabla T(x)|}.
\end{align*}
From here, fix $x$, and let $\Pi_{x}$ denote the conditional distribution of $A \mid X=x$, with density $\pi(\, \cdot\mid x)$ supported on $\mathcal{A}$. Since the target distribution is the point mass $\delta_{a_\star}$, any measurable map $T$ satisfying $T_{\#}\Pi_{x}=\delta_{a_\star}$ must satisfy $T(a)=a_\star$ for $\Pi_x$-almost every $a$. We may therefore choose the optimal transport map to be $T^*(a)=a_\star$. Next, recall by \cref{wasserstein_geodesic_definition} that the Wasserstein geodesic is given by the family of probability measures
\begin{align*}
    \Pi_{t, x} = \left\{(1-t)\operatorname{id}+tT^* \right\}_{\#}\Pi_x.
\end{align*}
Next, define $F_t(u) = (1-t)u+ta_\star$. Although $T^*$ itself is not invertible, for every $t\in[0,1)$, the interpolating map $F_t$ is an affine diffeomorphism of $\mathbb{R}$, with an inverse and derivative given by
\begin{align*}
    F_t^{-1}(a) = \frac{a-ta_\star}{1-t} \qquad \text{and} \qquad  \frac{\partial}{\partial a}F_t^{-1}(a) = \frac{1}{1-t},
\end{align*}
respectively. Furthermore, note that $F_t$ maps the original treatment support $\mathcal{A}$ onto the transported support $\mathcal{A}_{t} = F_t(\mathcal{A}) = \left\{(1-t)u+ta_\star:u\in\mathcal{A}\right\}$. Putting everything together and applying the change-of-variables formula, we find that the measure $\Pi_{t,x}$ has density
\begin{align*}
    \nu_t(a\mid x) = \pi(F_t^{-1}(a)\mid x) \left| \frac{\partial}{\partial a} F_t^{-1}(a) \right| \mathbb{I}(a\in\mathcal{A}_{t}) = \frac{1}{1-t} \pi\left( \frac{a-ta_\star}{1-t} \mid x \right) \mathbb{I}(a\in\mathcal{A}_{t}).
\end{align*}
Finally, note that this expression integrates to one: with the change of variables $a = (1 - t)u + t a_\star$ and $da = (1 - t)du$, it follows that
\begin{align*}
    \int_{\mathbb{R}}\nu_t(a\mid x)da &= \int_{\mathcal{A}_{t}} \frac{1}{1-t} \pi\left( \frac{a-ta_\star}{1-t} \mid x \right)da = \int_{\mathcal{A}}\pi(u\mid x)du =1.
\end{align*}
At the endpoint $t=1$, the map $F_t$ is constant and therefore is no longer invertible, so the resulting intervention becomes $\Pi_{1,x} = \delta_{a_\star},$ which does not possess a density with respect to Lebesgue measure.
\end{proof}

\subsection{Proof of \cref{curvature_decomposition}} \label{curvature_decomposition_proof}

\begin{proof}[\textbf{Proof:}]

To begin, note that we assume that there exists a function $M \in L_1(P_X)$ such that
\begin{align*}
    &\sup_{t\in (0, 1)}\int_0^1
    \Bigg[|\mu(x,a)| \left| \frac{\partial}{\partial t} \rho_t(a\mid x) \right| + \left| \frac{\partial}{\partial a}\mu(x,a) \right| |\rho_t(a\mid x)\vec{v}_t(a,x)| \, +
        \\
    &\phantom{\sup_{t\in (0, 1)}\int_0^1
    \Bigg[} \left| \frac{\partial}{\partial t}\vec{v}_t(a,x) \right|\left| \frac{\partial}{\partial a}\mu(x,a) \right| \rho_t(a\mid x) + \left| \vec{v}_t(a,x) \frac{\partial}{\partial a}\mu(x,a) \right| \left| \frac{\partial}{\partial t} \rho_t(a\mid x) \right| \, +
        \\
&\phantom{\sup_{t\in (0, 1)}\int_0^1
    \Bigg[}
        \left|
            \frac{\partial}{\partial a}
            \left\{
                \vec{v}_t(a,x)
                \frac{\partial}{\partial a}\mu(x,a)
            \right\}
        \right|
        |\rho_t(a\mid x)\vec{v}_t(a,x)|
    \Bigg]da
    \leq
    M(x).
\end{align*}
This assumption allows us to justify differentiation under the expectation by the dominated convergence theorem. We now differentiate $\psi_\rho(t)$ with respect to $t$. Observe that
    \begin{align*}
         \psi^\prime_\rho(t) &= \mathbb{E} \! \left[ \int_{\mathcal{A}} \frac{\partial}{\partial t} \left\{ \mu(X, a) \right\} \rho_t(a \! \mid \! X) da \right] + \mathbb{E} \! \left[ \int_{\mathcal{A}}  \mu(X, a) \frac{\partial}{\partial t} \left\{  \rho_t(a \! \mid \! X) \right\}  da \right] \\
        &\overset{(i)}{=} - \mathbb{E} \! \left[ \int_{\mathcal{A}}  \mu(X, a) \frac{\partial}{\partial a} \big\{ \rho_t(a \! \mid \! X) \vec{v}_t(a, X) \big\}  da \right] \\
        &\overset{(ii)}{=}  - \mathbb{E}\left[\left. \mu(X,a) \rho_t(a\mid X)\vec{v}_t(a,X) \right|_{a=0}^{a=1} \right] + \mathbb{E}\left[ \int_0^1 \frac{\partial}{\partial u}  \mu(X, u) \Big|_{u = a}
        \vec{v}_t(a,X) \rho_t(a\mid X) da \right] \\
        &\overset{(iii)}{=} \mathbb{E}_{\rho_t} \! \left[ \vec{v}_t(A, X) \frac{\partial}{\partial u}  \mu(X, u) \Big|_{u = A} \right],
    \end{align*}
    where $(i)$ follows since $\mu(x, a)$ has no dependence on $t$ and by applying the continuity equation $\frac{\partial}{\partial t} \rho_t(a \! \mid \! x) = - \frac{\partial}{\partial a} \{ \rho_t(a \! \mid \! x) \vec{v}_t(a, x) \}$ and $(ii)$ follows by integrating by parts, and $(iii)$ follows under the no-flux assumption, where we use the convention that for an integrable function $f$,
    \begin{align*}
        \mathbb{E}_{\rho_t}[f(X, A)] = \int_{\mathcal{X}}\int_{\mathcal{A}} f(x, a)\rho_t(a\mid x) da\,dP_X(x).
    \end{align*}
    From here, we differentiate again to find that
    \begin{align*}
        \psi^{\prime \prime}_\rho(t) = \mathbb{E}_{\rho_t}\! \left[ \frac{\partial}{\partial t} \vec{v}_t(A, X)  \frac{\partial}{\partial u}  \mu(X, u) \Big|_{u = A} \right] + \underbrace{\mathbb{E}\! \left[\int_{\mathcal{A}} \vec{v}_t(a, X) \frac{\partial}{\partial a}  \mu(X, a) \frac{\partial}{\partial t}\rho_t(a \! \mid \! X) da  \right]}_{:= \Delta }.
    \end{align*}
    Again, we make use of the continuity equation identity $\frac{\partial}{\partial t} \rho_t(a \! \mid \! x) = - \frac{\partial}{\partial a}\{ \rho_t(a \! \mid \! x) \vec{v}_t(a, x) \}$ to see that
    \begin{align*}
         \Delta &= - \mathbb{E}\! \left[\int_{\mathcal{A}} \vec{v}_t(a, X) \frac{\partial}{\partial a} \left\{ \mu(X, a) \right\} \frac{\partial}{\partial a}\{ \rho_t(a \! \mid \! X) \vec{v}_t(a, X) \}  da  \right] \\
         &\overset{(iv)}{=} \mathbb{E} \! \left[ \int_{\mathcal{A}} \frac{\partial}{\partial a}\{ \vec{v}_t(a, X) \frac{\partial}{\partial a} \left\{ \mu(X, a) \right\}  \} \vec{v}_t(a, X) \rho_t(a \! \mid \! X) da \right] \\
         &= \mathbb{E} \! \left[ \int_{\mathcal{A}} \left(\frac{\partial}{\partial a}\vec{v}_t(a,X) \frac{\partial}{\partial a}\mu(X,a) +\vec{v}_t(a,X) \frac{\partial^2}{\partial a^2}\mu(X,a) \right) \vec{v}_t(a, X) \rho_t(a \! \mid \! X) da \right],
    \end{align*}
    where in $(iv)$ we again integrate by parts and apply the no-flux condition. Consequently, after collecting like terms we can see that
    \begin{align*}
    \psi''_\rho(t) = \mathbb{E}_{\rho_t} \! \left[ \vec{v}_t(A, X)^2 \frac{\partial^2 }{\partial u^2} \mu(X, u) \Big|_{u = A} \right] + \mathbb{E}_{\rho_t} \!\left[ \vec{a}_t(A, X) \frac{\partial }{\partial u} \mu(X, u) \Big|_{u = A} \right],
\end{align*}
where $\vec{a}_t(a, x) = \frac{\partial}{\partial t} \vec{v}_t(a, x) + \vec{v}_t(a, x) \frac{\partial}{\partial a} \vec{v}_t(a, x)$.

\end{proof}

\subsection{Proof of \cref{effect_curvature}} \label{effect_curvature_proof}

\begin{proof}[\textbf{Proof:}] For $t\in[0,1]$, define $F_t(a) = (1-t)a+ta_\star$. Next fix $x$, and recall (as discussed in the proof of \cref{wasserstein_geodesic_density_point_mass}) that the conditional Wasserstein intervention measure is the pushforward $\Pi_{t,x} = (F_t)_{\#}\Pi_x$, where $\Pi_x$ denotes the conditional distribution of $A\mid X=x$. Therefore, by the defining property of a pushforward
measure,
\begin{align*}
    \int_{\mathcal{A}} \mu(x,a) d\Pi_{t,x}(a) = \int_{\mathcal{A}} \mu(x,F_t(a)) d\Pi_x(a).
\end{align*}
Averaging over $P_X$ gives us back
\begin{align*}
    \psi_\nu(t) &= \mathbb{E}\left[\mu(X,F_t(A))\right] = \mathbb{E}\left[ \mu(X,(1-t)A+ta_\star)\right].
\end{align*}
Note that this identity holds for every $t\in[0,1]$, including $t=1$, because $F_1(A)=a_\star$ almost surely. Now, for $j\in\{0,\ldots,k\}$, define
\begin{align*}
    G_j(t,X,A) = (a_\star-A)^j \left. \frac{\partial^j}{\partial u^j} \mu(X,u) \right|_{u=F_t(A)}.
\end{align*}
In particular, note that $G_0(t,X,A) = \mu(X,F_t(A))$. Recall that we assume there exists a measurable function $M:\mathcal{X}\to[0,\infty)$ satisfying $\mathbb{E}[M(X)]<\infty$ and
\begin{align} \label{envelope_assumption}
    \max_{0\leq j\leq k} \sup_{a\in\mathcal{A}} \left| \frac{\partial^j}{\partial a^j} \mu(x,a) \right| \leq M(x)
\end{align}
for $P_X$-almost every $x$. Since $A\in[0,1]$ and $a_\star \in[0, 1]$, it follows that $|a_\star-A|\leq 1$. Then, by \cref{envelope_assumption}, we have for each $j\in\{0,\ldots,k\}$ and $t \in [0, 1]$ that $|G_j(t,X,A)| \leq M(X)$. Thus, each $G_j(t,X,A)$ is dominated by an integrable random variable that does not depend on $t$ and consequently, differentiation under the expectation is justified by the dominated convergence theorem. From here, treating
$(X,A)$ as fixed, the chain rule gives
\begin{align*}
    \frac{\partial}{\partial t}
    \mu(X,F_t(A)) &= \left. \frac{\partial}{\partial u} \mu(X,u) \right|_{u=F_t(A)} \frac{\partial}{\partial t}F_t(A) = \underbrace{(a_\star-A) \left. \frac{\partial}{\partial u} \mu(X,u) \right|_{u=F_t(A)}}_{G_1(t,X,A)},
\end{align*}
which gives us that $ \psi_\nu'(t) = \mathbb{E}[G_1(t,X,A)]$. We now proceed by induction. Suppose that, for some $j\in\{1,\ldots,k-1\}$,
\begin{align} \label{wasserstein_induction_hypothesis}
    \psi_\nu^{(j)}(t) = \mathbb{E}[G_j(t,X,A)].
\end{align}
The factor $(a_\star-A)^j$ does not depend on $t$. Hence, another application of the chain rule gives
\begin{align*}
    \frac{\partial}{\partial t} G_j(t,X,A) &= (a_\star-A)^j \left. \frac{\partial^{j+1}}{\partial u^{j+1}} \mu(X,u) \right|_{u=F_t(A)} \frac{\partial}{\partial t}F_t(A)\\
    &= (a_\star-A)^{j+1} \left. \frac{\partial^{j+1}}{\partial u^{j+1}} \mu(X,u) \right|_{u=F_t(A)}\\
    &= G_{j+1}(t,X,A).
\end{align*}
Once again, \cref{envelope_assumption} provides an integrable envelope so we may therefore differentiate
\cref{wasserstein_induction_hypothesis} under the expectation to obtain
\begin{align*}
    \psi_\nu^{(j+1)}(t) = \mathbb{E}[G_{j+1}(t,X,A)],
\end{align*}
which completes the inductive proof. All that remains is to establish continuity of the derivatives. Let $t_m\to t$ for some $t\in[0,1]$. Since $a\mapsto\mu(x,a)$ is $k$-times continuously differentiable, $G_j(t_m,X,A) \longrightarrow G_j(t,X,A)$ almost surely for every $j\leq k$. By \cref{envelope_assumption} and the dominated convergence
theorem,
\begin{align*}
    \mathbb{E}[G_j(t_m,X,A)] \longrightarrow \mathbb{E}[G_j(t,X,A)].
\end{align*}
Thus, each derivative is continuous on $(0,1)$ and extends continuously to the endpoints. At $t=0$ and $t=1$, these limits give the corresponding right- and left-hand derivatives, respectively. In particular,
\begin{align*}
    \psi_\nu^{(j)}(0+) &= \mathbb{E}\left[ (a_\star-A)^j \frac{\partial^j}{\partial a^j} \mu(X,A) \right], \\
    \psi_\nu^{(j)}(1-) &= \mathbb{E}\left[ (a_\star-A)^j \frac{\partial^j}{\partial a^j} \mu(X,a_\star) \right].
\end{align*}
This completes the proof.

\end{proof}

\subsection{Proof of \cref{eif_theorem}} \label{eif_theorem_proof}

\begin{proof}[\textbf{Proof:}] Here, we derive the efficient influence function for the incremental effect under the Wasserstein intervention,
\begin{align*}
    \psi_\nu(t) = \int_{\mathcal{X}} \int_{\mathcal{A}} \mu(x, a) \nu_t(a \! \mid \! x) da \, dP_X(x),
\end{align*}
where the Wasserstein geodesic $\nu_t$ is defined as
\begin{align*}
     \nu_t(a \! \mid \! x) = \frac{1}{1 - t} \pi \left( \frac{a - t a_{\star}}{1 - t} \mid x\right) \mathbb{I}(a\in\mathcal{A}_{t})
\end{align*}
and $\mathcal{A}_t = \left\{(1-t)u+ta_\star:u\in\mathcal{A} \right\}$. Now, let $\{P_\varepsilon:\varepsilon\in(-\delta,\delta)\}$ be an arbitrary regular one-dimensional parametric submodel through $P$ at $\varepsilon=0$, with density $p_\varepsilon$ and score
\begin{align*}
    S(Z) = \left. \frac{\partial}{\partial\varepsilon} \log\left( p_\varepsilon(Z) \right) \right|_{\varepsilon=0}.
\end{align*}
We write $\mu_\varepsilon$, $\pi_\varepsilon$, and $\nu_{t,\varepsilon}$ to be the corresponding nuisance functions and Wasserstein intervention density under $P_\varepsilon$. Note that all quantities without an $\varepsilon$ subscript are evaluated at $\varepsilon=0$. Now, let $F_t(u) = (1-t)u+ta_\star$. Because $\nu_{t,\varepsilon}(\, \cdot\mid x)$ is the conditional distribution of $F_t(A)$ when $A\mid X=x\sim\pi_\varepsilon(\, \cdot\mid x)$, the pushforward identity gives the representation
\begin{align*}
    \psi_\nu(t;P_\varepsilon) = \int_{\mathcal{X}} \int_{\mathcal{A}} \mu_\varepsilon(x,F_t(a)) p_{X,A,\varepsilon}(x,a) da\,dx = \mathbb{E}_{P_\varepsilon} \! \left[ \mu_\varepsilon(X,F_t(A))\right],
\end{align*}
where $p_{X,A,\varepsilon}(x,a)=p_{X,\varepsilon}(x) \pi_\varepsilon(a\mid x)$. Next, define the score for the marginal distribution of $(X,A)$ by $S_{X,A}(X,A) = \mathbb{E}[S(Z)\mid X,A]$, and the conditional outcome score by $S_{Y\mid X,A}(Z) = S(Z)-S_{X,A}(X,A)$. Regularity of the submodel implies
\begin{align*}
    \left. \frac{\partial}{\partial\varepsilon} p_{X,A,\varepsilon}(x,a)\right|_{\varepsilon=0} = p_{X,A}(x,a)S_{X,A}(x,a).
\end{align*}
It also gives, for almost every $(x,a)$,
\begin{align*}
    \dot\mu(x,a) = \left. \frac{\partial}{\partial\varepsilon} \mu_\varepsilon(x,a) \right|_{\varepsilon=0} = \mathbb{E}\left[\{Y-\mu(x,a)\} S_{Y\mid X,A}(Z) \mid X=x,A=a \right]
\end{align*}
since differentiating the conditional mean under the integral gives
\begin{align*}
    \dot\mu(x,a) = \int y\,p(y\mid x,a)S_{Y\mid X,A}(x,a,y)dy,
\end{align*}
and since the conditional score has mean zero, $y$ may be replaced with $y-\mu(x,a)$. From here, differentiating $\mathbb{E}_{P_\varepsilon} \! \left[ \mu_\varepsilon(X,F_t(A))\right]$ at $\varepsilon=0$ yields
\begin{align*}
    \left. \frac{\partial}{\partial\varepsilon} \psi_\nu(t;P_\varepsilon) \right|_{\varepsilon=0} =\int_{\mathcal{X}} \int_{\mathcal{A}} \dot\mu(x,F_t(a)) p_{X,A}(x,a) da\,dx + \mathbb{E}\left[\mu(X,F_t(A))S_{X,A}(X,A) \right].
\end{align*}
We consider the two terms separately. For the first term, the change of variables $u=F_t(a)$ and the definition of the pushforward density imply
\begin{align*}
    \int_{\mathcal{X}} \int_{\mathcal{A}} \dot\mu(x,F_t(a)) \pi(a\mid x)p_X(x) da\,dx = \int_{\mathcal{X}} \int_{\mathcal{A}} \dot\mu(x,u) \nu_t(u\mid x)p_X(x) du\,dx.
\end{align*}
Substituting in $\dot\mu(x,a)$ therefore gives
\begin{align*}
    \int_{\mathcal{X}} \int_{\mathcal{A}} \dot\mu(x,u) \nu_t(u\mid x)p_X(x) du\,dx = \mathbb{E}\left[ \frac{\nu_t(A\mid X)}{\pi(A\mid X)} \{Y-\mu(X,A)\} S_{Y\mid X,A}(Z) \right].
\end{align*}
Moreover, note that
\begin{align*}
    \mathbb{E}\left[\frac{\nu_t(A\mid X)}{\pi(A\mid X)}\{Y-\mu(X,A)\}S_{X,A}(X,A)\right] =0,
\end{align*}
because the weighted residual has conditional mean zero given $(X,A)$. Hence the conditional outcome score in the preceding display may be replaced by the full score $S(Z)$. For the second term, since
$\mu(X,F_t(A))$ is measurable with respect to $(X,A)$ and $\mathbb{E}[S(Z)]=0$, it follows that
\begin{align*}
    \mathbb{E}\left[ \mu(X,F_t(A)) S_{X,A}(X,A)\right] =
    \mathbb{E}\left[\mu(X,F_t(A))S(Z)\right].
\end{align*}
Putting everything together, we now have that
\begin{align*}
    \left. \frac{\partial}{\partial\varepsilon} \psi_\nu(t;P_\varepsilon) \right|_{\varepsilon=0} = \mathbb{E}\left[\varphi_\nu(Z;t,P)S(Z) \right],
\end{align*}
where 
\begin{align*}
    \varphi_\nu(Z; t, P) =  \frac{\nu_t(A \mid X)}{\pi(A \mid X)}\big(Y - \mu(X, A) \big) +  \mu\big(X, F_t(A)\big) - \psi_\nu(t; P).
\end{align*}
Notably, $\varphi_\nu(Z; t)$ is mean zero, and by our integrability assumptions, $\varphi_\nu(\cdot;t,P)\in L_2^0(P)$. Since the tangent space of the nonparametric model is $L_2^0(P)$, this gradient is the canonical gradient and therefore the efficient influence function.
\end{proof}

\subsection{Proof of \cref{var_corollary}} \label{var_corollary_proof}

\begin{proof}[\textbf{Proof:}]
To begin, let $h = 1 - t \in (0, 1]$ and recall that the efficient influence function for the incremental Wasserstein effect is given by
\begin{align*}
    \varphi_\nu(Z; 1-h) =  \frac{\nu_{1-h}(A \! \mid \! X)}{\pi(A \! \mid \! X)}\big(Y - \mu(X, A) \big) +  \mu\big(X, a_\star + h(A - a_\star) \big) - \psi_\nu(1 - h).
\end{align*}
Our first goal is to derive the nonparametric efficiency bound, $\sigma^2_\nu(1-h) = \mathbb{V}(\varphi_\nu(Z; 1-h))$. It follows almost immediately that
\begin{align*}
     \sigma^2_\nu(1-h) = \mathbb{E}\left[\left(\frac{\nu_{1-h}(A \! \mid \! X)}{\pi(A \! \mid \! X)}\right)^2 \mathbb{V}(Y \mid X, A)  \right] + \mathbb{V}\big(\mu\big(X, a_\star + h\{A - a_\star\}\big)\big).
\end{align*}
To see this, note that for any function $f(X, A)$, by applying the law of iterated expectations it follows that $\mathbb{E}\left[f(X, A)\{Y - \mu(X, A)\}\right] = \mathbb{E}\left[ f(X, A)\mathbb{E}\left[Y - \mu(X, A)\mid X, A\right]\right] = 0$. Thus, the cross term drops out of the variance calculation. Furthermore, the first term simplifies to
\begin{align*}
    \mathbb{E}\left[\left(  \frac{\nu_{1-h}(A \! \mid \! X)}{\pi(A \! \mid \! X)}\big(Y - \mu(X, A) \big)\right)^2 \right] &= \mathbb{E}\Bigg[ \left( \frac{\nu_{1-h}(A \! \mid \! X)}{\pi(A \! \mid \! X)} \right)^2\underbrace{\mathbb{E} \left[ \big(Y - \mu(X, A) \big)^2 \mid X, A\right]}_{\mathbb{V}(Y \mid X, A)}\Bigg].
\end{align*}
Finally, note that $a_\star + h(A - a_\star) \mid X = x \sim \nu_{1-h}(\, \cdot  \mid x)$. Therefore, we can instead write the second variance term as $\mathbb{V}\big(\mu\big(X, a_\star + h\{A - a_\star\}\big)\big) = \mathbb{V}_{\nu_{1-h}}(\mu(X, A))$.
By applying the law of total variance, we can further say
\begin{align*}
    \mathbb{V}_{\nu_{1-h}}(\mu(X, A)) = \mathbb{E}\left[ \mathbb{V}_{\nu_{1-h}}(\mu(X,A)\mid X)\right] +\mathbb{V}\left[ \mathbb{E}_{\nu_{1-h}}(\mu(X,A)\mid X) \right].
\end{align*}
This completes the derivation of the nonparametric efficiency bound itself. Next, we lower and upper bound $\sigma^2_\nu(1-h)$ to obtain a growth rate in terms of $h$. \bigskip

To begin, we consider an upper bound for $\sigma^2_\nu(1-h)$. Recall that $\nu_{1-h}(a \! \mid \! x) = \frac{1}{h} \pi\left(a_\star + \frac{a - a_\star}{h} \mid x \right)$. Then, observe that
\begin{align*}
    \mathbb{E}\left[\left(\frac{\nu_{1-h}(A \! \mid \! X)}{\pi(A \! \mid \! X)}\right)^2 \mathbb{V}(Y \mid X, A)  \right] &= \mathbb{E}\left[\int_\mathcal{A} \frac{\nu_{1-h}(a \! \mid \! X)^2}{\pi(a \! \mid \! X)} \mathbb{V}(Y \mid X, a) da \right] \\
    &= \frac{1}{h^2} \mathbb{E}\left[\int_\mathcal{A} \frac{\pi(a_\star + \frac{a - a_\star}{h} \! \mid \! X)^2}{\pi(a \! \mid \! X)} \mathbb{V}(Y \mid X, a) da \right] \\
    &\overset{(i)}{=} \frac{1}{h} \mathbb{E}\left[\int_\mathcal{A}\pi(u \! \mid \! X)^2 \frac{\mathbb{V}(Y \mid X, a_\star + h(u - a_\star))}{\pi(a_\star + h(u - a_\star) \! \mid \! X)}  du \right], 
\end{align*}
where in $(i)$ we make the change of variables $a = a_\star + h(u - a_\star)$. From here, suppose that $0 < \pi_{\min} \leq \pi(a \! \mid \! x) \leq \pi_{\max} < \infty$ and $\mathbb{E}[Y^2 \mid X = x, A = a] \leq \sigma^2_{\max} < \infty$ for all $x \in \mathcal{X}$ and $a \in \mathcal{A}$. Then, because
\begin{align*}
    \mathbb{V}(Y \mid X = x, A = a) \leq \mathbb{E}[Y^2 \mid X = x, A = a] \leq \sigma^2_{\max}
\end{align*}
we can easily obtain the upper bound
\begin{align*}
    \frac{1}{h} \mathbb{E}\left[\int_\mathcal{A}\pi(u \! \mid \! X)^2 \frac{\mathbb{V}(Y \mid X, a_\star + h(u - a_\star))}{\pi(a_\star + h(u - a_\star) \! \mid \! X)}  du \right] \leq \frac{1}{h}\left(\sigma^2_{\max} \frac{\pi_{\max}}{\pi_{\min}} \right)
\end{align*}
since $\int_\mathcal{A}\pi(u \! \mid \! X)^2 du \leq \pi_{\max} \int_\mathcal{A}\pi(u \! \mid \! X)du = \pi_{\max}$. Additionally, again, under the boundedness of $\mathbb{E}(Y^2 \mid X = x, A = a)$ it follows that $\mathbb{V}(\mu(X, a_\star + h(A - a_\star))) \leq \sigma^2_{\max}$ and therefore,
\begin{align*}
    \sigma^2_\nu(1-h) \leq \frac{1}{h}\left(\sigma^2_{\max} \frac{\pi_{\max}}{\pi_{\min}} \right) + \sigma^2_{\max}.
\end{align*}
Next, to obtain a lower bound on $\sigma^2_\nu(1-h)$ we can immediately see that
\begin{align*}
    \sigma^2_\nu(1-h) \geq \mathbb{E}\left[\left(\frac{\nu_{1-h}(A \! \mid \! X)}{\pi(A \! \mid \! X)}\right)^2 \mathbb{V}(Y \mid X, A)  \right]
\end{align*}
since $\mathbb{V}(\mu(X, a_\star + h\{A - a_\star\})) \geq 0$, so we can simply drop it. Then, to further bound this expression, under the assumption that $\mathbb{V}(Y \mid X = x, A = a) \geq \sigma^2_{\min} > 0$ for all $x \in \mathcal{X}$ and $a \in \mathcal{A}$, it is easy to see that 
\begin{align*}
    \frac{1}{h} \mathbb{E}\left[\int_\mathcal{A}\pi(u \! \mid \! X)^2 \frac{\mathbb{V}(Y \mid X, a_\star + h(u - a_\star))}{\pi(a_\star + h(u - a_\star) \! \mid \! X)}  du \right] \geq \frac{1}{h}\left(\frac{\sigma^2_{\min}}{\pi_{\max}} \right) 
\end{align*}
since after applying the Cauchy-Schwarz inequality conditionally on $X = x$,
\begin{align*}
    1 = \left\{\int_{\mathcal{A}} \pi(u \! \mid \! x) du\right\}^2 \leq |\mathcal{A}| \int_{\mathcal{A}} \pi(u \! \mid \! x)^2 du,
\end{align*}
where $|\mathcal{A}|$ is the Lebesgue measure of the treatment support. Since we take $\mathcal{A} = [0, 1]$, $\int_{\mathcal{A}} \pi(u \! \mid \! x)^2 du \geq 1$, and the previous inequality follows. Putting everything together, it follows that
\begin{align*}
    \frac{1}{h}\left(\frac{\sigma^2_{\min}}{\pi_{\max}} \right) \leq \sigma^2_\nu(1-h) \leq \frac{1}{h}\left(\sigma^2_{\max} \frac{\pi_{\max}}{\pi_{\min}} \right) + \sigma^2_{\max}
\end{align*}
and so, for a sufficiently small $h$ we have that $\sigma^2_\nu(1-h) \asymp h^{-1}$. Finally, to complete the proof, next we explore the limit of $h \sigma^2_\nu(1-h)$ as $h \to 0$. \bigskip 

Now, further suppose that, for $P_X$-almost every $x$, that $a \mapsto \pi(a \! \mid \! x)$ and $a \mapsto \mathbb{V}(Y \mid X = x, A = a)$ are continuous at $a_\star$ (or assume one-sided continuity if $a_\star$ lies on a boundary of $\mathcal{A}$) and recall that
\begin{align*}
    \mathbb{E}\left[\left(\frac{\nu_{1-h}(A \! \mid \! X)}{\pi(A \! \mid \! X)}\right)^2 \mathbb{V}(Y \mid X, A)  \right] = \frac{1}{h} \mathbb{E}\left[\int_\mathcal{A}\pi(u \! \mid \! X)^2 \frac{\mathbb{V}(Y \mid X, a_\star + h(u - a_\star))}{\pi(a_\star + h(u - a_\star) \! \mid \! X)}  du \right].
\end{align*}
Then, under the continuity assumption, and by the boundedness conditions previously established, we may apply the dominated convergence theorem to obtain
\begin{align*}
     h \left(\mathbb{E}\left[\left(\frac{\nu_{1-h}(A \! \mid \! X)}{\pi(A \! \mid \! X)}\right)^2 \mathbb{V}(Y \mid X, A)  \right] \right) \longrightarrow \mathbb{E}\left[\frac{\mathbb{V}(Y \mid X, a_\star)}{\pi(a_\star \mid X)} \int_\mathcal{A} \pi(a \mid X)^2 da \right]
\end{align*}
as $h \to 0$. Finally, since $0 \leq \mathbb{V}(\mu(X, a_\star + h\{A - a_\star\})) \leq \sigma^2_{\max}$ and $h \sigma^2_{\max} \longrightarrow 0$ as $h \to 0$, it also follows that $h \mathbb{V}(\mu(X, a_\star + h\{A - a_\star\})) \longrightarrow 0$. Putting everything together,
\begin{align*}
    h \sigma^2_\nu(1-h) \longrightarrow \mathbb{E}\left[\frac{\sigma^2_\star(X)}{\pi_\star(X)} \| \pi( \, \cdot \mid X) \|^2_{L_2(\mathcal{A})} \right]
\end{align*}
as $h \to 0$, where we say $\sigma^2_\star(X) = \mathbb{V}(Y \mid X, A = a_\star)$ and $\pi_{\star}(X) = \pi(a_\star \mid X)$.

\end{proof}

\subsection{Proof of \cref{wasserstein_asymptotics}} \label{wasserstein_asymptotics_proof}

\begin{proof}[\textbf{Proof:}] In this proof we establish asymptotic normality for the one-step estimator of the incremental Wasserstein effect. Let $\phi_{\nu, t_n}(Z; P) = \varphi_{\nu, t_n}(Z; P) + \psi_{\nu, t_n}(P)$ where $\varphi_{\nu, t_n}(Z; P)$ is the efficient influence function defined in \cref{eif_theorem}. Then, the one-step estimator is
\begin{align*}
    \widehat{\psi}_{\nu, t_n} = P_n \{\phi_{\nu, t_n}(\, \cdot \, ; \widehat{P}) \}
\end{align*}
where $P_n$ denotes the empirical distribution. Suppose $t_n \to 1$ and let $\sigma^2_{\nu}(t_n) = \mathbb{V}(\varphi_{\nu, t_n}(Z; P))$, as defined in \cref{var_corollary}. Then, one can decompose the bias-corrected one-step estimator as
\begin{align*}
    \frac{\sqrt{n}}{\sigma_{\nu}(t_n)} \left(\widehat{\psi}_{\nu, t_n}- \psi_{\nu, t_n} \right) = \frac{\sqrt{n}}{\sigma_{\nu}(t_n)} \Big[ S_1 + S_2 + R_2(\widehat{P}, P) \Big]
\end{align*}
where $S_1 = (P_n - P) \{\varphi_{\nu, t_n}(Z; P) \}$, $S_2 = (P_n - P) \{ \varphi_{\nu, t_n}(Z; \widehat{P}) - \varphi_{\nu, t_n}(Z; P) \}$, and the remainder from the von Mises expansion is 
\begin{align*}
    R_2(\widehat{P}, P) = \psi_{\nu, t_n}(\widehat{P}) - \psi_{\nu, t_n}(P) + \int \varphi_{\nu, t_n}(z, \widehat{P}) dP_Z(z).
\end{align*}
For a more detailed background, readers can refer to \cite{kennedy2024semiparametric} for a review. We use $\widehat{P}$ throughout the proof, but assume cross-fitting has been used to learn each nuisance function --- for notational simplicity we suppress the fold index throughout. Conditional on the training sample, the estimated nuisance functions are fixed and independent of the observations in the corresponding validation fold. Since the number of folds is fixed, the following argument may simply be applied fold by fold \citep{Chernozhukov2018}. To facilitate ease of reading, we split this proof into three parts corresponding to each term in the decomposition. For the first term, we will show that Lindeberg's condition holds in order to apply the Central Limit Theorem. For the second term, i.e., the empirical process term, we show that it is of order $o_P\{(n(1-t_n))^{-1/2}\}$. Finally, for the third term we will show the bias $R_2(\widehat{P}, P)$ is a second-order product of errors that vanishes with $\sqrt{n} / \sigma_\nu(t_n)$ under certain rate conditions. In all parts, we make sure to establish an explicit dependence on the path parameter $t_n$. \bigskip

\textbf{Lindeberg's Condition:} For each $n$, let $\varphi_{n,i} = \varphi_{\nu,t_n}(Z_i;P)$. In this section we wish to show that for all $\varepsilon > 0$,
    \begin{align*}
        \lim_{n\to\infty}\frac{1}{s_n^2}\sum_{i=1}^n\mathbb{E}\left[
        \varphi^2_{n, i}  \mathbb{I}\{|\varphi_{n, i}|>\varepsilon s_n\}
        \right] = 0,
    \end{align*}
where $s^2_n = \sum^n_{i=1} \mathbb{V}(\varphi_{n, i})$. By applying \cref{var_corollary} we can immediately obtain the inequality
\begin{align} \label{indicator_ref}
    \mathbb{E}\left[
    \varphi^2_{n, i} \mathbb{I}\{|\varphi_{n, i}|>\varepsilon s_n\}
    \right] \leq \mathbb{E}\left[
    \varphi^2_{n, i} \mathbb{I}\left\{ |\varphi_{n, i}|> \varepsilon \sqrt{\frac{n}{1 - t_n}}\left(\frac{\sigma_{\text{min}}}{\sqrt{\pi_{\max}}}\right)\right\} 
    \right].
\end{align}
From here, our goal is to establish the conditions under which the indicator function in \cref{indicator_ref} converges to zero as $n \to \infty$ and $t_n \to 1$; after applying the dominated convergence theorem, this will imply Lindeberg's condition holds. Thus, we now establish upper bounds for each term in
\begin{align*}
    \varphi_{n, i} =  \frac{\nu_{t_n}(A_i \mid X_i)}{\pi(A_i \mid X_i)}\big(Y_i - \mu(X_i, A_i) \big) +  \mu\big(X_i, (1 - t_n)A_i + t_n a_{\star}\big) - \psi_\nu(t_n).
\end{align*}
First, we note that under the assumption that $|Y| \leq C$ with probability one, it follows that $|\mu(X,A)|=|\mathbb{E}[Y\mid X,A]|\leq C$, so any terms containing the regression function $\mu(X, A)$ can easily be bounded. Next, observe that under the assumption that $0 < \pi_{\min} \leq \pi(a \! \mid \! x) \leq \pi_{\max} < \infty$ for all $x \in \mathcal{X}$ and $a \in \mathcal{A}$ that
\begin{align*}
    \frac{\nu_{t_n}(A_i \mid X_i)}{\pi(A_i \mid X_i)} &= \frac{1}{1-t_n} \frac{\pi(\frac{A_i - t_n a_{\star}}{1 - t_n} \mid X_i)}{\pi(A_i \mid X_i)} \leq \frac{1}{1-t_n} \left(\frac{\pi_{\max}}{\pi_{\min}} \right).
\end{align*}
Consequently, it can be shown that $| \varphi_{n, i}| \lesssim \frac{1}{1 - t_n}$ and furthermore that
\begin{align*}
    \mathbb{I}\left\{\frac{1}{1 - t_n} > \varepsilon \sqrt{\frac{n}{1 - t_n}}\left(\frac{\sigma_{\text{min}}}{\sqrt{\pi_{\max}}}\right)\right\} \to 0
\end{align*}
as long as $n(1-t_n) \to \infty$ as $t_n \to 1$. Therefore, by applying the squeeze and dominated convergence theorems in conjunction, it follows that Lindeberg's condition holds, and thus
\begin{align*}
    \frac{\sqrt{n}}{\sigma_{\nu}(t_n)} (P_n - P) \{ \varphi_{\nu, t_n}(Z; P) \} \overset{d}{\longrightarrow} N(0, 1).
\end{align*} \smallskip

\textbf{Empirical Process Term:} Since $(P_n-P)$ annihilates constants, we may work with the uncentered score $\phi_{\nu,t_n}$ defined above. Specifically, we will show that
\begin{align*}
    \|\phi_{\nu, t_n}(Z; \widehat{P}) - \phi_{\nu, t_n}(Z; P) \|_2 = o_P\left( (1 - t_n)^{-1/2} \right),
\end{align*}
which will be sufficient for controlling the empirical process term. To proceed, observe that after grouping together like terms, $\phi_{\nu, t_n}(Z; \widehat{P}) - \phi_{\nu, t_n}(Z; P) = D_1(t_n) + D_2(t_n) + D_3(t_n)$  where
\begin{align*}
    D_1(t_n) &= \left(\frac{\widehat{\nu}_{t_n}(A \mid X)}{\widehat{\pi}(A \mid X)} - \frac{\nu_{t_n}(A \mid X)}{\pi(A \mid X)} \right) \Big(Y - \mu(X, A)\Big), \\
    D_2(t_n) &= - \frac{\widehat{\nu}_{t_n}(A \mid X)}{\widehat{\pi}(A \mid X)}\Big( \widehat{\mu}(X, A) - \mu(X, A) \Big),  \\
    D_3(t_n) &= \widehat{\mu}(X, (1-t_n)A + t_n a_{\star}) - \mu(X, (1-t_n)A + t_n a_{\star}).
\end{align*}
Then, applying the triangle inequality, it follows that
\begin{align*}
    \| \phi_{\nu, t_n}(Z; \widehat{P}) - \phi_{\nu, t_n}(Z; P) \|_2 \leq \|D_1(t_n)\|_2 + \|D_2(t_n)\|_2 + \|D_3(t_n)\|_2.
\end{align*}
We proceed by bounding each norm one by one. First, to handle $D_1(t_n)$, we can immediately see that under the assumption that $|Y| \leq C$ with probability one, then $|\mu(X,A)|\leq C$ and thus $|Y-\mu(X,A)|\leq 2C$. It therefore follows that
\begin{align*}
    \|D_1(t_n)\|_2 &\leq 2C \left\| \frac{\widehat\nu_{t_n}}{\widehat\pi} - \frac{\nu_{t_n}}{\pi} \right\|_2.
\end{align*}
From here, we further bound this norm as
\begin{align*}
    \left\| \frac{\widehat{\nu}_{t_n}}{\widehat{\pi}} - \frac{\nu_{t_n}}{\pi} \right\|_2 &= \left\|\frac{\widehat{\nu}_{t_n} - \nu_{t_n}}{\pi} + \widehat{\nu}_{t_n} \left(\frac{1}{\widehat{\pi}} - \frac{1}{\pi}\right)\right\|_2 \leq \left\|\frac{\widehat{\nu}_{t_n} - \nu_{t_n}}{\pi} \right\|_2 + \left\| \widehat{\nu}_{t_n} \left(\frac{1}{\widehat{\pi}} - \frac{1}{\pi}\right)\right\|_2.
\end{align*}
Now, we bound the first term. Observe that
\begin{align*}
    \left\|\frac{\widehat{\nu}_{t_n} - \nu_{t_n}}{\pi} \right\|_2  &= \mathbb{E}\left[ \int_{\mathcal{A}}\frac{(\widehat{\nu}_{t_n}(a \mid X) - \nu_{t_n}(a \mid X))^2}{\pi(a \mid X)} da \right]^{1/2} \\
    &\leq \frac{1}{\sqrt{\pi_{\min}}}\mathbb{E}\left[ \int_{\mathcal{A}}(\widehat{\nu}_{t_n}(a \mid X) - \nu_{t_n}(a \mid X))^2 da \right]^{1/2}  \\
    &= \frac{1}{(1-t_n)\sqrt{\pi_{\min}}}\mathbb{E}\left[ \int_{\mathcal{A}_{t_n}}\left\{\widehat{\pi}\left( \frac{a-t_na_\star}{1-t_n} \mid X\right) - \pi\left( \frac{a-t_na_\star}{1-t_n} \mid X\right)\right\}^2 da \right]^{1/2}  \\
    &\overset{(i)}{=} \frac{1}{\sqrt{\pi_{\min}(1 - t_n)}}\mathbb{E}\left[ \int_{\mathcal{A}}(\widehat{\pi}(u \mid X) - \pi(u \mid X))^2 du \right]^{1/2} \\
    &\overset{(ii)}{\leq} \frac{1}{\pi_{\min} \sqrt{1 - t_n}}\mathbb{E}\left[ \int_{\mathcal{A}}(\widehat{\pi}(u \mid X) - \pi(u \mid X))^2 \pi(u \mid X) du \right]^{1/2} \\
    &= \frac{1}{\pi_{\min} \sqrt{1 - t_n}} \left\| \widehat{\pi} - \pi \right\|_2,
\end{align*}
where in $(i)$ we apply the change of variables $a=(1-t_n)u+t_na_\star$ and in $(ii)$ we use the assumption $\pi(u\mid X)\geq\pi_{\min}$, which implies, pointwise,
\begin{align*}
    \left\{ \widehat{\pi}(u\mid X)-\pi(u\mid X) \right\}^2 \leq \frac{1}{\pi_{\min}} \left\{ \widehat{\pi}(u\mid X)-\pi(u\mid X) \right\}^2 \pi(u\mid X).
\end{align*}
Next, it follows that
\begin{align*}
    \left\|\widehat{\nu}_{t_n} \left(\frac{1}{\widehat{\pi}} - \frac{1}{\pi}\right) \right\|^2_2 &= \mathbb{E}\left[ \int_{\mathcal{A}} \widehat{\nu}_{t_n}(a \mid X)^2 \left(\frac{(\widehat{\pi}(a \mid X) - \pi(a \mid X))^2}{(\widehat{\pi}(a \mid X) \pi(a \mid X))^2} \right)\pi(a \mid X) da \right] \\
    &=\frac{1}{(1-t_n)^2} \mathbb{E}\left[ \int_{\mathcal{A}_{t_n}}\frac{\widehat{\pi} \left( \frac{a-t_na_\star}{1-t_n} \mid X \right)^2}{ \widehat{\pi}(a\mid X)^2 \pi(a\mid X)} \left( \widehat{\pi}(a\mid X) - \pi(a\mid X)\right)^2 da \right]\\
    &\leq  \frac{\pi_{\max}^2 }{ \pi_{\min}^3(1-t_n)^2} \mathbb{E}\left[\int_{\mathcal{A}_{t_n}}\left( \widehat{\pi}(a\mid X) - \pi(a\mid X) \right)^2 da \right]\\
    &\leq \frac{ \pi_{\max}^2}{\pi_{\min}^4(1-t_n)} \left\| \widehat{\pi}-\pi \right\|_{t_n}^2,
\end{align*}
where we say for some measurable function $f$ that $\|f\|^2_{t} =\frac{1}{1-t}\mathbb{E}\left[ f(X,A)^2\mathbb{I}(A\in\mathcal{A}_t) \right]$, where $\mathcal{A}_t = \left\{(1-t)u+ta_\star:u\in\mathcal{A} \right\}$. Combining both results, we find
\begin{align} \label{different_norms}
    \left\| \frac{\widehat{\nu}_{t_n}}{\widehat{\pi}} - \frac{\nu_{t_n}}{\pi} \right\|_2 \lesssim \frac{1}{\sqrt{1 - t_n}} \Big\{ \|\widehat\pi-\pi\|_{2} + \|\widehat\pi-\pi\|_{t_n} \Big\}.
\end{align}
This presents an interesting problem: \cref{different_norms} contains two different requirements on the nuisance estimation of $\pi$; one $L_2(P)$ norm and one local $L_2(P)$ seminorm. In order to control $\|D_1(t_n)\|_2$ typically one would impose consistency on the stricter of the norms, and in fact, we can directly relate the two norms since
\begin{align*}
    (1-t ) \|f\|^2_{t} = \mathbb{E}\left[ f(X,A)^2\mathbb{I}(A\in\mathcal{A}_t) \right] \leq \mathbb{E}\left[ f(X,A)^2 \right] = \| f \|^2_2
\end{align*}
and therefore, $\|f\|_{t_n} \leq \frac{1}{\sqrt{1 - t_n}} \| f \|_2$. If we were working with a fixed $t_n$ we could then assume $\left\| \widehat{\pi}-\pi \right\|_{2} = o_P\left(1 \right)$, which would then imply that
\begin{align*}
    \|\widehat\pi-\pi\|_{t_n} = o_P(1).
\end{align*}
However, since we take $t_n \to 1$, the $(1 - t_n)^{-1/2}$ term blows up as $t_n$ approaches one, and thus, assuming $\|f\|_2 = o_P(1)$ does not necessarily imply that $\|f\|_{t_n} = o_P(1)$. Therefore, we instead assume a collective 
\begin{align*}
    \left\| \widehat{\pi}-\pi \right\|_{t_n} + \left\| \widehat{\pi}-\pi \right\|_{2} = o_P\left(1\right).
\end{align*}
From here, we can follow analogous steps to see that
\begin{align*}
    \| D_2(t_n) \|^2_2 &= \mathbb{E}\left[ \int_{\mathcal{A}_{t_n}} \left\{ \frac{1}{1-t_n} \frac{ \widehat\pi\left(\frac{a - t_n a_{\star}}{1 - t_n} \mid X\right)}{\widehat\pi(a\mid X)}\right\}^2 \Big( \widehat{\mu}(X, a) - \mu(X, a) \Big)^2 \pi(a\mid X)da \right]\\
    &\lesssim \frac{1}{(1-t_n)^2} \mathbb{E}\left[\Big( \widehat{\mu}(X, A) - \mu(X, A) \Big)^2 \mathbb{I}(A\in\mathcal{A}_{t_n}) \right]\\
    &= \frac{1}{1-t_n} \|\widehat\mu-\mu\|_{t_n}^2.
\end{align*}
Therefore,  $\|D_2\|_2 = o_P((1-t_n)^{-1/2})$ under the assumption that $\|\widehat\mu-\mu\|_{t_n} = o_P(1)$. Finally, again following similar steps as before,
\begin{align*}
    \|D_3(t_n)\|_2^2 &= \mathbb{E}\left[ \int_{\mathcal{A}} \Big( \widehat{\mu}(X, (1-t_n)a + t_n a_{\star}) - \mu(X, (1-t_n)a + t_n a_{\star}) \Big)^2 \pi(a\mid X)da \right]\\
    &= \frac{1}{1-t_n} \mathbb{E}\left[ \int_{\mathcal{A}_{t_n}}\Big( \widehat{\mu}(X, u) - \mu(X,u) \Big)^2 \pi\left(a_{\star} + \frac{u-a_{\star}}{1-t_n}\mid X\right)du \right] \\
    &\leq \frac{1}{1-t_n} \frac{\pi_{\max}}{\pi_{\min}} \mathbb{E}\left[ \int_{\mathcal{A}_{t_n}}\Big( \widehat{\mu}(X, u) - \mu(X,u) \Big)^2 \pi\left(u\mid X\right)du \right] \\
    &=\frac{\pi_{\max}}{\pi_{\min}} \|\widehat\mu-\mu\|_{t_n}^2.
\end{align*}
Putting everything together, we now have 
\begin{align*}
    \sqrt{1 - t_n} \left\| \phi_{\nu, t_n}(Z; \widehat{P}) - \phi_{\nu, t_n}(Z; P) \right\|_2  \lesssim \|\widehat\pi-\pi\|_{2} + \|\widehat\pi-\pi\|_{t_n} + \|\widehat{\mu}-\mu\|_{t_n}.
\end{align*}
Conditional on the training sample, let $  g_n(Z) = \phi_{\nu, t_n}(Z; \widehat{P}) - \phi_{\nu, t_n}(Z; P)$. Cross-fitting implies that $g_n$ is fixed and independent of the validation observations. Therefore,
\begin{align*}
    \mathbb{E}\left[\left\{ \frac{\sqrt n}{\sigma_{\nu}(t_n)} (P_n-P)g_n \right\}^2 \mid  \widehat{P} \right] &\leq \frac{\|g_n\|_2^2}{\sigma^2_{\nu}(t_n)} \lesssim \underbrace{\Big[ \| \widehat\pi-\pi \|_{2} + \|\widehat\pi - \pi\|_{t_n} + \|\widehat\mu-\mu\|_{t_n} \Big]^2}_{=o_P(1)},
\end{align*}
where it follows that $\sigma^2_{\nu}(t_n) \asymp (1-t_n)^{-1}$ by \cref{var_corollary}. Conditional Chebyshev's inequality therefore gives
\begin{align*}
    \frac{\sqrt n}{\sigma_{\nu}(t_n)} (P_n-P) \left\{ \varphi_{\nu, t_n}(Z; \widehat{P}) - \varphi_{\nu, t_n}(Z; P) \right\} = o_P(1).
\end{align*}

\bigskip

\textbf{Remainder Bound:} Finally, in order to establish asymptotic normality we must bound the remainder term. From here, we want to show that the von Mises expansion is a second-order product of errors. Recall that for two distributions, $P$ and $\widehat{P}$, the remainder term in the von Mises expansion is given by
\begin{align*}
    R_2(\widehat{P}, P) &= \mathbb{E}\left[\phi_{\nu, t_n}(Z; \widehat{P}) - \psi_{\nu, t_n}(P)\right] \\[0.05in]
    &= \mathbb{E}\left[ \frac{\widehat{\nu}_{t_n}(A \mid X)}{\widehat{\pi}(A \mid X)}\Big(Y - \widehat{\mu}(X, A) \Big) + \widehat{\mu}(X, (1 - t_n)A + t_n a_{\star}) -  \psi_{\nu, t_n}(P)\right] \\
    &= \mathbb{E}\left[ \left(\mu(X, A) - \widehat{\mu}(X, A) \right)\left(  \frac{\widehat{\nu}_{t_n}(A \mid X)}{\widehat{\pi}(A \mid X)} - \frac{\nu_{t_n}(A \mid X)}{\pi(A \mid X)} \right)  \right].
\end{align*}
Therefore, following similar steps to the proof of the empirical process term, it follows by applying the Cauchy-Schwarz inequality that
\begin{align*}
    R_2(\widehat{P}, P) &\leq \mathbb{E}\left[\left(\widehat{\mu}(X, A) - \mu(X, A) \right)^2 \mathbb{I}(A\in\mathcal{A}_{t_n}) \right]^{1/2}
    \left\| \frac{\widehat{\nu}_{t_n}(A \mid X)}{\widehat{\pi}(A \mid X)} - \frac{\nu_{t_n}(A \mid X)}{\pi(A \mid X)} \right\|_2 \\
    &\lesssim \| \widehat{\mu} - \mu \|_{t_n} \Big( \|\widehat{\pi} - \pi\|_{t_n} + \|\widehat{\pi} - \pi\|_2 \Big).
\end{align*}
Consequently, since by \cref{var_corollary} we know that $\sigma^2_{\nu}(t_n) \asymp (1 - t_n)^{-1}$, it follows that
\begin{align*}
     \frac{\sqrt{n}}{\sigma_{\nu}(t_n)} |R_{2}(\widehat{P},P)| \lesssim \sqrt{n(1 - t_n)} \| \widehat{\mu} - \mu \|_{t_n} \Big( \|\widehat{\pi} - \pi\|_{t_n} + \|\widehat{\pi} - \pi\|_2 \Big) = o_P(1).
\end{align*}

\textbf{Asymptotics:} Combining each of the previous results and applying Slutsky's theorem, it now follows that
\begin{align*}
     \frac{\sqrt{n}}{\sigma_\nu(t_n)}\left(\widehat{\psi}_\nu(t_n) - \psi_\nu(t_n) \right)\overset{d}{\longrightarrow}N(0,1),
\end{align*}
thereby completing the proof of \cref{wasserstein_asymptotics}.
\end{proof}

\subsection{Proof of \cref{derivative_EIF}} \label{derivative_EIF_proof}

\begin{proof}[\textbf{Proof:}]

Fix an integer $k\geq1$ and $t\in [0,1)$. Then, we define $\theta_{k,t} = \psi_\nu^{(k)}(t)$ to be the $k$th derivative of the incremental Wasserstein effect. By \cref{effect_curvature}, we know that
\begin{align}
    \theta_{k,t} = \mathbb{E}\left[ (a_\star-A)^k \left. \frac{\partial^k}{\partial u^k}\mu(X,u) \right|_{u=F_t(A)} \right],
    \label{wasserstein_kth_derivative_representation}
\end{align}
where $F_t(A) = (1 - t)A + t a_\star$. For $u\in\mathcal{A}_t=F_t(\mathcal{A})$, let
\begin{align*}
    Q_{k,t}(u\mid x) = \left( \frac{a_\star-u}{1-t} \right)^k \nu_t(u\mid x).
\end{align*}
Next, let $\mathcal{A}_t=[\ell_t,r_t]$, where $\ell_t=ta_\star$ and $r_t=1-t+ta_\star$. Then, for $j=0,\ldots,k-1$, the change of variables $u=F_t(a)$ gives
\begin{align*}
    \left.\frac{\partial^j}{\partial u^j} Q_{k,t}(u\mid x)\right|_{u=F_t(a)} =\frac{1}{(1-t)^{j+1}} \frac{\partial^j}{\partial a^j}\left\{(a_\star-a)^k\pi(a\mid x)\right\}.
\end{align*}
Consequently, \cref{no_flux_assumption} implies
\begin{align} \label{zero_flux_assumption}
    \lim_{u\downarrow\ell_t} \left\{ \frac{\partial^j}{\partial u^j}Q_{k,t}(u\mid x) \right\} =\lim_{u\uparrow r_t} \left\{  \frac{\partial^j}{\partial u^j} Q_{k,t}(u\mid x) \right\} = 0, \qquad j=0,\ldots,k-1,
\end{align}
where the limits are taken from the interior of $\mathcal{A}_t$. Now, let $\{P_\varepsilon:\varepsilon\in(-\delta,\delta)\}$ be an arbitrary smooth regular parametric submodel through $P=P_0$ satisfying the hypotheses of \cref{derivative_EIF}, with score decomposition
\begin{align*}
    S(Z) = S_X(X) + S_{A\mid X}(A\mid X) + S_{Y\mid X,A}(Z).
\end{align*}
We write $S_{X,A}=S_X+S_{A\mid X}$ and define
\begin{align*}
    \mu^\dagger(x,a) = \left. \frac{\partial}{\partial\varepsilon} \mu_\varepsilon(x,a) \right|_{\varepsilon=0}.
\end{align*}
Then, the usual conditional-mean calculation gives
\begin{align} \label{mu_dagger}
    \mu^\dagger(x,a) = \mathbb{E}\left[ \{Y-\mu(X,A)\}S_{Y\mid X,A}(Z) \mid  X=x,A=a \right].
\end{align}
For notational convenience, let
\begin{align*}
    g_{k,t}(x,a) = (a_\star-a)^k \left. \frac{\partial^k}{\partial u^k}\mu(x,u) \right|_{u=F_t(a)}.
\end{align*}
Then, differentiating \cref{wasserstein_kth_derivative_representation} along the submodel gives
\begin{align*}
    \left. \frac{\partial}{\partial\varepsilon} \theta_{k,t}(P_\varepsilon) \right|_{\varepsilon=0} =  \mathbb{E}\left[ g_{k,t}(X,A)S_{X,A}(X,A) \right] + \mathbb{E}\left[ (a_\star-A)^k \left. \frac{\partial^k}{\partial u^k} \mu^\dagger(X,u) \right|_{u=F_t(A)} \right].
\end{align*}
Now, conditional on $X=x$, we make the change of variables $u=F_t(a)$. Note that the conditional distribution of $F_t(A)$ has density $\nu_t(\,\cdot\mid x)$ and
\begin{align*}
    a_\star-F_t^{-1}(u) = \frac{a_\star-u}{1-t}.
\end{align*}
Consequently, 
\begin{align*} 
    \mathbb{E}\left[ (a_\star-A)^k \left. \frac{\partial^k}{\partial u^k} \mu^\dagger(X,u) \right|_{u=F_t(A)} \right] = \mathbb{E}\left[ \int_{\mathcal{A}_t} Q_{k,t}(u\mid X) \frac{\partial^k}{\partial u^k} \mu^\dagger(X,u)du \right].
\end{align*}
Integrating $\int_{\mathcal{A}_t} Q_{k,t}(u\mid x) \frac{\partial^k}{\partial u^k} \mu^\dagger(x,u)du$ by parts $k$ times yields
\begin{align*}
    \sum_{j=0}^{k-1} (-1)^j\left[ \frac{\partial^j}{\partial u^j}Q_{k,t}(u\mid x) \frac{\partial^m}{\partial u^m}\mu^\dagger(x,u)\right]_{\ell_t}^{r_t} +(-1)^k\int_{\mathcal{A}_t} \frac{\partial^k}{\partial u^k}Q_{k,t}(u\mid x) \mu^\dagger(x,u)du,
\end{align*}
where $m = k-1-j$. Now, note that every term in the boundary sum vanishes under the no-flux assumption in \cref{zero_flux_assumption}. Now, define
\begin{align}
    H_{k,t}(x,a) = \frac{(-1)^k \frac{\partial^k }{\partial a^k} Q_{k,t}(a\mid x)}{\pi(a\mid x)},
    \label{wasserstein_kth_riesz_weight}
\end{align}
where $H_{k,t}(x,a)=0$ for $a\notin\mathcal{A}_t$. It then follows that
\begin{align*}
     \mathbb{E}\left[ \int_{\mathcal{A}_t} Q_{k,t}(u\mid X) \frac{\partial^k}{\partial u^k} \mu^\dagger(X,u)du \right] &= \mathbb{E}\left[H_{k,t}(X,A)\mu^\dagger(X,A) \right] \\
     &\overset{(i)}{=} \mathbb{E}\Big[H_{k,t}(X,A) \{Y-\mu(X,A)\}S_{Y\mid X,A}(Z) \Big],
\end{align*}
where $(i)$ follows by \cref{mu_dagger}. Since the weighted residual has conditional mean zero given $(X,A)$, the conditional-outcome score may be replaced by the full score $S(Z)$. Similarly, because $g_{k,t}$ is measurable with respect to $(X,A)$, $\mathbb{E}[S_{Y\mid X,A}(Z)\mid X,A]=0$, and $\mathbb{E}[S(Z)]=0$, we have
\begin{align*}
    \mathbb{E}\left[ g_{k,t}(X,A)S_{X,A}(X,A)\right] = \mathbb{E}\left[ (g_{k,t}(X,A)-\theta_{k,t})S(Z) \right].
\end{align*}
Putting everything together, 
\begin{align*}
    \left. \left\{  \frac{\partial}{\partial\varepsilon} \theta_{k,t}(P_\varepsilon) \right\}\right|_{\varepsilon=0} = \mathbb{E}\left[ \varphi_{\nu,k}(Z;t)S(Z) \right],
\end{align*}
where
\begin{align*}
    \varphi_{\nu,k}(Z;t) =  H_{k,t}(X,A)\{Y-\mu(X,A)\} + (a_\star-A)^k \left. \frac{\partial^k}{\partial u^k}\mu(X,u) \right|_{u=F_t(A)} - \theta_{k,t}.
\end{align*}
This expression is mean zero. Under the assumed square-integrability conditions, it belongs to $L_2(P)$ and since the tangent space of the smooth nonparametric model is dense in $L_2^0(P)$, it is the efficient influence function. \bigskip

For completeness, we establish two alternative representations of $H_{k,t}$. Let $f$ be a smooth test function. Then,
\begin{align*}
    \frac{\partial^k}{\partial t^k} \int f(u)\nu_t(u\mid x)du &= \frac{\partial^k}{\partial t^k} \int f(F_t(a))\pi(a\mid x)da \\
    &= \int (a_\star-a)^k f^{(k)}(F_t(a)) \pi(a\mid x)da \\
    &= \int_{\mathcal{A}_t}  Q_{k,t}(u\mid x)f^{(k)}(u)du \\
    &= (-1)^k \int_{\mathcal{A}_t} \frac{\partial^k}{\partial u^k} Q_{k,t}(u\mid x)f(u)du,
\end{align*}
where the final equality follows from \cref{zero_flux_assumption}. Hence, 
\begin{align*}
    \frac{\partial^k}{\partial t^k} \nu_t(u\mid x) = (-1)^k \frac{\partial^k}{\partial u^k} Q_{k,t}(u\mid x).
\end{align*}
The boundary-cancellation condition ensures that this derivative has
no boundary point masses or derivatives thereof. Thus,
\begin{align*}
    H_{k,t}(x,a) =  \frac{\frac{\partial^k}{\partial t^k} \nu_t(a\mid x)}{\pi(a\mid x)}.
\end{align*}
Finally, define $z_t(a) = F_t^{-1}(a)-a_\star = \frac{a-a_\star}{1-t}$. Then, on the interior of $\mathcal{A}_t$, it follows that
\begin{align*}
    \nu_t(a\mid x) = \frac{\pi(a_\star+z_t(a)\mid x)}{1-t} \qquad \text{and} \qquad  Q_{k,t}(a\mid x) = \frac{(-z_t(a))^k \pi(a_\star+z_t(a)\mid x)}{1-t}.
\end{align*}
Thus, we can see that 
\begin{align*}
    (-1)^k \frac{\partial^k}{\partial a^k} Q_{k,t}(a\mid x) = (1-t)^{-(k+1)} \frac{\partial^k}{\partial z^k} \left[z^k\pi(a_\star+z\mid x) \right]_{z=z_t(a)}.
\end{align*}
From here, applying Leibniz's rule gives
\begin{align*}
    \frac{\partial^k}{\partial z^k} \left[ z^k\pi(a_\star+z\mid x) \right] = \sum_{j=0}^{k} \binom{k}{j} \frac{k!}{j!} z^j \pi^{(j)}(a_\star+z\mid x),
\end{align*}
and consequently,
\begin{align*}
    H_{k,t}(x,a) = \frac{\mathbb{I}(a\in\mathcal{A}_t)}{(1-t)^{k+1}\pi(a\mid x) } \sum_{j=0}^{k} \binom{k}{j} \frac{k!}{j!} z_t(a)^j \pi^{(j)} (F_t^{-1}(a)\mid x),
\end{align*}
which completes the proof.
\end{proof}

\subsection{Proof of \cref{minimax_elbow}} \label{minimax_elbow_proof}

\begin{proof}[\textbf{Proof:}] To begin, recall that our goal is to show that for some $t_0 \in [0, 1)$, and for all sufficiently large $n$ and every $t\in[t_0,1]$,
\begin{align*} 
\mathcal{R}_n(t) =  \inf_{\widehat\psi_t} \sup_{\mu\in\mathcal{H}_q^\beta(L)} \mathbb{E}_\mu\left[\left| \widehat\psi_t -  \psi_t(\mu) \right| \right] \asymp \frac{1}{\sqrt{n(1-t)^q}} \wedge n^{-\beta/(2\beta+q)}.
\end{align*}
Here, we assume that $Y=\mu(A)+\varepsilon$ where $\varepsilon\sim N(0,\sigma^2)$, $\varepsilon\ind A$, and $A$ has a known density $\pi$. Let $K$ be a bounded probability density with compact support $\mathcal{U}$ so that $\psi_t(\mu) = \int_{\mathcal{A}}\mu(a)\rho_t(a)da$, under the local density
\begin{align*}
    \rho_t(a) &= \frac{1}{(1-t)^q} K\left(\frac{a-a_\star}{1-t}\right).
\end{align*}
Now, suppose that $a_\star+(1-t)\mathcal{U}\subseteq\mathcal{A}$ for every $t\in[t_0,1)$. Furthermore, suppose that there exists $r_0 > 0$ such that $0< \pi_{\min} \leq \pi(a) \leq \pi_{\max} <\infty$ for Lebesgue-almost every $a\in\mathcal{A}\cap B(a_\star,r_0)$, where $B(a, r)$ denotes the open Euclidean ball of radius $r$ centered at $a$. Throughout the proof, we let $h = 1 - t$ denote the bandwidth, and similarly, we let $h_0 = 1 - t_0$. We use $c$ and $C$ to denote positive constants independent of both $n$ and $h$. Throughout, we assume $n$ is large enough such that $b_n \leq h_0$ where $b_n=n^{-1/(2\beta+q)}$. Finally, let
\begin{align*}
     r_n(h)=\frac{1}{\sqrt{n(h\vee b_n)^{q}}}.
\end{align*}
With these assumptions and definitions in place, we proceed with establishing the lower bound. 
\medskip

\textbf{Lower Bound:} For the lower bound, we use Le Cam's two-point method: we restrict the H\"older class to two regression functions $\mu_0$ and $\mu_1$. The goal is to show that if their target values are well separated but their induced data distributions are difficult to distinguish, then no estimator can be accurate under both models. More specifically, for each regression function $\mu$, let $P_\mu$ denote the distribution of one observation and $P_\mu^{\otimes n}$ the corresponding product distribution of the full sample. Le Cam's two-point lemma \citep{tsbakov2009} implies that, for any $\mu_0,\mu_1\in\mathcal{H}_q^\beta(L)$,
\begin{align*}
    \mathcal{R}_n(t) \geq \frac{|\psi_t(\mu_1)-\psi_t(\mu_0)|}{4} \left\{ 1- \operatorname{TV}\!\left( P_{\mu_1}^{\otimes n}, P_{\mu_0}^{\otimes n} \right) \right\}.
\end{align*}
Consequently, a lower bound of order $r_n(h)$ follows if we can
construct two regression functions that both belong to the H\"older class, whose target values are separated by order $r_n(h)$, but whose $n$-sample laws remain close in total variation. \medskip

To begin, we proceed with constructing a smooth perturbation. Let
\begin{align*}
    R_{\mathcal{U}} = \sup_{u\in\mathcal{U}}\|u\|_2
\end{align*}
denote the radius of the smallest Euclidean ball centered at zero that contains $\mathcal{U}$. Since $\mathcal{U}$ is compact and $a_\star+h_0\mathcal{U}\subset B(a_\star,r_0)$, it follows that $h_0 \|u \|_2 < r_0$ for every $u \in \mathcal{U}$. Thus, $h_0 R_{\mathcal{U}} < r_0$, and furthermore, the interval $(R_{\mathcal{U}}, r_0 / h_0)$ is nonempty. Thus, there exists some radius $R_{\mathcal{V}} \in (R_{\mathcal{U}}, r_0 / h_0)$, such that $R_{\mathcal{V}} > R_{\mathcal{U}}$ and $R_{\mathcal{V}} < r_0 / h_0$. Now, we choose an infinitely differentiable, compactly supported function $G:\mathbb{R}^q \to [0,1]$ such that $0\leq G\leq1$, $ G(u)=1$ for $\|u\|_2\leq R_{\mathcal{U}}$, and $\operatorname{supp}(G) \subseteq \overline{B}(0,R_{\mathcal{V}})$. This gives us the smooth transition
\begin{align} \label{smooth_transition}
    \begin{cases}
G(u)=1, & \|u\|_2\leq R_{\mathcal{U}},\\
0<G(u)<1, & R_{\mathcal{U}} < \|u\|_2<R_{\mathcal{V}},\\
G(u)=0, & \|u\|_2\geq R_{\mathcal{V}}.
\end{cases}
\end{align}
Now, for $h\in[0,h_0]$, set $s=h\vee b_n$ and $\eta=(ns^q)^{-1/2} = r_n(h)$, and define the null and fluctuated alternative regression functions as
\begin{align*}
      \mu_0(a)=0, \qquad \mu_1(a) = \delta\eta G\!\left(\frac{a-a_\star}{s}\right),
\end{align*}
where $\delta > 0$ is some fixed constant. We now proceed with proving H\"older admissibility, target separation, and statistical indistinguishability. \medskip

\textit{H\"older admissibility:} Here, we must show that $\mu_0, \mu_1 \in  \mathcal{H}_q^\beta(L)$. Trivially, it follows that $\mu_0 \in \mathcal{H}_q^\beta(L)$. To address $\mu_1$, we first define
\begin{align*}
      [G]_{\beta} = \sup_{x,y\in\mathbb{R}^q} \frac{|G(x)-G(y)|}{\|x-y\|_2^\beta}
\end{align*}
for $x \neq y$. Note that because $G$ is bounded and infinitely differentiable, it follows that $[G]_{\beta} < \infty$ for all $0 < \beta \leq 1$. Then, for any $a, a^\prime \in \mathcal{A}$,
\begin{align*}
    |\mu_1(a)-\mu_1(a^\prime)| \leq \delta\eta [G]_\beta \left\| \frac{a-a^\prime}{s} \right\|_2^\beta = \delta\eta s^{-\beta}[G]_\beta \|a-a^\prime\|_2^\beta.
\end{align*}
Then, since $s\geq b_n$,
\begin{align*}
    \eta s^{-\beta} &= n^{-1/2}s^{-(\beta+q/2)} \leq n^{-1/2}b_n^{-(\beta+q/2)} = 1.
\end{align*}
Therefore, the H\"older constant of $\mu_1$ is at most
$\delta[G]_\beta$. Moreover, because $0\leq G\leq1$,
\begin{align*}
      \|\mu_1\|_{\infty,\mathcal{A}} \leq \delta\eta \leq \delta
\end{align*}
for all sufficiently large $n$. Hence $\mu_0,\mu_1\in\mathcal{H}_q^\beta(L)$ whenever $\delta>0$ is chosen to be sufficiently small. \medskip

\textit{Target separation:} Here, our goal is to show that $|\psi_t(\mu_1) - \psi_t(\mu_0)| \asymp \eta$. First, suppose that $h > 0$. Observe that since $s \geq h$, it follows that for every $u \in \mathcal{U}$,
\begin{align*}
      \left\|\frac{h}{s}u\right\|_2 \leq \|u\|_2 \leq R_{\mathcal{U}}.
\end{align*}
Then, by our construction of $G$ outlined in \cref{smooth_transition}, it follows that $G((h/s)u)=1$. Consequently,
\begin{align*}
     \psi_t(\mu_1)-\psi_t(\mu_0) = \delta\eta \int_{\mathcal{U}} G\!\left(\frac{h}{s}u\right)K(u)du = \delta\eta \int_{\mathcal{U}}K(u) du = \delta\eta.
\end{align*}
At $h = 0$ it similarly follows that
\begin{align*}
     \psi_1(\mu_1)-\psi_1(\mu_0) = \mu_1(a_\star) = \delta\eta G(0) = \delta\eta.
\end{align*}
Thus, $|\psi_t(\mu_1)-\psi_t(\mu_0)| = \delta\eta$ for every $h\in[0,h_0]$. \medskip

\textit{Statistical indistinguishability:} First, because $s \leq h_0$, observe that the support of $\mu_1$ is contained within $\overline{B}(a_\star,sR_{\mathcal{V}}) \subseteq \overline{B}(a_\star,h_0R_{\mathcal{V}}) \subset B(a_\star,r_0)$. Then, since the design distribution is the same under both alternatives and the conditional outcome distributions are Gaussian with a common variance, it follows that
\begin{align*}
    \operatorname{KL}\!\left(P_{\mu_1}^{\otimes n}, P_{\mu_0}^{\otimes n}\right) &= \frac{n}{2\sigma^2} \int_{\mathcal{A}} (\mu_1(a)-\mu_0(a))^2 \pi(a)da \\
    &\leq \frac{\pi_{\max}\delta^2n\eta^2}{2\sigma^2} \int_{\mathbb{R}^q} G\!\left(\frac{a-a_\star}{s}\right)^2 da \\
    &= \frac{\pi_{\max}\delta^2n\eta^2s^q}{2\sigma^2} \int_{\mathbb{R}^q}G(u)^2du \\
    &\leq C\delta^2,
\end{align*}
where the last equality uses $n\eta^2s^q=1$. From here, choose $\delta > 0$ sufficiently small such that the H\"older admissibility conditions hold and that $C\delta^2 \leq 1/8$. Then, Pinsker's inequality \citep{tsbakov2009} yields
\begin{align*}
        \operatorname{TV}\!\left( P_{\mu_1}^{\otimes n}, P_{\mu_0}^{\otimes n} \right) \leq \sqrt{
        \frac{1}{2} \operatorname{KL}\!\left( P_{\mu_1}^{\otimes n}, P_{\mu_0}^{\otimes n} \right)}\leq \frac{1}{4}.
\end{align*}
Putting everything together, Le Cam's two-point lemma gives
\begin{align*}
     \mathcal{R}_n(t) \geq c\eta = c\,r_n(h)
\end{align*}
uniformly over $h\in[0,h_0]$, which completes the lower bound. 

\medskip

\textbf{Upper Bound:} To prove the upper bound, let $\overline h=h\vee b_n$. When $h\geq b_n$, this is simply the original intervention bandwidth; however, when $h<b_n$, we deliberately estimate the smoother target at bandwidth $b_n$ and control the resulting approximation bias using H\"older continuity. For $\overline{h}>0$, we define
\begin{align*}
    \rho_{\overline{h}}(a) = \overline{h}^{-q} K\!\left(\frac{a-a_\star}{\overline{h}}\right)
\end{align*}
so that $\widehat\psi_t = \frac{1}{n} \sum_{i=1}^n \frac{\rho_{\overline{h}}(A_i)}{\pi(A_i)}Y_i,$ where the summand is defined to be zero off the support of $\rho_{\overline{h}}$ or whenever the displayed ratio is undefined. Note that since $\overline{h}\leq h_0$, the support of $\rho_{\overline{h}}$ is contained in $\mathcal{A}\cap B(a_\star,r_0)$, on which $\pi$ is bounded away from zero. Because the design density is known,
\begin{align*}
    \mathbb{E}_\mu(\widehat\psi_t) = \int_{\mathcal{A}} \mu(a)\rho_{\overline{h}}(a)da = \int_{\mathcal{U}} \mu(a_\star+\overline h u)K(u)du = \Psi_{\overline{h}}(\mu).
\end{align*}
Thus the estimator is exactly unbiased when $h\geq b_n$, because then $\overline{h}=h$. If $h<b_n$, its bias relative to the desired target is controlled by H\"older continuity:
\begin{align*}
    \left| \mathbb{E}_\mu(\widehat\psi_t)-\psi_t(\mu) \right| &= |\Psi_{\overline{h}}(\mu)-\Psi_h(\mu)| \\
    &\leq \int_{\mathcal{U}} \left| \mu(a_\star+\overline{h} u) - \mu(a_\star+hu) \right| K(u)du \\
    &\leq L|\overline{h}-h|^\beta \int_{\mathcal{U}} \|u\|_2^\beta K(u)du.
\end{align*}
Consequently, it follows that
\begin{align*}
    \sup_{\mu\in\mathcal{H}_q^\beta(L)} \left| \mathbb{E}_\mu(\widehat\psi_t) -  \psi_t(\mu) \right| \leq
    \begin{cases}
        0, & h\geq b_n,\\
        Cb_n^\beta, & 0\leq h<b_n.
    \end{cases}
\end{align*}
Next, we bound the variance. Note that $\mathbb{E}_\mu(Y^2\mid A=a)=\mu(a)^2+\sigma^2\leq L^2+\sigma^2$, it follows that
\begin{align*}
    \mathbb{V}_\mu(\widehat\psi_t) &\leq \frac{L^2+\sigma^2}{n} \int_{\mathcal{A}} \frac{\rho_{\overline{h}}(a)^2}{\pi(a)}da \\
    &\leq \frac{L^2+\sigma^2}{n\pi_{\min}\overline{h}^{2q}} \int_{\mathcal{A}} K\!\left(\frac{a-a_\star}{\overline{h}} \right)^2 da \\
    &= \frac{L^2+\sigma^2}{n\pi_{\min}\overline{h}^q} \int_{\mathbb{R}^q}K(u)^2 du \\
    &\leq \frac{C}{n\overline{h}^q}.
\end{align*}
Therefore,
\begin{align*}
     \mathbb{E}_\mu \left[\left| \widehat\psi_t-\mathbb{E}_\mu(\widehat\psi_t) \right| \right] \leq \sqrt{ \mathbb{V}_\mu(\widehat\psi_t)} \leq \frac{C}{\sqrt{n(h\vee b_n)^q}} = C\,r_n(h).
\end{align*}
Finally, note that $b_n^\beta = (nb_n^q)^{-1/2} = n^{-\beta/(2\beta+q)}.$ Consequently, when $h<b_n$, both the approximation bias and the standard deviation are of order $r_n(h)$. When $h\geq b_n$, the bias is zero and the standard deviation is again of order $r_n(h)$. Thus,
\begin{align*}
    \sup_{\mu\in\mathcal{H}_q^\beta(L)} \mathbb{E}_\mu \left| \widehat\psi_t-\psi_t(\mu) \right| &\leq \sup_{\mu\in\mathcal{H}_q^\beta(L)} \left| \mathbb{E}_\mu(\widehat\psi_t)-\psi_t(\mu) \right| + \sup_{\mu\in\mathcal{H}_q^\beta(L)} \mathbb{E}_\mu \left| \widehat\psi_t-\mathbb{E}_\mu(\widehat\psi_t) \right| \leq C\,r_n(h)
\end{align*}
uniformly over $h\in[0,h_0]$. Then, combining the lower and upper bounds yields the final result,
\begin{align*}
        r_n(h) = \begin{cases}
        (nh^q)^{-1/2}, & h\geq b_n,\\[1mm]
        n^{-\beta/(2\beta+q)}, & 0\leq h<b_n,
    \end{cases}
\end{align*}
thereby completing the proof.

\end{proof}
\subsection{Proof of \cref{higher_order_normality}} \label{higher_order_normality_proof}

\begin{proof}[\textbf{Proof:}] 
     To begin, we derive the kernel identity used in the theorem statement. Then, we split the proof into oracle, empirical process, and remainder sections similar to the proof of \cref{wasserstein_asymptotics}. Again, we assume all nuisance functions are cross-fit, so the rate conditions should be understood to hold foldwise, although we suppress this notation. \medskip

     \textbf{Kernel Identity:} To begin, let $F_t(a) = (1 - t)a + t a_\star$. Then, for $a \in \mathcal{A}_t$, set $u=F_t^{-1}(a)$ and $z=u-a_\star=\frac{a-a_\star}{1-t}$. Next, recall that we define
     \begin{align*}
           \eta_{k,t}(x,a) = \frac{\mathbb{I}(a\in\mathcal{A}_t)}{(1-t)^{k+1}} \sum_{j=0}^{k} \binom{k}{j} \frac{k!}{j!} \left( \frac{a-a_\star}{1-t}\right)^j \pi^{(j)}\left(\frac{a-ta_\star}{1-t}\mid x\right)
     \end{align*}
     such that $\eta_{0, t} = \nu_t$. Then, observe that
     \begin{align*}
     \omega_{p,t}(x,a) &= \sum_{k=0}^{p}\frac{(1-t)^k}{k!}\eta_{k,t}(x,a)\\ 
     &= \frac{\mathbb{I}(a\in\mathcal{A}_t)}{1-t} \sum_{k=0}^{p}\sum_{j=0}^{k} \binom{k}{j}\frac{z^j}{j!}\pi^{(j)}(u\mid x)\\
     &= \frac{\mathbb{I}(a\in\mathcal{A}_t)}{1-t} \sum_{j=0}^{p} \left(\sum_{k=j}^{p}\binom{k}{j}\right) \frac{(u-a_\star)^j}{j!}\pi^{(j)}(u\mid x)\\
     &= \frac{\mathbb{I}(a\in\mathcal{A}_t)}{1-t}K_p(u\mid x),
\end{align*}
where the last equality uses the identity $ \sum_{k=j}^{p}\binom{k}{j} = \binom{p+1}{j+1}$. 
Thus, it follows that
\begin{align} \label{omega_kernel_def}
    \omega_{p,t}(x,a) = \frac{\mathbb{I}(a\in\mathcal{A}_t)}{1-t} K_p\!\left(F_t^{-1}(a)\mid x\right).
\end{align}

\textbf{Central Limit Theorem:} In this section, our goal is to establish asymptotic normality of the oracle term $(P_n - P) \{\Phi_{\nu, p}(Z; t)\}$. First, we establish limiting behavior of the variance. Let $\varepsilon = Y-\mu(X,A)$ and $\sigma^2(x,a)=\mathbb{V}(Y\mid X=x,A=a)$. Since $\mathbb{E}(\varepsilon \mid X,A)=0$, the residual term is uncorrelated with every function of $(X,A)$. Consequently,
\begin{align*}
    \sigma_{\nu,p}^2(t_n) = \mathbb{V}(\Phi_{\nu, p}(Z; t_n)) = \mathbb{E} \! \left[ \left(\frac{\omega_{p,t_n}(X,A)}{\pi(A\mid X)} \right)^2\sigma^2(X,A) \right] + \mathbb{V}\!\left(\Gamma_{p,t_n}(X,A)\right).
\end{align*}
Using \cref{omega_kernel_def} and the change of variables $a=F_n(u)=a_\star+(1 - t_n)(u-a_\star)$,
\begin{align*}
    (1 - t_n) \mathbb{E}\!\left[\left(\frac{\omega_{p,t_n}(X,A)}{\pi(A\mid X)} \right)^2\sigma^2(X,A) \right] = \mathbb{E}\!\left[ \int_{\mathcal{A}} K^2_p(u\mid X) \left(\frac{\sigma^2(X,F_n(u))}{\pi(F_n(u)\mid X)} \right) du \right].
\end{align*}
Recall that we assume for $P_X$-almost every $x$, $a\mapsto\pi(a\mid x)$ and $a\mapsto \mathbb{V}(Y\mid X=x,A=a)$ are continuous at $a_\star$. Furthermore, we assume $\sigma_\star^2(X)=\mathbb{V}(Y\mid X,A=a_\star)\geq\sigma_{\min}^2>0$ almost surely. Thus, following the same arguments as in the proof of \cref{var_corollary}, it follows that, as $t_n \to 1$,
\begin{align*}
    (1 - t_n) \mathbb{E}\!\left[\left(\frac{\omega_{p,t_n}(X,A)}{\pi(A\mid X)} \right)^2\sigma^2(X,A) \right] \longrightarrow \underbrace{\mathbb{E} \! \left[ \frac{\sigma_\star^2(X)}{\pi_\star(X)} \| K_p(\, \cdot \mid X) \|^2_{L_2(\mathcal{A})} \right]}_{ V_{p,\star}}.
\end{align*}
Furthermore, we can easily obtain scaling of $\sigma_{\nu,p}^2(t_n)$ as $t_n \to 1$ by noting that $V_{p,\star} \in (0, \infty)$ and $ \mathbb{V}\!\left(\Gamma_{p,t_n}(X,A)\right) = O(1)$. Thus,
\begin{align*}
    \sigma_{\nu,p}^2(t_n) \sim \frac{V_{p,\star}}{1 - t_n} \asymp (1 - t_n)^{-1}.
\end{align*}
Next, we prove the oracle central limit theorem. By condition $(ii)$, there exist $\delta>0$, $C<\infty$, and a measurable
$M_\mu\in L_{2+\delta}(P_X)$ such that, $P_X$-almost surely,
\begin{align*}
    \max_{0\leq k\leq p+1} \sup_{a\in\mathcal{A}}|\mu^{(k)}(X,a)| \leq M_\mu(X), \qquad \max_{0\leq j\leq p} \sup_{a\in\mathcal{A}}|\pi^{(j)}(a\mid X)| \leq C,
\end{align*}
and
\begin{align*}
    \sup_{a\in\mathcal{A}_{t_0}} \mathbb{E}\!\left[ |Y-\mu(X,A)|^{2+\delta} \mid X,A=a \right] \leq C.
\end{align*}
Now, we verify Lindeberg's condition. For each $n$, let $\Phi_{n,i} = \Phi_{\nu,p}(Z_i;t_n)$ and define
\begin{align*}
    s_n^2 = \sum_{i=1}^{n}\mathbb{V}(\Phi_{n,i}) = n\sigma_{\nu,p}^2(t_n).
\end{align*}
For each fixed $n$, the variables $\Phi_{n,1},\ldots,\Phi_{n,n}$ are independent and identically distributed, have mean zero, and have variance $\sigma_{\nu,p}^2(t_n)$. We will show that, for every $\varsigma > 0$,
\begin{align*}
      \frac{1}{s_n^2} \sum_{i=1}^{n} \mathbb{E}\!\left[\Phi_{n,i}^2 \mathbb{I}\!\left( |\Phi_{n,i}|>\varsigma s_n \right) \right] \longrightarrow 0.
\end{align*}
From here, write $\theta_{p,n} = P \{\Gamma_{p,t_n}\}$ and
\begin{align*}
      H_{p,n}(x,a) = \frac{\omega_{p,t_n}(x,a)}{\pi(a\mid x)}
\end{align*}
so that $\Phi_{n,i} = H_{p,n}(X_i,A_i)\varepsilon_i + \Gamma_{p,t_n}(X_i,A_i) - \theta_{p,n}$, where $\varepsilon_i=Y_i-\mu(X_i,A_i)$. Now, let $q = 2+\delta$ for some $\delta > 0$, as supplied by the moment-envelope condition. We first establish a $q$th-moment bound with the correct dependence on $t_n$. By \cref{omega_kernel_def},
\begin{align*}
     H_{p,n}(x,a) = \frac{\mathbb{I}(a\in\mathcal A_{t_n})}{1-t_n} \frac{K_p\!\left(F_n^{-1}(a)\mid x\right)}{\pi(a\mid x) }.
\end{align*}
Let $m_q(x,a) = \mathbb{E}\!\left[ |\varepsilon|^q \mid X=x,A=a \right]$. Then, using the conditional distribution of $A$ given $X$ and the change of
variables $a=F_n(u)=a_\star+(1-t_n)(u-a_\star)$ and $da=(1-t_n)du$, we obtain
\begin{align*}
    \mathbb{E}\left[ \left| H_{p,n}(X,A)\varepsilon \right|^q \right] &= \frac{1}{(1-t_n)^q} \mathbb{E}\!\left[ \int_{\mathcal{A}_{t_n}} \frac{ \left| K_p(F_n^{-1}(a)\mid X) \right|^q m_q(X,a)}{\pi(a\mid X)^{q-1}} da \right] \\
    &= \frac{1}{(1-t_n)^{q-1}} \mathbb{E}\!\left[
        \int_{\mathcal{A}} \frac{ |K_p(u\mid X)|^q m_q(X,F_n(u))}{ \pi(F_n(u)\mid X)^{q-1} } du \right].
\end{align*}
Under our assumed moment-envelope conditions, the expectation on the right-hand side is uniformly bounded. Hence
\begin{align*}
    \mathbb{E}\left[ \left| H_{p,n}(X,A)\varepsilon \right|^q \right] \lesssim
    (1-t_n)^{-(q-1)}.
\end{align*}
The same envelope argument used above for the second moment of
$\Gamma_{p,t}$ gives
\begin{align*}
    \sup_{t\in[t_0,1)} \mathbb{E} \left[ \left| \Gamma_{p,t}(X,A) \right|^q \right] <\infty.
\end{align*}
Then, by Jensen's inequality, $|\theta_{p,n}|^q = |P\{\Gamma_{p,t_n}\}|^q \leq P \{|\Gamma_{p,t_n}|^q \} = O(1)$. Thus, by the elementary inequality $|x+y+z|^q\lesssim |x|^q+|y|^q+|z|^q$, we have that
\begin{align*}
       \mathbb{E}\left[ |\Phi_{n,i}|^q \right] \lesssim (1-t_n)^{-(q-1)}.
\end{align*}
From here, note that on the event $\{|\Phi_{n,1}|> \varsigma s_n\}$ it follows that
\begin{align*}
    |\Phi_{n,1}|^2 \leq \frac{ |\Phi_{n,1}|^q }{(\varsigma s_n)^{q-2}}.
\end{align*}
Therefore,
\begin{align*}
    \frac{1}{s_n^2} \sum_{i=1}^{n} \mathbb{E}\!\left[ \Phi_{n,i}^2 \mathbb{I}\!\left( |\Phi_{n,i}|>\varsigma s_n \right) \right] \leq  \frac{ \mathbb{E}\left[ |\Phi_{n,1}|^q \right]}{\varsigma^{q-2}s_n^{q-2} \sigma_{\nu,p}^2(t_n)} =  \frac{ \mathbb{E}\left[ |\Phi_{n,1}|^q \right]}{ \varsigma^{q-2} n^{(q-2)/2} \sigma_{\nu,p}^q(t_n)}.
\end{align*}
Thus, putting everything together, we have that
\begin{align*}
    \frac{ \mathbb{E}\left[ |\Phi_{n,1}|^q \right]}{ n^{(q-2)/2} \sigma_{\nu,p}^q(t_n)} \lesssim \frac{ (1-t_n)^{-(q-1)} }{ n^{(q-2)/2} (1-t_n)^{-q/2} } = \left\{ n(1-t_n) \right\}^{-(q-2)/2} \longrightarrow 0,
\end{align*}
which follows under the assumption that $n(1-t_n) \to \infty$. Thus, Lindeberg's condition holds and consequently,
\begin{align*}
     \frac{\sqrt{n}}{\sigma_{\nu,p}(t_n)} (P_n-P) \left\{ \Phi_{\nu,p}(Z;t_n) \right\} \overset{d}{\longrightarrow} N(0,1).
\end{align*}

\medskip 

\textbf{Empirical Process Term:} We next consider $(P_n-P) \{  \widehat{\phi}_{\nu, p}(Z; t_n) -\phi_{\nu, p}(Z; t_n) \}$. Again, we assume all nuisance functions are cross-fitted. Then, conditional on the corresponding training sample, they may be regarded as fixed functions that are independent of the validation observations. Since the number of folds is fixed, it suffices to prove the following bounds within a generic validation fold. Thus, we suppress the fold index notation. Now, we define $\Delta_\mu=\widehat\mu-\mu$, $\Delta_H=\widehat H_{p,n}-H_{p,n}$, and $ \Delta_\Gamma = \widehat\Gamma_{p,t_n}-\Gamma_{p,t_n}.$ After grouping like terms, we have the exact decomposition $\widehat{\phi}_{\nu, p}(Z; t_n) -\phi_{\nu, p}(Z; t_n) = D_{1,n}(Z) + D_{2,n}(Z) + D_{3,n}(Z) + D_{4,n}(Z)$ where
\begin{align*}
    D_{1, n}(Z) &= \Delta_H(X,A)(Y-\mu(X,A)), \\ 
    D_{2, n}(Z) &= - H_{p,n}(X,A)\Delta_\mu(X,A),  \\
    D_{3, n}(Z) &= \Delta_\Gamma(X,A), \\
    D_{4, n}(Z) &= - \Delta_H(X,A)\Delta_\mu(X,A).
\end{align*}
The first three terms are first order in the nuisance-estimation errors and can be controlled through their conditional second moments. The fourth term is a second-order product, which we handle separately through its conditional first moment. We begin with bounding the $L_2$ norm of $D_{1,n}$. Since $\mathbb{E}[Y-\mu(X,A)\mid X,A]=0$, it follows that
\begin{align*}
    \|D_{1,n}\|_2^2 = \mathbb{E}\!\left[ \Delta_H(X,A)^2 \sigma^2(X,A) \right] \lesssim \|\Delta_H\|_2^2.
\end{align*}
Next, because
\begin{align*}
      |H_{p,n}(X,A)| \lesssim \frac{\mathbb{I}(A\in\mathcal A_{t_n})}{1-t_n},
\end{align*}
it follows that
\begin{align*}
    \|D_{2,n}\|_2^2 &= \mathbb{E}\!\left[H_{p,n}(X,A)^2 \Delta_\mu(X,A)^2 \right] \\
    &\lesssim \frac{1}{(1-t_n)^2} \mathbb{E}\!\left[ \Delta_\mu(X,A)^2 \mathbb{I}(A\in\mathcal A_{t_n}) \right] \\
    &= \frac{1}{1-t_n} \|\widehat\mu-\mu\|_{t_n}^2.
\end{align*}
Finally, by definition we have $ \|D_{3,n}\|_2 = \|\widehat\Gamma_{p,t_n}-\Gamma_{p,t_n}\|_2.$ Now, define the linear component
\begin{align*}
     g_n(Z) = D_{1,n}(Z) + D_{2,n}(Z) + D_{3,n}(Z).
\end{align*}
Then, applying the triangle inequality and our previously established bounds, it immediately follows that
\begin{align*}
    \sqrt{1-t_n}\,\|g_n\|_2 \lesssim \sqrt{1-t_n}\, \left\| \frac{\widehat\omega_{p,t_n}}{\widehat\pi} - \frac{\omega_{p,t_n}}{\pi} \right\|_2 + \|\widehat\mu-\mu\|_{t_n} + \sqrt{1-t_n}\, \|\widehat\Gamma_{p,t_n}-\Gamma_{p,t_n}\|_2,
\end{align*}
which we assume is $o_P(1)$. Again, we let $\mathcal{T}_n$ denote the training sample associated with the generic validation fold. Conditional on $\mathcal T_n$, the function $g_n$ is fixed and independent of the validation observations. Therefore,
\begin{align*}
    \mathbb{E}\!\left[\left(\frac{\sqrt{n}}{\sigma_{\nu,p}(t_n)} (P_n-P)g_n \right)^2 \mid \mathcal{T}_n \right] =  \frac{\mathbb{V}(g_n(Z)\mid\mathcal{T}_n)}{\sigma_{\nu,p}^2(t_n)} \leq \frac{\|g_n\|_2^2}{\sigma_{\nu,p}^2(t_n)} \lesssim (1-t_n)\|g_n\|_2^2 = o_P(1),
\end{align*}
where the penultimate inequality uses $\sigma_{\nu,p}^2(t_n)\asymp(1-t_n)^{-1}$. The conditional Chebyshev inequality now yields
\begin{align*}
    \frac{\sqrt{n}}{\sigma_{\nu,p}(t_n)} (P_n-P)g_n = o_P(1).
\end{align*}
From here, all that is left is to control $D_{4,n}$. Applying the Cauchy--Schwarz inequality, we find that
\begin{align*}
    P\{|D_{4,n}|\} &= \mathbb{E}\!\left[ |\Delta_H(X,A)\Delta_\mu(X,A)| \right] \\
    &\leq \mathbb{E}\!\left[ \Delta_\mu(X,A)^2 \mathbb{I}(A\in\mathcal A_{t_n})\right]^{1/2} \|\Delta_H\|_2 \\
    &= \|\widehat\mu-\mu\|_{t_n} \left[ \sqrt{1-t_n} \left\| \frac{\widehat\omega_{p,t_n}}{\widehat\pi} - \frac{\omega_{p,t_n}}{\pi} \right\|_2 \right] \\
    &= o_P\!\left\{ (n(1-t_n))^{-1/2} \right\}.
\end{align*}
Again conditioning on $\mathcal T_n$, note the elementary inequality $\mathbb{E}\!\left[ |(P_n-P)f| \mid  \mathcal{T}_n \right] \leq 2P|f|$ holds for every fixed integrable function $f$. Applying this inequality with $f=D_{4,n}$ gives
\begin{align*}
    \mathbb{E}\!\left[ \left| \frac{\sqrt{n}}{\sigma_{\nu,p}(t_n)} (P_n-P)D_{4,n} \right| \mid \mathcal{T}_n \right] \leq \frac{2\sqrt n}{\sigma_{\nu,p}(t_n)} P\{|D_{4,n}| \} \lesssim \sqrt{n(1-t_n)}\,P\{|D_{4,n}|\} = o_P(1),
\end{align*}
where we again used $\sigma_{\nu,p}^2(t_n)\asymp(1-t_n)^{-1}$.
Conditional Markov's inequality therefore implies
\begin{align*}
     \frac{\sqrt{n}}{\sigma_{\nu,p}(t_n)} (P_n-P)D_{4,n} = o_P(1).
\end{align*}
Putting everything together, it now follows that
\begin{align*}
      \frac{\sqrt{n}}{\sigma_{\nu,p}(t_n)} (P_n-P) \left\{ \widehat{\phi}_{\nu, p}(Z; t_n) -\phi_{\nu, p}(Z; t_n) \right\} = o_P(1).
\end{align*}
\medskip 

\textbf{Remainder Bound:} Finally, we must control the remainder of the von Mises decomposition. Observe that
\begin{align*}
      R_{p,n} = \mathbb{E}\!\left[ \widehat\phi_{\nu,p}(Z;t_n) - \phi_{\nu,p}(Z;t_n)\right] = \mathbb{E}\!\left[ \widehat\phi_{\nu,p}(Z;t_n) - \theta_{p, n}\right]
\end{align*}
because $\mathbb{E}\!\left[\phi_{\nu,p}(Z;t_n) \right] = \mathbb{E}\!\left[ \Gamma_{p,t_n}(X,A) \right] = \theta_{p,n}$. Next, let $ \Delta_\mu(x,a) = \widehat\mu(x,a)-\mu(x,a)$ and note that by iterated expectation 
\begin{align*}
    \mathbb{E}\!\left[\widehat H_{p,n}(X,A)(Y-\widehat\mu(X,A))\right] = - \mathbb{E}\!\left[ \widehat H_{p,n}(X,A) \Delta_\mu(X,A)\right].
\end{align*}
Similarly, because $\mathbb{E}\!\left[H_{p,n}(X,A)(Y-\mu(X,A))\right] = 0$, it follows that
\begin{align*}
    R_{p,n} = - \mathbb{E}\!\left[ \widehat H_{p,n}(X,A) \Delta_\mu(X,A)\right] + \mathbb{E}\!\left[ \widehat\Gamma_{p,t_n}(X,A) - \Gamma_{p,t_n}(X,A) \right].
\end{align*}
Next, recall that 
\begin{align*}
     \Gamma_{p,t_n}(x,a) = \sum_{k=0}^{p} \frac{(1-t_n)^k}{k!} (a_\star-a)^k \mu^{(k)} \!\left(x,(1-t_n)a + t_n a_\star \right).
\end{align*}
Under the assumption that the derivative estimators are obtained from the same fitted function $\widehat\mu$, we may write $\widehat\mu^{(k)}-\mu^{(k)} = (\widehat\mu-\mu)^{(k)} = \Delta_\mu^{(k)}$. Thus,
\begin{align*}
    \mathbb{E}\!\left[ \widehat\Gamma_{p,t_n}(X,A) - \Gamma_{p,t_n}(X,A)\right] = \sum_{k=0}^{p} \frac{(1-t_n)^k}{k!} \mathbb{E}\!\left[(a_\star-A)^k \Delta_\mu^{(k)}\!\left( X,(1-t_n)A+t_na_\star\right) \right].
\end{align*}
From here, to rewrite each term in this sum, we fix $x$ and again let $F_{t_n}(a) = (1-t_n)a+t_na_\star$. For $u\in\mathcal{A}_{t_n}$ we define
\begin{align*}
     Q_{k,t_n}(u\mid x) = \left(\frac{a_\star-u}{1-t_n} \right)^k \nu_{t_n}(u\mid x).
\end{align*}
Applying the change of variables $u=F_{t_n}(a)$ we can then see that 
\begin{align*}
    \int_{\mathcal{A}} (a_\star-a)^k \Delta_\mu^{(k)} \!\left(x,F_{t_n}(a)\right) \pi(a\mid x) da &= \int_{\mathcal{A}_{t_n}} \Delta_\mu^{(k)}(x,u) Q_{k,t_n}(u\mid x)du \\
    &\overset{(i)}{=}  \int_{\mathcal{A}} \Delta_\mu(x,u) \eta_{k,t_n}(x,u)du,
\end{align*}
where $(i)$ follows after applying integration by parts $k$ times, and using the fact that
\begin{align*}
     \eta_{k,t_n}(x,u) = (-1)^k \frac{\partial^k}{\partial u^k} Q_{k,t_n}(u\mid x)
\end{align*}
as shown in the proof of \cref{derivative_EIF}. Consequently, it follows that
\begin{align*}
    \mathbb{E}\!\left[ \widehat\Gamma_{p,t_n}(X,A) - \Gamma_{p,t_n}(X,A)\right] &= \mathbb{E}\!\left[\int_{\mathcal{A}} \Delta_\mu(X,u) \left(\sum_{k=0}^{p} \frac{(1-t_n)^k}{k!} \eta_{k,t_n}(X,u) \right) du \right] \\
    &=  \mathbb{E}\!\left[ \int_{\mathcal{A}} \Delta_\mu(X,u) \omega_{p,t_n}(X,u) du \right] \\
    &=  \mathbb{E}\!\left[\frac{\omega_{p,t_n}(X,A)}{\pi(A\mid X)} \Delta_\mu(X,A) \right] \\
    &= \mathbb{E}\!\left[ H_{p,n}(X,A) \Delta_\mu(X,A) \right].
\end{align*}
Then, substituting this result into our definition of $R_{p, n}$ yields
\begin{align*}
    R_{p,n} &= - \mathbb{E}\!\left[\left(\widehat H_{p,n}(X,A)-H_{p,n}(X,A)\right) \Delta_\mu(X,A) \right] \\
    &= \mathbb{E}\!\left[(\mu(X,A)-\widehat\mu(X,A))\left(\frac{\widehat\omega_{p,t_n}(X,A)}{\widehat\pi(A\mid X)} - \frac{\omega_{p,t_n}(X,A)}{\pi(A\mid X)} \right)\right].
\end{align*}
Now, all that remains is to establish a bound on $R_{p,n}$. Note that by construction, $\omega_{p,t_n}$ and $\widehat\omega_{p,t_n}$ both vanish outside
$\mathcal{A}_{t_n}$. Thus, applying the Cauchy--Schwarz inequality, it follows that
\begin{align*}
     |R_{p,n}| &\leq \mathbb{E}\!\left[ \Delta_\mu(X,A)^2 \mathbb{I}(A\in\mathcal A_{t_n})\right]^{1/2} \mathbb{E}\!\left[\left(\widehat H_{p,n}(X,A)-H_{p,n}(X,A) \right)^2 \right]^{1/2} \\
    &= \|\widehat\mu-\mu\|_{t_n} \left(\sqrt{1-t_n} \left\| \frac{\widehat\omega_{p,t_n}}{\widehat\pi} - \frac{\omega_{p,t_n}}{\pi} \right\|_2 \right).
\end{align*}
Finally, because $\sigma_{\nu,p}^2(t_n)\asymp(1-t_n)^{-1}$,
\begin{align*}
    \frac{\sqrt{n}}{\sigma_{\nu,p}(t_n)} |R_{p,n}| &\lesssim
    \sqrt{n(1-t_n)}|R_{p,n}| =o_P(1),
\end{align*}
thereby completing the proof of the remainder bound. 

\medskip 

\textbf{Sharp Intervention:} Now, we consider the approximation argument for the sharp intervention. Recall that
\begin{align*}
    \theta_{p,n} &=\mathbb{E}\!\left[\Gamma_{p,t_n}(X,A)\right] \\
    &= \sum_{k=0}^{p} \frac{(1-t_n)^k}{k!} \mathbb{E}\!\left[ (a_\star-A)^k \mu^{(k)} \!\left(X,(1-t_n)A+t_na_\star\right) \right] \\
    &= \sum_{k=0}^{p} \frac{(1-t_n)^k}{k!} \psi_\nu^{(k)}(t_n).
\end{align*}
Thus, $\theta_{p,n}$ is the order-$p$ Taylor approximation to $\psi_\nu(1)=\psi_\star$, expanded around $t_n$. Then, observe that under our smoothness envelope assumption,
\begin{align*}
    \left|\psi_\nu^{(p+1)}(s)\right| &= \left| \mathbb{E}\!\left[ (a_\star-A)^{p+1} \mu^{(p+1)}\!\left(X,(1-s)A+sa_\star\right) \right] \right| \leq \underbrace{\sup_{a\in\mathcal A}|a_\star-a|^{p+1}\mathbb{E}[M_\mu(X)]}_{C_{p+1}} <\infty.
\end{align*}
The assumed smoothness of $\mu$ along with dominated convergence also gives
\begin{align*}
    \psi_\nu(s) = \mathbb{E}\!\left[\mu\!\left(X,(1-s)A+sa_\star\right) \right] \longrightarrow \mathbb{E}[\mu(X,a_\star)] = \psi_\star.
\end{align*}
Consequently, Taylor's theorem with an integral remainder yields
\begin{align*}
    \psi_\star-\theta_{p,n} = \frac{1}{p!}
    \int_{t_n}^{1} (1-s)^p\psi_\nu^{(p+1)}(s)ds.
\end{align*}
Then, for all sufficiently large $n$,
\begin{align*}
    |\theta_{p,n}-\psi_\star| &\leq \frac{1}{p!} \int_{t_n}^{1} (1-s)^p \left|\psi_\nu^{(p+1)}(s)\right|ds \leq \frac{C_{p+1}}{p!} \int_{t_n}^{1}(1-s)^p ds = \frac{C_{p+1}}{(p+1)!} (1-t_n)^{p+1}.
\end{align*}
Since $\sigma_{\nu,p}^2(t_n)\asymp(1-t_n)^{-1}$, it follows that
\begin{align*}
    \frac{\sqrt{n}}{\sigma_{\nu,p}(t_n)} |\theta_{p,n}-\psi_\star| \lesssim \sqrt{n(1-t_n)}(1-t_n)^{p+1} = \left( n(1-t_n)^{2p+3} \right)^{1/2} \longrightarrow 0.
\end{align*}
With this approximation-bias result complete, we now combine all parts to conclude the proof.

\medskip
\noindent\textbf{Conclusion:}
We now combine the oracle, empirical-process, remainder, and approximation-bias arguments. To now make the aggregation across cross-fitting folds explicit, let $\mathcal{I}_1,\ldots,\mathcal{I}_J$ denote the validation folds, where $J$ is fixed, and write $n_j=|\mathcal{I}_j|$. Let $\mathcal{T}_{n,j}$ denote the training sample for fold $j$, and let $\widehat\phi_{\nu,p}^{(j)}$ be the score constructed from the nuisance functions fitted on $\mathcal{T}_{n,j}$. Define
\begin{align*}
     d_{n,j}(z) = \widehat\phi_{\nu,p}^{(j)}(z;t_n) - \phi_{\nu,p}(z;t_n),
\end{align*}
and $R_{p,n}^{(j)} = \mathbb{E}\!\left[d_{n,j}(Z) \mid \mathcal{T}_{n,j}\right]$, where $Z$ in the conditional expectation is a fresh observation from the true distribution, independent of the training sample. Thus, $R_{p,n}^{(j)}$ is the fold-specific version of the remainder controlled above. Then, define the full empirical-process error and averaged remainder as
\begin{align*}
    \mathcal{E}_{p,n} = \frac{1}{n} \sum_{j=1}^{J} \sum_{i\in\mathcal{I}_j}\left( d_{n,j}(Z_i)-R_{p,n}^{(j)} \right) \qquad \text{and} \qquad \overline{R}_{p,n} = \sum_{j=1}^{J} \frac{n_j}{n}R_{p,n}^{(j)}.
\end{align*}
Because $\widehat\psi_{\nu,p}^{\,\mathrm{BC}}(t_n) =\frac{1}{n} \sum_{j=1}^{J} \sum_{i\in\mathcal{I}_j} \widehat\phi_{\nu,p}^{(j)}(Z_i;t_n)$ and
$\Phi_{\nu,p}(Z;t_n)=\phi_{\nu,p}(Z;t_n)-\theta_{p,n}$, we have the exact decomposition
\begin{align*}
    \widehat\psi_{\nu,p}^{\,\mathrm{BC}}(t_n)-\psi_\star = \frac{1}{n} \sum_{i=1}^{n}\Phi_{\nu,p}(Z_i;t_n) + \mathcal{E}_{p,n} + \overline{R}_{p,n} + \theta_{p,n}-\psi_\star.
\end{align*}
The preceding empirical-process and remainder bounds imply
\begin{align*}
    \frac{\sqrt{n}}{\sigma_{\nu,p}(t_n)} \mathcal{E}_{p,n} = o_P(1),\qquad \frac{\sqrt{n}}{\sigma_{\nu,p}(t_n)} \overline{R}_{p,n} = o_P(1).
\end{align*}
Therefore, it then follows that
\begin{align*}
    \frac{\sqrt{n}}{\sigma_{\nu,p}(t_n)} \left( \widehat\psi_{\nu,p}^{\,\mathrm{BC}}(t_n)-\psi_\star \right) = \frac{1}{\sqrt{n}\sigma_{\nu,p}(t_n)} \sum_{i=1}^{n}\Phi_{\nu,p}(Z_i;t_n) + o_P(1) \overset{d}{\longrightarrow} N(0,1),
\end{align*}
where the final step follows from the oracle central limit theorem and Slutsky's theorem. Finally, the variance calculation gives
\begin{align*}
     (1-t_n)\sigma_{\nu,p}^2(t_n) \longrightarrow V_{p,\star}\in(0,\infty).
\end{align*}
Therefore,
\begin{align*}
    \sqrt{n(1-t_n)} \left( \widehat\psi_{\nu,p}^{\,\mathrm{BC}}(t_n)-\psi_\star \right) = \sqrt{1-t_n}\sigma_{\nu,p}(t_n) \frac{\sqrt{n}}{\sigma_{\nu,p}(t_n)} \left( \widehat\psi_{\nu,p}^{\,\mathrm{BC}}(t_n)-\psi_\star \right) \overset{d}{\longrightarrow} N(0,V_{p,\star})
\end{align*}
by another application of Slutsky's theorem.

\end{proof}

\subsection{Proof of \cref{fr_geodesic}} \label{fr_geodesic_proof}

\begin{proof}[\textbf{Proof:}] \cref{fr_geodesic} is the standard great-circle interpolation on the unit sphere in square-root coordinates; see, for example, \citet{kurtek2015}. In this proof, we give a short verification to show the geodesic also applies to probability measures that are not Lebesgue absolutely continuous. To begin, write
\begin{align*}
     s_P=\sqrt{\frac{dP}{d\lambda}}, \qquad s_Q=\sqrt{\frac{dQ}{d\lambda}},
\end{align*}
and $\langle s_P,s_Q\rangle_{L^2(\lambda)}=\cos(\theta)$. Because $P\neq Q$, it follows that $\theta\in(0,\pi/2]$. Next, define
\begin{align*}
     \zeta = \frac{s_Q-\cos(\theta)s_P}{\sin(\theta)}.
\end{align*}
Then $\{s_P, \zeta\}$ is an orthonormal set and $s_Q=\cos(\theta)s_P+\sin(\theta)\zeta$. Using the angle-addition identity,
\begin{align*}
    s_t = \frac{\sin\{(1-t)\theta\}}{\sin(\theta)}s_P + \frac{\sin(t\theta)}{\sin(\theta)}s_Q = \cos(t\theta)s_P+\sin(t\theta) \zeta,
\end{align*}
it then follows immediately that $\|s_t\|_{L^2(\lambda)}=1$. Moreover, the two coefficients are nonnegative for $t\in[0,1]$, and hence $s_t \geq 0$ $\lambda$-almost everywhere. Consequently, $s_t^2\,d\lambda$ defines a probability measure, which we denote by $\nu_t^{\mathrm{SH}}$. Finally, for any $r,t\in[0,1]$, orthonormality gives
\begin{align*}
    \left\langle s_r,s_t\right\rangle_{L^2(\lambda)} &= \cos(r\theta)\cos(t\theta) + \sin(r\theta)\sin(t\theta) = \cos\{(t-r)\theta\}.
\end{align*}
Since $|t-r|\theta\in[0,\pi/2]$, it follows that
\begin{align*}
     \theta\!\left(\nu_r^{\mathrm{SH}}, \nu_t^{\mathrm{SH}} \right) = \arccos\!\left( \left\langle s_r,s_t\right\rangle_{L^2(\lambda)} \right) = |t-r|\theta.
\end{align*}
Thus the path is a constant-speed minimizing geodesic from $P$ to $Q$. The construction is independent of the choice of $\lambda$ because both the Hellinger affinity and the square-root measure obtained by squaring $s_t$ are invariant under a change of common dominating measure.
    
\end{proof}

\subsection{Proof of \cref{fr_pointmass}} \label{fr_pointmass_proof}

\begin{proof}[\textbf{Proof:}]

The assumption that $\Pi_x(\{a_\star\})=0$ implies that $\Pi_x$ and $\delta_{a_\star}$ are mutually singular. Therefore their
square-root representations are orthogonal under any common
dominating measure, so their Hellinger affinity is zero and consequently,
\begin{align*}
     \theta(\Pi_x,\delta_{a_\star}) = \arccos(0) = \frac{\pi}{2}.
\end{align*}
Then, applying \cref{fr_geodesic} gives
\begin{align*}
     \sqrt{\frac{ d\nu_t^{\mathrm{SH}}(\,\cdot\mid x)}{d(\Pi_x+\delta_{a_\star})}} = \cos\left(\frac{\pi t}{2}\right) \sqrt{\frac{d\Pi_x}{d(\Pi_x+\delta_{a_\star})}} + \sin\left(\frac{\pi t}{2}\right) \sqrt{\frac{d\delta_{a_\star}}{d(\Pi_x+\delta_{a_\star})}}.
\end{align*}
Note that these two terms have disjoint supports; thus, squaring produces no cross term and gives us
\begin{align*}
    \nu_t^{\mathrm{SH}}(\,\cdot\mid x) = \cos^2\left(\frac{\pi t}{2}\right)\Pi_x + \sin^2\left(\frac{\pi t}{2}\right)\delta_{a_\star}.
\end{align*}
\end{proof}

\end{document}